\pdfoutput=1
\documentclass[final,5p,twocolumn,authoryear]{elsarticle}
\usepackage{amsmath,amssymb,amsfonts}
\usepackage{bm}
\usepackage{graphicx}
\usepackage{booktabs}
\usepackage{multirow}
\usepackage{siunitx}
\usepackage{xcolor}
\usepackage{placeins}
\usepackage{float}
\usepackage{balance}
\usepackage[hidelinks]{hyperref}

\newcommand{\AsCovDeployedNongross}{91.4\%}
\newcommand{\AsCovScaledNongross}{91.9\%}
\newcommand{\AsRatioMedianLater}{2\%}
\newcommand{\AsRatioMedianShotOne}{32\%}
\newcommand{\AsRatioPFiveLater}{10\%}
\newcommand{\CabAlphaOneCovNongross}{72\%}
\newcommand{\CabAlphaOneFalseStop}{25/52}
\newcommand{\CabAlphaOneFalseStopStable}{3/37}
\newcommand{\CabAlphaOneMinimaxCorrect}{4.87}
\newcommand{\CabAlphaOneRatio}{0.66}
\newcommand{\CabBlockRelMax}{6.3\%}
\newcommand{\CabBlockRelPNinetyNine}{3.0\%}
\newcommand{\CabFlipMaxMm}{0.03}
\newcommand{\CabFlips}{5}
\newcommand{\CabFullCovNongross}{91\%}
\newcommand{\CabFullFalseStop}{11/38}
\newcommand{\CabFullFalseStopStable}{1/29}
\newcommand{\CabFullMinimaxCorrect}{4.88}
\newcommand{\CabValMedian}{0.04\%}
\newcommand{\CalibAlpha}{2.41}
\newcommand{\DiagDepCertN}{10}
\newcommand{\DiagDepCertOptClear}{7}
\newcommand{\DiagDepEstClear}{21}
\newcommand{\DiagDepEstClearEarly}{13}
\newcommand{\DiagDepN}{56}
\newcommand{\DiagDepOptClear}{26}
\newcommand{\DiagDepTie}{9}
\newcommand{\FdSOneAtFlagsSpecAmbiguity}{40\%}
\newcommand{\FdSOneAtFlagsSpecPriorDistance}{45\%}
\newcommand{\FdSOneAtFlagsSpecShift}{25\%}
\newcommand{\FdSOneAtInstabSpecAmbiguity}{81\%}
\newcommand{\FdSOneAtInstabSpecShift}{86\%}
\newcommand{\FdSOneAucMax}{0.91}
\newcommand{\FdSOneAucMin}{0.83}
\newcommand{\FdSOneBpCaught}{90\%}
\newcommand{\FdSOneBpCorrectN}{1434}
\newcommand{\FdSOneBpFalseAlarmN}{304}
\newcommand{\FdSOneBpFalseAlarmPct}{21\%}
\newcommand{\FdSOneBpGross}{13\%}
\newcommand{\FdSOneBpN}{1652}
\newcommand{\FdSOneConfPct}{32\%}
\newcommand{\FdSOneFlagsPpv}{76\%}
\newcommand{\FdSOneFlagsSens}{41\%}
\newcommand{\FdSOneFlagsSensCI}{[38, 44]}
\newcommand{\FdSOneFlagsSpec}{98\%}
\newcommand{\FdSOneGross}{14\%}
\newcommand{\FdSOneInstabSens}{78\%}
\newcommand{\FdSOneInstabSensCI}{[76, 81]}
\newcommand{\FdSOneInstabSpec}{79\%}
\newcommand{\FdSOneInstabSpecCI}{[78, 81]}
\newcommand{\FdSOneInstabSpecLater}{89\%}
\newcommand{\FdSOneInstabSpecShotTwo}{31\%}
\newcommand{\FdSOneInstabilityAuc}{0.85}
\newcommand{\FdSOneLocalBoundAuc}{0.84}
\newcommand{\FdSOneLocalMisfitAuc}{0.75}
\newcommand{\FdSOneMmAucMax}{0.92}
\newcommand{\FdSOneMmAucMin}{0.84}
\newcommand{\FdSOneMmN}{1105}
\newcommand{\FdSOneShiftAuc}{0.91}
\newcommand{\FdSOneShiftAucCI}{[0.90, 0.92]}
\newcommand{\FdSOneStratAucMax}{0.94}
\newcommand{\FdSOneStratAucMin}{0.87}
\newcommand{\FdSOneStratFlagsSens}{32\%}
\newcommand{\FdSOneStratGrossN}{679}
\newcommand{\FdSOneStratInstabSens}{86\%}
\newcommand{\FdSOneStratInstabSpec}{84\%}
\newcommand{\FdSOneStratLocalBoundAuc}{0.81}
\newcommand{\FdSOneStratLocalMisfitAuc}{0.84}
\newcommand{\FdSOneStratN}{10846}
\newcommand{\FdSOneStratShiftAuc}{0.94}
\newcommand{\FdTSOneFail}{19\%}
\newcommand{\FdTSOneShiftAuc}{0.90}
\newcommand{\FdTSOneStratLocalBoundAuc}{0.83}
\newcommand{\FdTSOneStratShiftAuc}{0.93}
\newcommand{\FigMethodCaseBoundShotOne}{23.7}
\newcommand{\FigMethodDepthPct}{0.3\%}
\newcommand{\FigMethodFlagTag}{F3}
\newcommand{\FigMethodLocalBound}{1.93}
\newcommand{\FigMethodLocalShot}{3}
\newcommand{\FigMethodShift}{2.29}
\newcommand{\FigMethodShiftBound}{1.73}
\newcommand{\FigMethodShiftTag}{F4}
\newcommand{\FigMethodStableShot}{4}
\newcommand{\FigMethodTrueStable}{0.67}
\newcommand{\FigTwoBoundDopt}{2.6}
\newcommand{\FigTwoBoundFixed}{4.8}
\newcommand{\FigTwoBoundMinimax}{2.4}
\newcommand{\FigTwoCase}{031}
\newcommand{\FigTwoFixedRank}{300}
\newcommand{\FigTwoFixedWorstBound}{4.7}
\newcommand{\FigTwoFixedWorstMinimax}{2.0}
\newcommand{\FigTwoFixedWorstTag}{F2}
\newcommand{\FigTwoFixedWorstVol}{45}
\newcommand{\FlagsSOneBlocked}{93}
\newcommand{\FlagsSOneBlockedFalse}{85}
\newcommand{\GrossSOneShotOne}{40\%}
\newcommand{\GrossSOneShotTwoPlus}{14\%}
\newcommand{\HwMaxDiff}{17.7}
\newcommand{\HwSameCertN}{13}
\newcommand{\HwSameTrueN}{14}
\newcommand{\HwTraj}{14}
\newcommand{\HzObCostFixed}{0.83}
\newcommand{\HzObCostMax}{1.43}
\newcommand{\HzObCostMin}{0.43}
\newcommand{\HzObCostMinimax}{0.87}
\newcommand{\HzObDiffTrueDoptHolm}{0.160}
\newcommand{\HzObDiffTrueFixed}{+0.10}
\newcommand{\HzObDiffTrueFixedCI}{[$-$0.47, 0.63]}
\newcommand{\HzObDiffTrueFixedHolm}{1.000}
\newcommand{\HzObDoptTrue}{2.57}
\newcommand{\HzObFixedTrue}{2.97}
\newcommand{\HzObMinimaxTrue}{3.07}
\newcommand{\HzObVsFixedTrueDoptHolm}{0.502}
\newcommand{\HzSFourMinimaxCert}{2.20}
\newcommand{\HzSFourMinimaxCertStable}{2.70}
\newcommand{\HzSFourMinimaxTrue}{1.40}
\newcommand{\HzSFourRandomCert}{4.32}
\newcommand{\HzSOneDiffCertFixed}{$-$0.52}
\newcommand{\HzSOneDiffCertFixedCI}{[$-$0.73, $-$0.32]}
\newcommand{\HzSOneDiffCertFixedHolm}{$<$0.001}
\newcommand{\HzSOneDiffCertStableFixed}{$-$0.30}
\newcommand{\HzSOneDiffCertStableFixedHolm}{0.184}
\newcommand{\HzSOneDiffCorrectFixed}{$-$0.37}
\newcommand{\HzSOneDiffCorrectFixedCI}{[$-$0.62, $-$0.13]}
\newcommand{\HzSOneDiffCorrectFixedHolm}{0.043}
\newcommand{\HzSOneDiffTrueFixed}{$-$0.02}
\newcommand{\HzSOneDiffTrueFixedCI}{[$-$0.38, 0.37]}
\newcommand{\HzSOneDiffTrueFixedHolm}{1.000}
\newcommand{\HzSOneDoptCorrect}{4.80}
\newcommand{\HzSOneDoptTrue}{3.50}
\newcommand{\HzSOneFixedCert}{4.72}
\newcommand{\HzSOneFixedCorrect}{5.25}
\newcommand{\HzSOneFixedTrue}{3.70}
\newcommand{\HzSOneGradientTrue}{3.32}
\newcommand{\HzSOneMinimaxCert}{4.20}
\newcommand{\HzSOneMinimaxCorrect}{4.88}
\newcommand{\HzSOneMinimaxTrue}{3.68}
\newcommand{\HzSOneMinimaxTrueAtStopPNinety}{18.6}
\newcommand{\HzSOneVsFixedTrueGradient}{$-$0.38}
\newcommand{\HzSOneVsFixedTrueGradientCI}{[$-$0.70, $-$0.10]}
\newcommand{\HzSOneVsFixedTrueMinHolm}{0.142}
\newcommand{\HzSThreeDiffCertFixed}{$-$0.73}
\newcommand{\HzSThreeDiffCertFixedHolm}{0.011}
\newcommand{\HzSThreeDiffCorrectFixed}{$-$0.33}
\newcommand{\HzSThreeDiffTrueFixed}{+0.27}
\newcommand{\HzSTwoDiffCertFixed}{$-$0.60}
\newcommand{\HzSTwoDiffCertFixedHolm}{0.060}
\newcommand{\HzSTwoDiffCorrectFixed}{$-$0.33}
\newcommand{\HzSTwoDiffTrueFixed}{+0.13}
\newcommand{\MCAwareTraceMax}{1.08}
\newcommand{\MCAwareTraceMin}{0.95}
\newcommand{\MCCrbTraceMax}{3.3}
\newcommand{\MCCrbTraceMin}{1.9}
\newcommand{\MCPlainTraceMin}{0.69}
\newcommand{\ObCovNongross}{89\%}
\newcommand{\ObDepFalsePct}{13\%}
\newcommand{\ObDepGrossShotThree}{20\%}
\newcommand{\ObDepGrossShotTwo}{26\%}
\newcommand{\ObFalsePct}{5\%}
\newcommand{\ObGrossShotThree}{2\%}
\newcommand{\ObGrossShotTwo}{6\%}
\newcommand{\ObPredMinimaxCenterTrueRatio}{1.13}
\newcommand{\ObPredMinimaxEdgeTrueRatio}{1.26}
\newcommand{\OrcCases}{30}
\newcommand{\OrcDepGrossRate}{18\%}
\newcommand{\OrcGrossRate}{20\%}
\newcommand{\PredAoptTrueRatio}{1.70}
\newcommand{\PredCrbKappaCert}{17}
\newcommand{\PredCrbKappaMedianBound}{1.81}
\newcommand{\PredCrbWhiteCert}{21}
\newcommand{\PredCrbWhiteMedianBound}{1.58}
\newcommand{\PredCvCovMax}{94\%}
\newcommand{\PredCvCovMin}{93\%}
\newcommand{\PredDoptPostRatio}{1.03}
\newcommand{\PredDoptTrueRatio}{1.07}
\newcommand{\PredMinimaxCenterTrueRatio}{1.18}
\newcommand{\PredMinimaxEdgeFrac}{41\%}
\newcommand{\PredMinimaxEdgeTrueRatio}{1.54}
\newcommand{\PredMinimaxPostRatio}{1.04}
\newcommand{\PredMinimaxTrueRatio}{1.32}
\newcommand{\PredSpecCert}{26}
\newcommand{\PredSpecCovAlphaOne}{74\%}
\newcommand{\PredSpecFalseCert}{6}
\newcommand{\PredSpecMedianBound}{1.53}
\newcommand{\RealAlphaMax}{9.39}
\newcommand{\RealAlphaMin}{3.76}
\newcommand{\RealRandomCertified}{67\%}
\newcommand{\RealSpecDoptFewer}{3}
\newcommand{\RealSpecMaxAngleFewer}{3}
\newcommand{\RealSpecN}{5}
\newcommand{\RealSpecRandomFewer}{5}
\newcommand{\RealTwoShotMinimaxCert}{80\%}
\newcommand{\RealTwoShotMinimaxTrue}{100\%}
\newcommand{\RealTwoShotPolyPoseCert}{40\%}
\newcommand{\RealTwoShotPolyPoseTrue}{80\%}
\newcommand{\RealismLargeMedian}{0.45}
\newcommand{\RealismRealMedian}{0.45}
\newcommand{\RealismRealRelVar}{0.15}
\newcommand{\RealismRelVar}{0.19}
\newcommand{\SFiveCases}{15}
\newcommand{\SFiveCovGapLate}{3--4}
\newcommand{\SFiveCovGapShotTwo}{2}
\newcommand{\SFiveCovNongross}{86.4\%}
\newcommand{\SFiveDiceMedian}{0.89}
\newcommand{\SFiveDicePTen}{0.83}
\newcommand{\SFiveGrossRate}{18\%}
\newcommand{\SFiveLocalDiffCI}{$+1$ to $+25$}
\newcommand{\SFiveLocalDiffPts}{$+11$}
\newcommand{\SFiveLocalFalsePct}{30\%}
\newcommand{\SFivePairBeyond}{8.2\%}
\newcommand{\SFivePairFromGross}{41}
\newcommand{\SFivePairGross}{20.1\%}
\newcommand{\SFivePairMedian}{0.69}
\newcommand{\SFivePairN}{1384}
\newcommand{\SFivePairToGross}{39}
\newcommand{\SFiveRefCovNongross}{90.1\%}
\newcommand{\SFiveRefGrossRate}{19\%}
\newcommand{\SFiveRefLocalFalsePct}{19\%}
\newcommand{\SFiveRefPairBeyond}{6.1\%}
\newcommand{\SFiveRefPairGross}{20.2\%}
\newcommand{\SFiveRefPairMedian}{0.63}
\newcommand{\SFiveRefShiftAuc}{0.88}
\newcommand{\SFiveRefStableCert}{31}
\newcommand{\SFiveRefStableFalse}{3}
\newcommand{\SFiveRefStableFalseMid}{0}
\newcommand{\SFiveRefStableFalsePct}{10\%}
\newcommand{\SFiveShiftAuc}{0.88}
\newcommand{\SFiveStableCert}{28}
\newcommand{\SFiveStableClassPath}{2}
\newcommand{\SFiveStableClassPresent}{3}
\newcommand{\SFiveStableDiffCI}{$-1$ to $+28$}
\newcommand{\SFiveStableDiffPts}{$+12$}
\newcommand{\SFiveStableFalse}{6}
\newcommand{\SFiveStableFalseMid}{4}
\newcommand{\SFiveStableFalsePct}{21\%}
\newcommand{\SFiveStableNewDice}{0.77}
\newcommand{\SFiveStableNewErr}{2.06}
\newcommand{\SFiveStableNewRefErr}{1.16}
\newcommand{\SFiveStablePresentCaughtInRef}{2}
\newcommand{\SFourCovUnflagged}{96\%}
\newcommand{\SOneCases}{60}
\newcommand{\SOneCertRegs}{1559}
\newcommand{\SOneCovNongross}{91\%}
\newcommand{\SOneCovUnflagged}{83\%}
\newcommand{\SOneDiffCertFixed}{$-$0.67}
\newcommand{\SOneDiffCertFixedHolm}{$<$0.001}
\newcommand{\SOneFalseRegs}{187}
\newcommand{\SOneFalseViolBodies}{204}
\newcommand{\SOneFalseViolGross}{146}
\newcommand{\SOneFalseViolRatio}{6.3}
\newcommand{\SOneGrossRate}{15\%}
\newcommand{\SOneVolADopt}{2.97}
\newcommand{\SOneVolAMinimax}{3.22}
\newcommand{\SOneVolDDopt}{1.39}
\newcommand{\SOneVolDMinimax}{1.49}
\newcommand{\SThreeCovUnflagged}{88\%}
\newcommand{\STwoCovUnflagged}{76\%}
\newcommand{\SegDiceMedian}{0.88}
\newcommand{\SrSOneFromThreeFalsePct}{21\%}
\newcommand{\SrSOneLocalCert}{468}
\newcommand{\SrSOneLocalCorrect}{362}
\newcommand{\SrSOneLocalEqFalsePct}{25\%}
\newcommand{\SrSOneLocalFalse}{106}
\newcommand{\SrSOneLocalFalseGross}{78}
\newcommand{\SrSOneLocalFalseMid}{28}
\newcommand{\SrSOneLocalFalsePct}{23\%}
\newcommand{\SrSOneLocalFalsePctCI}{16--29\%}
\newcommand{\SrSOneLocalMinimaxCorrect}{27}
\newcommand{\SrSOneLocalMinimaxFalse}{11/38}
\newcommand{\SrSOneLocalMinimaxFalseN}{11}
\newcommand{\SrSOneLocalMinimaxNone}{22}
\newcommand{\SrSOneLocalNone}{372}
\newcommand{\SrSOneLocalPolicyRange}{11--57\%}
\newcommand{\SrSOneLocalTraj}{840}
\newcommand{\SrSOneMOtherRange}{10--15\%}
\newcommand{\SrSOneMShiftCert}{368}
\newcommand{\SrSOneMShiftFalse}{19}
\newcommand{\SrSOneMShiftFalsePct}{5\%}
\newcommand{\SrSOneRandomTraj}{300}
\newcommand{\SrSOneStableCert}{368}
\newcommand{\SrSOneStableCorrect}{345}
\newcommand{\SrSOneStableEqFalsePct}{7\%}
\newcommand{\SrSOneStableFalse}{23}
\newcommand{\SrSOneStableFalseGross}{15}
\newcommand{\SrSOneStableFalseMid}{8}
\newcommand{\SrSOneStableFalsePct}{6\%}
\newcommand{\SrSOneStableFalsePctCI}{4--10\%}
\newcommand{\SrSOneStableFixedCorrect}{21}
\newcommand{\SrSOneStableMinimaxCorrect}{28}
\newcommand{\SrSOneStableMinimaxFalse}{1/29}
\newcommand{\SrSOneStableMinimaxFalseCP}{0.1--17.8\%}
\newcommand{\SrSOneStableMinimaxFalseN}{1}
\newcommand{\SrSOneStableMinimaxNone}{31}
\newcommand{\SrSOneStableNone}{472}
\newcommand{\SrSOneStablePolicyRange}{0--20\%}
\newcommand{\SrSOneTwiceCorrect}{294}
\newcommand{\SrSOneTwiceFalsePct}{10\%}
\newcommand{\SrSThreeLocalFalsePct}{10\%}
\newcommand{\SrSThreeStableFalsePct}{3\%}
\newcommand{\SrSTwoLocalFalsePct}{21\%}
\newcommand{\SrSTwoStableFalsePct}{9\%}
\newcommand{\SuppAlphaDeepdrr}{0.49}
\newcommand{\SuppAlphaLight}{0.10}
\newcommand{\SuppAlphaReal}{3.65}
\newcommand{\SuppCalibBodies}{424}
\newcommand{\SuppCalibCert}{26}
\newcommand{\SuppCalibCvCov}{94\%}
\newcommand{\SuppCalibCvCovCI}{[89\%, 98\%]}
\newcommand{\SuppCalibFalseCert}{6}
\newcommand{\SuppCalibGrossFlagged}{33\%}
\newcommand{\SuppCalibGrossRate}{18\%}
\newcommand{\SuppCalibGrossVolA}{52\%}
\newcommand{\SuppCalibGrossVolB}{15\%}
\newcommand{\SuppCalibGrossVolC}{23\%}
\newcommand{\SuppCalibGrossVolD}{2\%}
\newcommand{\SuppCalibMedianBound}{1.67}
\newcommand{\SuppCalibNonGrossFlag}{3\%}
\newcommand{\SuppCalibRegs}{120}
\newcommand{\SuppFirstTestCommit}{\texttt{b8f2d20}}
\newcommand{\SuppFirstTestRun}{2026-10-29 20:04}
\newcommand{\SuppKHighCases}{13}
\newcommand{\SuppKHighCertDiffFixed}{$-$0.23}
\newcommand{\SuppKHighCertDiffFixedCI}{[$-$0.77, 0.31]}
\newcommand{\SuppKLowCertDiffFixed}{$-$0.81}
\newcommand{\SuppKLowCertDiffFixedCI}{[$-$1.19, $-$0.46]}
\newcommand{\SuppMFourCases}{20}
\newcommand{\SuppMFourDeepdrrMedian}{0.151}
\newcommand{\SuppMFourDeepdrrSuccess}{97\%}
\newcommand{\SuppMFourInverseMedian}{0.0002}
\newcommand{\SuppMFourLightCleanMedian}{0.015}
\newcommand{\SuppMFourLightNoisyMedian}{0.049}
\newcommand{\SuppMFourSeconds}{36}
\newcommand{\SuppOvBodies}{221}
\newcommand{\SuppOvHighDiffDopt}{+0.28}
\newcommand{\SuppOvHighDiffDoptHolm}{0.562}
\newcommand{\SuppOvHighGross}{26\%}
\newcommand{\SuppOvLargeGrossHigh}{27\%}
\newcommand{\SuppOvLargeGrossLow}{1\%}
\newcommand{\SuppOvLowGross}{5\%}
\newcommand{\SuppOvSmallGrossHigh}{25\%}
\newcommand{\SuppOvSmallGrossLow}{14\%}
\newcommand{\SuppOvSpearman}{-0.41}
\newcommand{\SuppRealCovSynthAlpha}{39\%}
\newcommand{\SuppRealGridSetAlpha}{3.61}
\newcommand{\SuppRealGridSetMax}{11.5}
\newcommand{\SuppRealGridSetMedian}{0.44}
\newcommand{\SuppRealRunSetAlpha}{4.08}
\newcommand{\SuppRealRunSetGrossMax}{63}
\newcommand{\SuppRealRunSetGrossMin}{35}
\newcommand{\SuppRealRunSetGrossN}{3}
\newcommand{\SuppRealRunSetMax}{62.8}
\newcommand{\SuppRealRunSetMedian}{0.45}
\newcommand{\SuppRunsScanned}{350}
\newcommand{\SuppSOneStableDiffFixed}{$-$0.57}
\newcommand{\SuppSOneStableDiffFixedHolm}{0.006}
\newcommand{\SuppSTwoRunOneCov}{65\%}
\newcommand{\SuppSTwoRunOneFalsePct}{22\%}
\newcommand{\SuppSTwoRunTwoCov}{76\%}
\newcommand{\SuppSTwoRunTwoFalsePct}{9\%}
\newcommand{\SuppTestUseBefore}{none}
\newcommand{\SvBest}{68\%}
\newcommand{\SvBestRandom}{53\%}
\newcommand{\SvFixed}{43\%}
\newcommand{\SvGradient}{47\%}
\newcommand{\SvMinimax}{32\%}
\newcommand{\SvPerCase}{12}
\newcommand{\SvSepFar}{21\%}
\newcommand{\SvSepMid}{41\%}
\newcommand{\TimePlanMedian}{4.7}
\newcommand{\TimePostMedian}{1.1}
\newcommand{\TimeRegMedian}{67}

\newcommand{\T}{\mathsf{T}}
\newcommand{\R}{\mathbb{R}}
\newcommand{\vxi}{\bm{\xi}}

\graphicspath{{figures/}}
\makeatletter\def\@journal{Medical Image Analysis}\makeatother

\begin{document}

\begin{frontmatter}

\title{When can multi-fragment 2D/3D registration be trusted? Calibrated certification and uncertainty-driven view planning for pelvic fracture fluoroscopy}

\author[vu]{Bowen Xiang\corref{cor1}}
\ead{bowen.xiang@vanderbilt.edu}
\cortext[cor1]{Corresponding author.}
\affiliation[vu]{organization={Department of Biomedical Engineering, Vanderbilt University},
            addressline={2301 Vanderbilt Place},
            city={Nashville},
            state={TN 37235},
            country={USA}}

\begin{abstract}
Registering a segmented CT to fluoroscopy can locate every fracture fragment but can guide surgery only when trusted. After each shot, our closed loop registers all fragments, bounds their errors with a calibrated sandwich covariance, flags implausible or ambiguous ones, and stops or plans the next view. On simulated fluoroscopy of \SOneCases{} held-out fracture CTs with realistic image uncertainty, 95\% bounds covered \SOneCovUnflagged{} of unflagged fragments and \SOneCovNongross{} of those within 5\,mm, but from the second view on \FdSOneGross{} failed by over 5\,mm, a third with bounds of at most 2\,mm. A fragment's shift when a view was added separated tightly bounded failures best (exploratory; AUC \FdSOneStratShiftAuc, bound \FdSOneStratLocalBoundAuc). Requiring that the newest view moved no fragment beyond its bound, preferred post hoc, lowered false stops at 2\,mm from \SrSOneLocalFalsePct{} to \SrSOneStableFalsePct{} overall, with \SrSOneStableCorrect{} instead of \SrSOneLocalCorrect{} correct ones but more trajectories uncertified; delaying the stop did not, and other signals at equal acceptance left \SrSOneMOtherRange{} false. With a segmentation error shared by all views (median Dice \SFiveDiceMedian), \SFiveStableFalsePct{} of this rule's stops were false against \SFiveRefStableFalsePct{} without it (\SFiveCases{} cases; exploratory). Minimax planning certified sooner than the clinical protocol without the check (\HzSOneMinimaxCert{} versus \HzSOneFixedCert{} shots, Holm $p$ \HzSOneDiffCertFixedHolm), but no criterion reached the pre-specified endpoint, a true 2\,mm error, significantly sooner, even in the true basin. On real fluoroscopy of five intact pelves no 2\,mm certificate was false. Trust takes evidence across views, not only a calibrated local bound.
\end{abstract}


\end{frontmatter}

\section{Introduction}
\label{sec:intro}

\begin{figure*}[t]
\centering
\includegraphics[width=\textwidth]{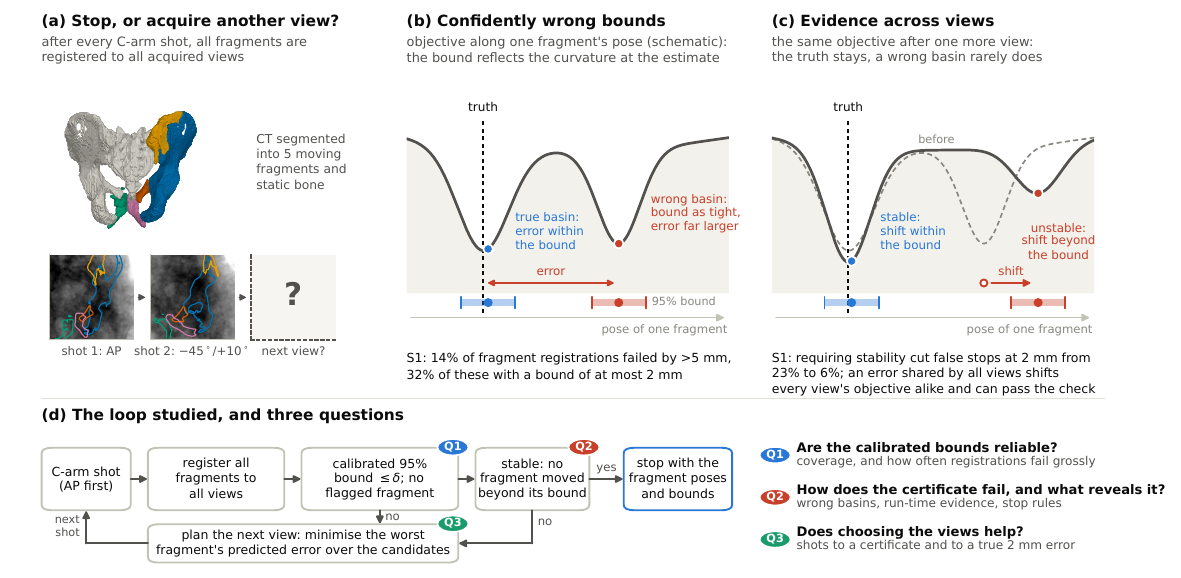}
\caption{When can a multi-fragment registration be trusted? (a) After every C-arm shot, the first anterior--posterior (AP), all fragments are registered to all acquired views (PENGWIN case \FigTwoCase, five moving fragments; outlines: the estimated fragments), and the system must stop or acquire another view. (b) A calibrated local bound reflects the curvature of the objective at the estimate, which can be as sharp in a wrong basin as at the truth (schematic): in the main study (S1), \FdSOneGross{} of fragment registrations from the second view on failed by more than 5\,mm, \FdSOneConfPct{} of them with a bound of at most 2\,mm. (c) All views agree at the true pose, but a wrong basin rarely survives a further view unchanged, so an estimate that moves beyond its bound is suspect: requiring stability lowered the false stops at 2\,mm from \SrSOneLocalFalsePct{} to \SrSOneStableFalsePct{} of first certificates (S1, all policies; preferred post hoc). A model error shared by all views shifts every view's objective alike and can pass this check (setting S5). (d) The loop studied and its three questions: Q1 is answered in Section~\ref{sec:res-bounds}, Q2 in Sections~\ref{sec:res-detect}--\ref{sec:res-s5}, Q3 in Section~\ref{sec:res-views}. Fig.~\ref{fig:method} details the method. The curves in (b) and (c) and the diagram in (d) are schematic; panel (a) and the numbers come from the study's data.}
\label{fig:overview}
\end{figure*}

Percutaneous fixation of pelvic fractures is guided by fluoroscopy: the surgeon reduces the displaced fragments and places screws in narrow bony corridors, checking them in inlet, outlet and lateral views, which with conventional imaging has taken a median of more than a hundred shots per screw \citep{alemdaroglu2014pedicle}, with the corresponding dose \citep{prost2024navigation}. Registering the pre-operative CT, segmented into its fragments, to the intra-operative images shows where every fragment is in three dimensions \citep{han2021fracture,grupp2020pose}. Such a registration can guide the next step only if the system also knows when its result is accurate, and which view to acquire when it is not.

Both are questions about uncertainty, and both are harder for several fragments than for one bone. One image barely constrains a fragment's depth, and small fragments overlapped by large ones are weakly constrained in any view, so whether a further view resolves a fragment depends on the fracture. A decision to stop (``every fragment is within 2\,mm'') needs a per-fragment error bound that holds in practice, not a residual or a similarity score. A bound derived from the registration's local fit, however, inherits its blind spot: intensity-based registration can converge to a wrong local optimum, a \emph{wrong basin}, where the local fit can look as well determined as at the truth.

We study when multi-fragment 2D/3D registration can be trusted, in a closed loop that after each shot registers all fragments to all acquired views, computes the covariance of the fragment poses from rendered image derivatives, calibrates it against held-out errors and certifies the case once the 95\% bound on the worst fragment's error is at most a threshold; otherwise it acquires the candidate view that minimises the worst fragment's predicted error (Fig.~\ref{fig:overview}; method in Fig.~\ref{fig:method}). Flags for implausible and ambiguous fragments and a check of stability under the next view add evidence from beyond the local fit. We ask three questions (Fig.~\ref{fig:overview}d): are the calibrated bounds reliable; how does the certificate fail, and which evidence available at run time reveals it; and does choosing the views help? Our contributions are:
\begin{itemize}
\item a certificate for intensity-based multi-fragment 2D/3D registration: per-fragment bounds from the sandwich covariance of a tempered maximum-a-posteriori (MAP) estimator with a gradient-correlation loss, computed from rendered derivatives and calibrated by one factor, combined with failure flags and a sequential stability check, with ablations that separate what makes the bounds valid from what makes the stop decision reliable;
\item an analysis of how the certificate fails: registrations into wrong basins, many with confident bounds, and their causes, separated by oracle ablations into optimisation failures and errors of the estimator; how well the signals available at run time detect them; and how stop rules built on these signals compare, with controls for stopping later;
\item sequential view planning on the calibrated posterior, with a worst-fragment (minimax) criterion as one instance, compared with A-, D- and E-optimal design, expected information gain with full and diagonal Fisher information, an image-gradient heuristic, the clinical protocol, maximal angular separation and random views;
\item an evaluation fixed before the test data were used, on \SOneCases{} held-out fracture cases with per-view image uncertainty matched to real fluoroscopy, with a tracked and an untracked C-arm and at lower noise, and on real fluoroscopy of cadaver specimens; and, added later with its readings fixed in advance, a stress test with a segmentation error shared by all views.
\end{itemize}

The calibrated bounds covered \SOneCovUnflagged{} of unflagged fragment registrations and \SOneCovNongross{} of those within 5\,mm of the truth, against a nominal 95\%. From the second view on \FdSOneGross{} of fragment registrations failed by more than 5\,mm, \FdSOneConfPct{} of them with a bound of at most 2\,mm. Among such tightly bounded fragments, a fragment's shift when a view was added separated failures best (AUC \FdSOneStratShiftAuc, the bound itself \FdSOneStratLocalBoundAuc), also when every error above 2\,mm counted as a failure. Over all view policies, \SrSOneLocalFalsePct{} of first certificates were false under the bound and flags alone and \SrSOneStableFalsePct{} with the stability check (planner: \SrSOneLocalMinimaxFalse{} and \SrSOneStableMinimaxFalse), at the cost of more trajectories without a certificate; delaying the stop did not reduce false certificates, requiring two consecutive certificates left \SrSOneTwiceFalsePct{} with fewer correct ones, and rules thresholding the other signals at the same acceptance left \SrSOneMOtherRange{} false (exploratory). When the registration model's fragment boundaries were off by about 2\,mm in every view, the stable rule's false share was \SFiveStableFalsePct{} against \SFiveRefStableFalsePct{} on the same \SFiveCases{} cases (exploratory; within sampling variation at these counts), in the direction expected of an error that need not move the estimate when a view is added. Planning certified sooner than the clinical protocol under the bound and flags alone, and on real images of intact pelves it certified sooner than random views on every specimen, without a false certificate. After correction for multiple comparisons, however, no criterion reached the pre-specified primary endpoint, a true worst-fragment error of 2\,mm, significantly sooner than the protocol, even when registrations were kept in the true basin. We report these negative results with the positive ones: gross failures, not wide posteriors, limited both accuracy and certification, and trusting a multi-fragment registration took evidence across views, not only a calibrated local bound; model errors that all views share may weaken both.
\section{Related work}
\label{sec:related}

\paragraph*{Multi-body registration in orthopaedic trauma}
Intensity-based 2D/3D registration of a CT to fluoroscopy is well established for single rigid bodies \citep{markelj2012review}. Several works register several bones or fragments at once: for reducing pelvic dislocations \citep{han2020multibody} and pelvic fractures \citep{han2021fracture}, periacetabular osteotomy fragments \citep{grupp2020pose} and the ankle joint \citep{vijayan2024robot}. Differentiable renderers have made such registrations fast \citep{gopalakrishnan2023fast,gopalakrishnan2024intraoperative}, including polyrigid registration from few views \citep{gopalakrishnan2025polypose} and patient-specific networks for initialisation \citep{gopalakrishnan2026rapid}. These methods estimate fragment poses from given views; they neither choose the views nor state, after each shot, how large each fragment's remaining error could be.

\paragraph*{Views and registration accuracy}
The effect of the angular separation of two views on 2D/3D registration accuracy has been analysed \citep{uneri2014registration}, and an analytic estimate of the target registration error of landmark-based registration has been used to choose the C-arm orientation before an endoscopic procedure \citep{hu2016registration}. For the ankle, \citet{vijayan2024robot} select two views once, before reduction, by a global image-gradient score with at least 20$^\circ$ separation. Learning-based systems choose a second view from the first image for K-wire placement \citep{killeen2023autonomous} or position the C-arm from landmarks with conformal uncertainty \citep{arrabi2025automated}, and a Bayesian reconstruction of the femur has guided viewing conditions that minimise its shape uncertainty \citep{thusini2020uncertainty}. None of these selects views sequentially from the posterior of several coupled fragments, or certifies when to stop.

\paragraph*{Registration uncertainty and error bounds}
For point-based rigid registration, the target registration error can be predicted from the fiducial localisation error \citep{fitzpatrick1998predicting,danilchenko2011general}. For intensity-based registration, the Cram\'er--Rao bound limits the attainable accuracy \citep{robinson2004fundamental}, and Fisher information has been used to relate registration accuracy to image quality \citep{ketcha2017effects}. Such bounds describe an efficient estimator under a correct noise model; registration with a normalised cross-correlation loss is neither, and real images add model mismatch. Bootstrap resampling \citep{kybic2010bootstrap} and posterior sampling \citep{risholm2013bayesian} estimate registration uncertainty without these assumptions, at the cost of many registrations or samples per estimate. We use the sandwich covariance of the estimator actually used, calibrate it against held-out errors, and temper the posterior in the sense of generalised Bayesian inference \citep{syring2019calibrating}.

\paragraph*{Detecting registration failure}
An error bound describes a registration near its optimum; whether that optimum is the right one is a separate question. \citet{varnavas2015fully} verify automated CT-to-fluoroscopy registration by similarity features and by the agreement of the poses found for several vertebrae, and \citet{sokooti2019quantitative} predict the local error of deformable registration with regression forests trained on annotated correspondences. For sparse-data elastic registration of the liver, \citet{heiselman2022registration} compute registration-uncertainty metrics and assess by ROC analysis how well they flag regions of large registration error. In visual tracking, running a tracker forward and backward in time and measuring the discrepancy detects tracking failures \citep{kalal2010forward}. Alongside the calibrated bound, we evaluate the evidence that a sequential multi-fragment registration provides at run time: competing optima found by the optimiser, the distance from the prior, and the consistency of each fragment's estimate as views are added.

\paragraph*{Optimal experimental design and active view selection}
Choosing measurements to minimise a functional of the posterior covariance is the subject of optimal and Bayesian experimental design \citep{pukelsheim2006optimal,chaloner1995bayesian}, with the classical A-, D- and E-optimal criteria and expected information gain. In X-ray imaging, task-driven source--detector trajectories optimise detectability for a given imaging task in cone-beam CT \citep{stayman2019taskdriven,capostagno2019taskdriven}. In computer vision, Fisher information with a diagonal approximation selects views for radiance fields \citep{jiang2024fisherrf}, and active view selection has been extended to sparse-view CT \citep{wu2026active}. We compare a worst-fragment (minimax) criterion with these whole-model criteria, computed from the same full-block predicted covariances.

\paragraph*{Clinical context: views and dose}
With conventional inlet, outlet and lateral imaging, a median of 132 fluoroscopy shots per iliosacral screw has been reported, against 29.5 with a pedicle-axis view and sacral mapping \citep{alemdaroglu2014pedicle}; a recent series reports a dose--area product of about 469\,cGy\,cm$^2$ per screw and a 10.9\% screw misplacement rate under fluoroscopy \citep{prost2024navigation}. Reduction of acetabular fractures is graded by the largest residual gap or step on plain radiographs: 1\,mm or less anatomical, 2--3\,mm imperfect and more than 3\,mm poor \citep{matta1996fractures,verbeek2017ct}; pelvic ring reductions are graded more coarsely (4\,mm or less excellent; \citealp{matta1996pelvic}). Our 1, 2 and 3\,mm thresholds span the acetabular grades, although the mTRE, a three-dimensional error averaged over the fragment surface, is not the same quantity as a radiographic gap.
\section{Methods}
\label{sec:methods}

\begin{figure*}[t]
\centering
\includegraphics[width=\textwidth]{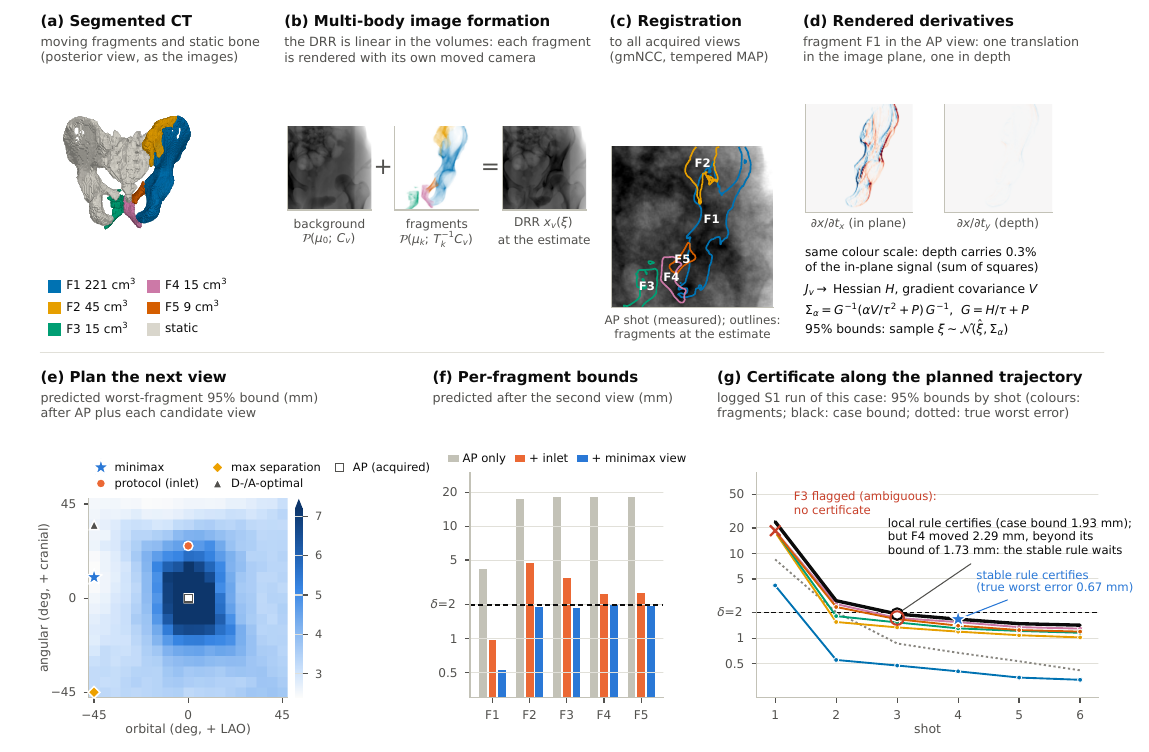}
\caption{The method on the case of Fig.~\ref{fig:overview}. (a) The CT is segmented into moving fragments (F1--F5 by volume) and static bone. (b) The digitally reconstructed radiograph (DRR) is the sum of a static background and one excess volume per fragment, each rendered with its own moved camera (Section~\ref{sec:render}). (c) All fragments are registered to all acquired views by the tempered MAP estimate~\eqref{eq:map}; here the AP shot. (d) Rendered derivatives give the Hessian $H$ and the gradient covariance $V$, hence the calibrated covariance $\Sigma_\alpha$ and the 95\% bounds (Sections~\ref{sec:cov} and~\ref{sec:calib}). One view barely observes depth: the depth derivative of F1 carries \FigMethodDepthPct{} of the in-plane signal, and after the AP shot the case bound is \FigMethodCaseBoundShotOne\,mm. (e) Model prediction (not the achieved error) of the 95\% bound on the worst fragment's mTRE after each of the 360 candidate views in addition to AP (Section~\ref{sec:predict}): \FigTwoBoundMinimax\,mm for the minimax choice, \FigTwoBoundDopt\,mm for the D- and A-optimal choice and \FigTwoBoundFixed\,mm for the protocol's inlet view (rank \FigTwoFixedRank{} of 360). (f) Predicted per-fragment bounds: the inlet view leaves \FigTwoFixedWorstTag{} (\FigTwoFixedWorstVol\,cm$^3$) at \FigTwoFixedWorstBound\,mm (minimax view: \FigTwoFixedWorstMinimax\,mm). Over all S1 cases the protocol's inlet view nevertheless brought the true worst-fragment error within 2\,mm after AP more often than the minimax choice (\SvFixed{} against \SvMinimax{} of cases; Table~\ref{tab:secondview}). (g) The logged S1 trajectory of this case under minimax: after the AP shot \FigMethodFlagTag{} is flagged as ambiguous; at shot \FigMethodLocalShot{} the case bound falls to \FigMethodLocalBound\,mm and the local rule certifies, but \FigMethodShiftTag{} moved \FigMethodShift\,mm, beyond its bound of \FigMethodShiftBound\,mm, so the stable rule certifies at shot \FigMethodStableShot{} (true worst-fragment error \FigMethodTrueStable\,mm).}
\label{fig:method}
\end{figure*}

\subsection{Problem setting}
A pre-operative CT of the fractured pelvis is segmented into $K$ moving fragments and a static anchor (the main sacral piece and the main piece of the least-injured hemipelvis). Fragment $k$ has an unknown rigid pose $T_k\in SE(3)$ relative to its CT position. We parameterise a pose perturbation by $\vxi_k=(\bm\omega_k,\bm t_k)\in\R^6$ about the fragment's current centroid $\bm c_k$: a point $\bm q$ moves to $R(\bm\omega_k)(\bm q-\bm c_k)+\bm c_k+\bm t_k$, so to first order by $A_{\bm q}\vxi_k$ with $A_{\bm q}=[-[\bm q-\bm c_k]_\times,\ I_3]$. The stacked parameter is $\vxi\in\R^{6K}$. The error of fragment $k$ is its mean target registration error (mTRE) over $M_k=300$ surface points $\bm q_{k,m}$ (farthest-point sampled from the fragment's surface):
\begin{equation}
\mathrm{mTRE}_k=\frac{1}{M_k}\sum_{m}\big\|\hat T_k\bm q_{k,m}-T_k\bm q_{k,m}\big\|.
\end{equation}
A C-arm view $v$ is given by its orbital and angular rotation about an isocentre. After each shot the system must (i) estimate all fragment poses from the views acquired so far, (ii) state how large each fragment's error could be, and (iii) either stop, because every fragment is certified below a threshold $\delta$ (2\,mm in our main analysis), or choose the next view.

\subsection{Multi-body image formation}
\label{sec:render}
We split the CT attenuation volume $\mu$ exactly into a static background $\mu_0$ and one excess volume $\mu_k$ per moving fragment. In $\mu_0$, the voxels of every moving fragment and of a one-voxel ring of adjacent soft tissue are set to the local soft-tissue level (the median attenuation 3--5\,mm from labelled bone); $\mu_k$ holds fragment $k$'s attenuation above that level, cropped around the fragment, so that $\mu_0+\sum_k\mu_k=\mu$ voxelwise. Because the line integral is linear, the digitally reconstructed radiograph (DRR) of the scene with fragment poses $T_{1..K}$ in a view with camera-to-world transform $C_v$ is
\begin{equation}
x_v(\vxi)=\mathcal P(\mu_0;C_v)+\sum_{k=1}^K\mathcal P(\mu_k;T_k^{-1}C_v),
\end{equation}
where $\mathcal P$ is a differentiable ray-marching projector (one call per volume, batched over views and poses). Each fragment is thus rendered with its own moved camera (Fig.~\ref{fig:method}b), and derivatives with respect to every fragment's pose are available by finite differences of the renderer or by automatic differentiation.

\subsection{Multi-fragment registration}
Given measured images $y_v$ (negative log of the detected intensity), the per-view loss is $L_v(\vxi)=1-\mathrm{gmNCC}(x_v(\vxi),y_v)$, a gradient and multi-scale normalised cross-correlation (NCC; patch NCC of Sobel gradients, $11\times11$ patches, and of intensities, global and $9\times9$ patches) computed after standardising both images globally, which makes it invariant to unknown detector gain and offset. Fragment poses are estimated as the mode of a tempered (generalised Bayesian) posterior (Fig.~\ref{fig:method}c)
\begin{equation}
\hat\vxi=\arg\min_{\vxi}\ \frac{1}{\tau}\sum_{v}L_v(\vxi)+\tfrac12(\vxi-\vxi_0)^\T P(\vxi-\vxi_0),\qquad \tau=\gamma s,
\label{eq:map}
\end{equation}
with a Gaussian pose prior $P$ centred on the CT pose (5$^\circ$ and 10\,mm per axis), the residual temperature $s$ estimated from the residual at the current estimate (Section~\ref{sec:cov}), and a tempering factor $\gamma=4$ \citep{syring2019calibrating}. Without the prior, directions that the acquired views do not observe (most prominently depth after a single view) are undetermined; without tempering, structured image noise creates spurious optima of small fragments that beat the true pose. The optimiser runs coarse to fine (32, 64 and 128\,px working resolution; Adam with a per-fragment trust region; the prior is included at every resolution) from two starts, the previous estimate and the prior mean. Each start is polished by a quasi-Newton method on~\eqref{eq:map}, and the start with the lowest objective is kept. A per-fragment global search over translations ($\pm30$\,mm at 64\,px) then proposes alternative optima, which replace the estimate only if they lower~\eqref{eq:map}. Settings are listed in \ref{app:config}.

\subsection{Posterior covariance from rendered derivatives}
\label{sec:cov}
Let $J_v=\partial x_v/\partial\vxi$ be the Jacobian of the rendered view at the estimate (finite differences of the renderer, $6K$ columns) and $H_v=J_v^\T\,\nabla^2_{xx}L_v\,J_v$ the Gauss--Newton Hessian of the loss, computed in closed form for gmNCC (Fig.~\ref{fig:method}d). Linearising~\eqref{eq:map}, the estimator's covariance is the sandwich
\begin{equation}
\Sigma=G^{-1}\Big(\frac{V}{\tau^{2}}+P\Big)G^{-1},\qquad G=\frac{H}{\tau}+P,
\label{eq:sandwich}
\end{equation}
where $H=\sum_v H_v$ and $V=\sum_v J_v^\T\nabla^2_{xy}L_v\,\mathrm{Cov}(n_v)\,\nabla^2_{yx}L_v\,J_v$ is the covariance of the loss gradient under image noise $n_v$. To first order the error of the estimate is $\hat\vxi-\vxi\approx-G^{-1}\big(g/\tau+P(\vxi-\vxi_0)\big)$, with $g$ the gradient of the data loss at the true poses; averaging over the image noise ($\mathrm{Cov}\,g=V$) and over true poses drawn from the prior ($\mathrm{Cov}\,P(\vxi-\vxi_0)=P$) gives~\eqref{eq:sandwich}. $\Sigma$ is thus the estimator's error covariance averaged over the prior, not the covariance $G^{-1}$ of the tempered posterior; for one fixed true pose the prior term is a bias rather than a variance. The simulated displacements are narrower than the prior (per axis about 3.3$^\circ$ and 5\,mm, against 5$^\circ$ and 10\,mm), so this term is conservative; we use the calibrated $\Sigma$ as a Gaussian approximation of the error. We model $\mathrm{Cov}(n_v)$ as stationary with the power spectrum of the view's own registration residual, and we linearise NCC around the noisy measurement (adding the residual variance to the measured image's patch variances), which removes the systematic over-prediction of a linearisation at the noise-free image. The residual temperature is $s=\mathrm{tr}(H^{+}V)/\mathrm{rank}(H)$, the value at which $V=sH$ would hold for an efficient estimator; it absorbs the overall noise level. In Monte-Carlo simulations with known white noise (five development cases, three views, relative noise 2--20\%, 100--200 draws, untempered estimator started at the truth), the empirical covariance of the registration was \MCAwareTraceMin--\MCAwareTraceMax{} times this prediction (trace); linearising at the noise-free image instead over-predicted it (ratio down to \MCPlainTraceMin), and the Cram\'er--Rao bound under-predicted it by a factor of \MCCrbTraceMin--\MCCrbTraceMax, because gmNCC is not an efficient estimator.

\subsection{Calibration, certified bounds and failure flags}
\label{sec:calib}
The linearised covariance ignores model mismatch (simulator versus reality, interpolation, residual structure). We therefore calibrate one scalar $\alpha$ on the data term, $\Sigma_\alpha=G^{-1}(\alpha V/\tau^{2}+P)G^{-1}$ (the prior is known exactly and is not scaled). The 95\% bound of fragment $k$ is the 0.95-quantile of $\mathrm{mTRE}_k$ under $\vxi\sim\mathcal N(\hat\vxi,\Sigma_\alpha)$ (2000 samples, first-order target motion), and the case bound is the 0.95-quantile of $\max_k\mathrm{mTRE}_k$. $\alpha$ is the squared 0.95-quantile of true error over bound among the fragments within 5\,mm of the truth in registrations without flagged fragments, on registrations of 2--3 views of the 30 calibration cases (four view sets per case: AP with one or two random views, and AP with the clinical inlet and outlet views); this gave $\alpha=\CalibAlpha$, with cross-validated coverage of \SuppCalibCvCov. This fit assumes that the bound scales with $\sqrt\alpha$; because $P$ is not scaled, the deployed bound is narrower where the prior contributes: on the S1 registrations from the second view on by a median \AsRatioMedianLater{} (by more than \AsRatioPFiveLater{} for 5\% of fragments; at the first shot, where the prior sets the depth, by \AsRatioMedianShotOne), which lowered the coverage of correctly registered fragments from \AsCovScaledNongross{} to \AsCovDeployedNongross. The 95\% level is that of the calibrated local model; it is not a guarantee for the error at the shot at which a sequential procedure chooses to stop, which we measure instead (Section~\ref{sec:results}). A fragment registration more than 5\,mm off is a \emph{gross} failure. From the second view on most gross failures are \emph{wrong basins}, optima of~\eqref{eq:map} other than the one around the true poses, reached because the optimiser missed the true basin or because the objective prefers the other optimum; kept in the true basin, few registrations were gross (Section~\ref{sec:results}). No finite $\alpha$ covers such errors; they are measured instead.

A local covariance cannot detect a registration that converged to a wrong basin. Two flags address this. A fragment is \emph{implausible} if its estimate lies more than $\sqrt{\chi^2_{6}(0.99)}=4.10$ prior standard deviations from the prior mean. It is \emph{ambiguous} if another polished optimum (another start, or an alternative from the global search) at least 2\,mm away (mean TRE) has an objective~\eqref{eq:map} within $\ln 20$ of the selected one, i.e.\ a posterior density ratio above $1/20$. The threshold is the largest value on a fixed grid that flags at most 2\% of correctly registered calibration fragments, and $\ln 20$ if none does; none did ($\ln 20$ flagged \SuppCalibNonGrossFlag). A shot \emph{certifies} the case when the case bound, a joint 95\% bound on the worst fragment's error that is at least as large as every fragment's own bound, is at most $\delta$ and no fragment is flagged (\emph{local} rule). The \emph{stable} rule additionally requires that adding the newest view moved no fragment by more than its own bound (the \emph{stability check}; the shift is the mean TRE between the previous and the current estimate), since a correct estimate moves within its uncertainty while a wrong basin tends to jump (Fig.~\ref{fig:method}g). The study protocol fixed both rules, to be reported side by side.

\subsection{Predicting a candidate view}
\label{sec:predict}
For a candidate view $v$ not yet acquired, we render the scene at the current estimate in that view and compute its Hessian $H_v$ from rendered derivatives exactly as for an acquired view. Its gradient covariance is unknown before acquisition; we model it as $V_v=sH_v$ (the efficient case at the current residual temperature), which gives, with $G_v=G+H_v/\tau$ and $\alpha$ scaling the whole data term including the candidate's, the predicted error covariance
\begin{equation}
\Sigma(v)=G_v^{-1}\Big(\alpha\,\frac{V+sH_v}{\tau^{2}}+P\Big)G_v^{-1}.
\end{equation}
Fragments are coupled through the full $6K\times6K$ blocks: a view that pins down one fragment also constrains its neighbours wherever their projections overlap (in our scenes this coupling proved weak; Section~\ref{sec:results}). The predicted root-mean-square (RMS) TRE of fragment $k$ is $e_k(v)=\sqrt{\mathrm{tr}(W_k\Sigma_{kk}(v))}$ with $W_k=\frac{1}{M_k}\sum_m A_{\bm q_{k,m}}^\T A_{\bm q_{k,m}}$.
\subsection{View-selection criteria}
The proposed \emph{minimax} criterion selects the candidate that minimises the worst fragment's predicted error, $\arg\min_v\max_k e_k(v)$, so the next shot goes where the least-determined fragment needs it (Fig.~\ref{fig:method}e,f). We compare it with whole-model optimal-design criteria computed from the same predicted covariances: A-optimal ($\sum_k e_k^2$), D-optimal ($\log\det\Sigma(v)$), E-optimal (largest eigenvalue of $\Sigma(v)$ with rotations scaled to a 50\,mm lever arm), and expected information gain in its first-order form $\mathrm{tr}(F_v\Sigma)$ with $F_v=H_v/\tau$, with full blocks or with diagonal Fisher information (FisherRF-style; \citealp{jiang2024fisherrf}).

\subsection{Sequential loop}
The first shot is anterior--posterior (AP). After every shot all acquired views are registered (warm-started from the previous estimate, with a restart from the prior mean), the calibrated posterior, bounds and flags are computed, and, unless the case is certified, every remaining candidate is scored and the best is acquired next (in our studies every trajectory continues to the budget and certification is evaluated offline). Candidate Hessians for all 361 candidates (orbital and angular angles in $[-45^\circ,45^\circ]$ in 5$^\circ$ steps) are computed in one batched pass per shot.

\subsection{Untracked C-arm}
When the executed C-arm pose is not measured, each acquired view gets an unknown camera correction (a rigid scene motion about the isocentre, prior 1$^\circ$ and 10\,mm per axis) that is estimated jointly with the fragments by block-coordinate descent; while $s$ is unknown, camera corrections are confined to 4.10 prior standard deviations. The posterior is over fragments and cameras; all bounds and criteria use the fragments' marginal covariance. A candidate view brings its own unknown camera, which we eliminate by the Schur complement of its Hessian with the camera prior. A final camera estimate beyond 4.10 prior standard deviations flags every fragment of the shot.
\section{Experiments}
\label{sec:experiments}

\subsection{Data}
\emph{PENGWIN} \citep{sang2026benchmark,sang2024pengwin} provides 100 pelvic CTs with fracture fragments segmented (sacrum and both hemipelves, each split into its fragments). We split the cases into development (10, used for the design choices and pilots; one revision of the registration was prompted by a calibration case, \ref{app:protocol}), calibration (30, used to fit $\alpha$ and the ambiguity threshold) and test (60, untouched until the studies below). \emph{DeepFluoro} \citep{grupp2020automatic} provides CTs and 24--111 real fluoroscopy images per specimen for six cadaveric specimens with ground-truth C-arm poses; it contains no fractures.

\subsection{Synthetic fracture scenes}
For each PENGWIN case, fragments smaller than 2\,cm$^3$ are merged into the fragment they touch most, the anchor is held static, and every other fragment receives one random rigid displacement (rotation uniform in $[0^\circ,10^\circ]$ about a random axis, translation uniform in $[0,15]$\,mm in a random direction, about its centroid). The C-arm has 1020\,mm source--detector and 769\,mm source--isocentre distances and a $1436\times1436$ detector of 0.194\,mm pixels (a 210\,mm field at the isocentre); registration works at up to 128\,px (1.6\,mm pixels at the isocentre). The isocentre is the centre of the moving fragments' bounding box; fragments mostly off the detector in all but one clinical view are held static.

\emph{Measured images} must not come from the registration's own forward model (the inverse crime). They are rendered from a CT physically resampled with the fragments moved, at full detector resolution with half the sampling step, passed through a light X-ray simulator (Poisson noise at $10^4$ photons per pixel, scatter as 0.5 times a 50\,mm Gaussian blur of the primary, detector blur, electronic noise, random gain, offset and gamma), area-averaged in intensity to the working resolution and log-transformed. Before the studies, with the three clinical views on \SuppMFourCases{} calibration cases, the registration reached a median fragment mTRE of \SuppMFourLightNoisyMedian\,mm on such images and \SuppMFourDeepdrrMedian\,mm on images from the physics-based simulator DeepDRR \citep{unberath2018deepdrr} (\SuppMFourDeepdrrSuccess{} of fragments within 2\,mm). Synthetic images nonetheless give per-view registration errors several times smaller than real fluoroscopy (the calibration factor fitted on DeepDRR images covered only \SuppRealCovSynthAlpha{} of real fragment registrations), which makes almost any second view sufficient. We therefore add a stationary Gaussian random field with the autocorrelation of real DeepFluoro registration residuals and $\eta$ times their relative variance; with $\eta=1$ the median error of large fragments ($\ge$100\,cm$^3$) with 2--3 views was \RealismLargeMedian\,mm, against \RealismRealMedian\,mm for real hemipelves. Every image is deterministic per (case, view), so all policies see identical measurements of the same view.

\subsection{Real-image study}
On DeepFluoro we register the left and right hemipelvis as two independent bodies relative to the sacrum and lumbar vertebrae (true relative displacement zero; femurs at their per-view ground-truth poses). Candidates are each specimen's real recorded views, except four whose ground truth the converted release of the data \citep{gopalakrishnan2026rapid} flags as questionable; the first shot is the most frontal view, and each specimen is run from four starting poses drawn from the pose prior about the truth. The calibration factor is fitted leave-one-specimen-out on registrations of 57 sets of 2--3 real views ($\alpha=\RealAlphaMin$--\RealAlphaMax). Specimen 03 is excluded from planning (its views span 10$^\circ$). We also compare with the pairs of real views used in the evaluation of PolyPose \citep{gopalakrishnan2025polypose}, registered with our pipeline: this compares view pairs, not PolyPose's registration method.

\subsection{Settings, policies and hypotheses}
S1 (main): tracked C-arm, $\eta=1$, all 60 test cases. S2: untracked C-arm with a per-view repositioning error of 1$^\circ$ and 10\,mm per axis, the first 15 test cases. S3: tracked, $\eta=0.5$, the same 15 cases. S4: real images, 5 specimens $\times$ 4 starting poses. Policies: minimax (ours); A-, D- and E-optimal design; expected information gain (EIG) with full and diagonal Fisher blocks; a gradient heuristic after \citet{vijayan2024robot} (largest mean image gradient of the predicted DRR among views at least 20$^\circ$ from all acquired ones); the clinical protocol (AP, inlet $+25^\circ$, outlet $-40^\circ$, left and right anterior oblique (LAO, RAO) $45^\circ$); maximal angular separation (largest minimum angle to the acquired views); and uniformly random views (5 seeds in S1, 3 in S2--S4). S2 and S3 run minimax, A-optimal design, the protocol, maximal separation and random views; S4 runs minimax, A- and D-optimal design, the gradient heuristic, maximal separation, random views and the PolyPose view pairs. S5, added after the main results (\ref{app:protocol}): S1's conditions plus a model error that is the same in every view, on the S2/S3 cases and policies. The registration model is built from a segmentation whose fragment boundaries are displaced by a smooth Gaussian random field (standard deviation 2\,mm, correlation length 10\,mm; growth never into another moving fragment), while the measured images come from the true segmentation; sized on the development cases, this gives a median Dice of \SegDiceMedian{} with the true fragments (on the S5 cases \SFiveDiceMedian, 10th percentile \SFiveDicePTen). Calibration, flags and rules are those of S1. Its design and readings were fixed before it ran: the certificate is robust to such errors if the stable rule's false share is at most 10\% and coverage within 5\,mm at least 85\%, and degraded otherwise; S5 is compared with S1 on the same cases and policies. Every trajectory runs to the budget (6 shots, 5 in S4; the protocol stops after its five views); certification is evaluated offline.

Hypotheses, fixed in advance: (H1) minimax brings every fragment below 2\,mm true mTRE with fewer shots than the clinical protocol and maximal angular separation; (H2) minimax beats the whole-model criteria on the worst fragment, most for small or overlapped fragments; (H3) the calibrated 95\% bounds cover 85--99\% of fragments and false certification is rare; (H4) full-block Fisher information beats diagonal approximations.

\subsection{Endpoints and analysis}
The primary endpoint is the number of shots until the true worst-fragment mTRE is at most 2\,mm; secondary endpoints are shots to certification under the local and the stable rule (reported side by side, as fixed before the test runs), false certifications, coverage of the 95\% bounds, and thresholds of 1 and 3\,mm. Unreached endpoints count as the horizon plus one, so mean shots to an endpoint combine speed and success; we also report how many trajectories reached it. The pre-specified analysis used the budget as horizon, which scores a protocol trajectory that fails by its fifth and last view as 7 while the other policies get a sixth shot; we therefore compare all policies on a common five-shot horizon (pre-specified results in \ref{app:prespec}; conclusions differ only where stated). Policies are compared per case (random views averaged over seeds) by the mean paired difference with a case-bootstrap 95\% confidence interval (CI; unadjusted) and the Wilcoxon signed-rank test, Holm-adjusted over the baselines of a setting and endpoint. In S4 the starting poses of a specimen are not independent, so we summarise per specimen without $p$-values.

A \emph{false certification} is a certified registration whose true worst-fragment mTRE exceeds the threshold. Defined after the results, and exploratory: \emph{false stops} (trajectories whose first certificate is a false certification, over the trajectories that certified; exact binomial intervals per policy) and \emph{shots to a correct certification} (a false first certificate counts as none); the comparison of every policy with the protocol; the certificate ablations; the detection of wrong registrations and the stop-rule controls (below); and the analyses of predicted versus true error, the second view, wrong basins and projected overlap. Rates over registrations count each registration (case and acquired views) once. A fragment registration is \emph{gross} if it ends more than 5\,mm from the truth; \emph{unflagged} fragments are not flagged as implausible or ambiguous. The protocol was committed before the test split was used (\ref{app:protocol}). Deviations from the original design are listed in \ref{app:protocol}; one affected a reported run: the first S2 run did not restart the untracked registration from the prior mean or confine camera corrections; after its results exposed this, both were fixed (with the camera flag) and S2 was re-run on the same cases.

\subsection{Detecting failed registrations}
To ask which evidence available at run time reveals a wrong basin, we took every fragment registration from the second view on (each case and set of acquired views once, whichever policies reached it) and scored seven signals that the planner logs: the \emph{prior distance} (the implausibility statistic); the \emph{ambiguity} (minus the objective gap to the nearest polished alternative at least 2\,mm away); the \emph{shift} of the fragment when the newest view was added (mean TRE between the previous and the current estimate), and the \emph{instability}, the shift over the fragment's calibrated bound, which the stability check thresholds at 1; the calibrated \emph{bound} itself; the \emph{start disagreement} (the distance between the fragment's estimates from the two registration starts); and the \emph{local misfit} (minus the gradient correlation of rendered and measured images in the fragment's projected footprint, pooled over the acquired views). For each signal we report the area under the receiver operating characteristic (ROC) curve (AUC) for a gross failure, and again with every error above 2\,mm, the certification threshold, counted as a failure, with a 95\% confidence interval from resampling cases, and for the rules of the run (the flags; the stability check) sensitivity and specificity with case-bootstrap intervals, together with every signal's sensitivity at each rule's specificity. A gross failure whose calibrated bound is at most 2\,mm is a \emph{confident failure}; because the instability divides by the bound, we also report every signal among the fragments with bounds of at most 2\,mm, and the shift without the division. To separate the stability check from simply stopping later, we compare the first certificates of all trajectories under the local rule, the local rule without stops before shot 3 (a pure delay at least as long as the stable rule's, which cannot stop at shot 1), two consecutive local certificates and the stable rule, as well as the local rule plus a threshold on each other signal, set on these data so that as many trajectories certify as under the stable rule (a comparison at equal acceptance, not a deployable rule). First certificates are pooled over the policies of a setting with case-bootstrap intervals, and also averaged with equal weight per policy; trajectories of different policies share registrations, so the pool is descriptive. Because these analyses use the test split, we fit no combined detector and no new rule.

\subsection{Oracle ablations}
Two ablations replace the deployable registration at every shot on the first 30 S1 test cases (random views with 3 seeds), keeping measurements, calibration, flags and planning; neither is deployable. The \emph{oracle start} runs the full registration from the true poses with one start, without the prior-mean restart and the global search; it was fixed before it was run. The \emph{oracle basin}, designed after the oracle-start results, runs only the finest stage (128\,px) and the polish from the true poses, which keeps the estimate in the true basin unless that basin's own optimum is far from the truth.

\subsection{Implementation}
\label{sec:implementation}
Rendering uses a fused GPU ray-marching kernel (float32) for registration and a float64 reference renderer for the polish and the covariance; experiments ran on single RTX 5070 and RTX 5090 GPUs, and every result records its code version, configuration and seeds. The software is written in Python. Every experiment ran from version-controlled code with a logged configuration; every number, table and data figure in this paper is computed by these scripts from the logged runs, and the conventions of the libraries used were checked against their source code.
\section{Results}
\label{sec:results}

\begin{figure*}[t]
\centering
\includegraphics[width=\textwidth]{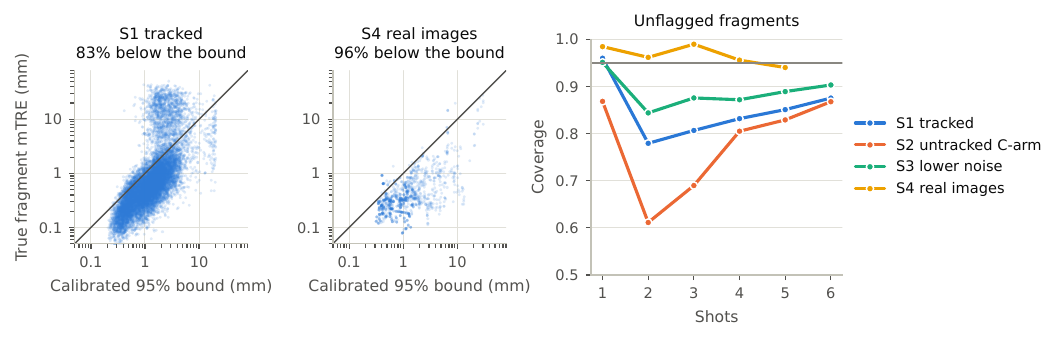}
\caption{Calibration ($\alpha=\CalibAlpha$; S4: leave-one-specimen-out). Left, middle: true fragment mTRE against the calibrated 95\% bound for fragment registrations not flagged as implausible or ambiguous (S1, S4); points far above the diagonal are gross failures, mostly wrong basins. Right: coverage of the bounds per shot (line: 95\%).}
\label{fig:calib}
\end{figure*}

\subsection{Calibrated bounds and gross failures}
\label{sec:res-bounds}
Rates over registrations count each case and set of acquired views once, whichever policies reached it. The calibrated 95\% bounds covered \SOneCovUnflagged{} of unflagged fragment registrations in S1, \SThreeCovUnflagged{} in S3, \STwoCovUnflagged{} in S2 and \SFourCovUnflagged{} on real images (Fig.~\ref{fig:calib}); H3's coverage range was met at lower noise and on real images. The shortfall in S1 comes from gross failures: \SOneGrossRate{} of fragment registrations ended more than 5\,mm off (\GrossSOneShotOne{} after the first view, whose depth only the prior constrains; \GrossSOneShotTwoPlus{} from the second view on), and the bounds covered \SOneCovNongross{} of the others, close to but below the nominal 95\% (\ObCovNongross{} in the true basin). A gross failure is not an honest but wide posterior. From the second view on, \FdSOneConfPct{} of the gross failures had a calibrated bound of at most 2\,mm (\emph{confident failures}): no calibration factor covers such errors without inflating every correct bound.

Most gross failures from the second view on are wrong basins, with two causes (exploratory). For up to two random gross failures per case from the second shot on, we compared objective~\eqref{eq:map} at the estimate and at the true basin's optimum (the oracle-basin registration). The true basin was better by more than $\ln 20$ in \DiagDepOptClear{} of \DiagDepN{} failures (the optimiser missed it), within $\ln 20$ in \DiagDepTie{} and worse in \DiagDepEstClear{} (the objective preferred the wrong pose; \DiagDepEstClearEarly{} of these with two or three views). Optimisation failures caused \DiagDepCertOptClear{} of the \DiagDepCertN{} sampled false certifications. Kept in the true basin (oracle basin, first 30 cases), \ObGrossShotTwo{} of fragment registrations were gross at shot 2 and \ObGrossShotThree{} at shot 3, against \ObDepGrossShotTwo{} and \ObDepGrossShotThree{} with the deployable registration on the same cases; started at the truth but free to leave (oracle start), \OrcGrossRate{} were gross over all shots (deployable \OrcDepGrossRate): the coarse stages moved small fragments out of the basin they started in.

\begin{figure*}[t]
\centering
\includegraphics[width=\textwidth]{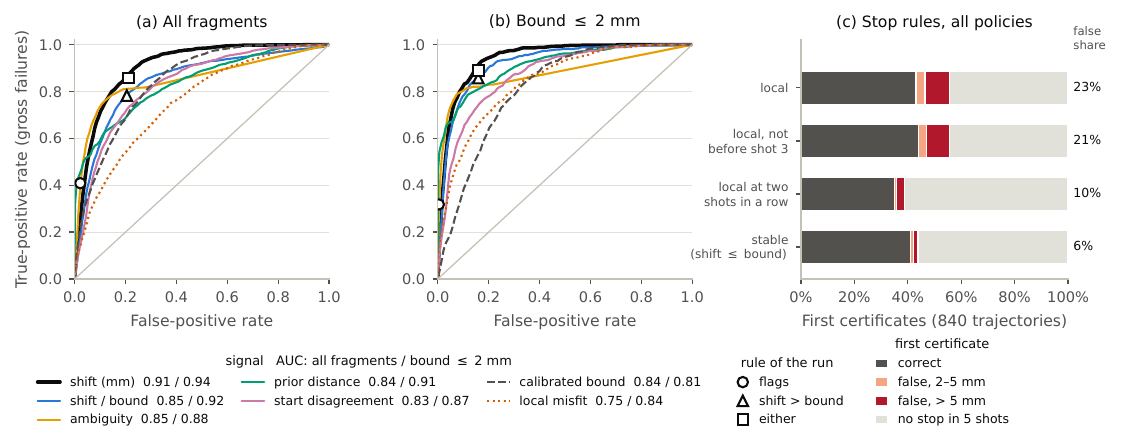}
\caption{Detecting failed registrations and stopping on the certificate (S1, exploratory). (a) ROC curves of the run-time signals for a gross failure ($>$5\,mm) among fragment registrations from the second view on, with the operating points of the run's rules: the flags, the stability check (shift $>$ bound) and either. (b) The same among fragments whose calibrated bound is at most 2\,mm, which the bound passes. (c) First 2\,mm certificate of each of the 840 trajectories of all ten policies under the local rule, the local rule without stops before shot 3, two consecutive local certificates, and the stable rule: correct, false (true worst-fragment mTRE 2--5\,mm or above 5\,mm) or none within five shots; right, the false share of first certificates.}
\label{fig:detection}
\end{figure*}

\begin{table*}[t]
\centering
\caption{Detecting failed registrations in S1 (exploratory): fragment registrations from the second view on, each case and set of acquired views once, with a failure defined as a true mTRE above 5\,mm (gross) or above 2\,mm (the certification threshold). For all fragments and for those with a calibrated bound $\le$2\,mm: AUC of each run-time signal (higher = more suspicious) and sensitivity / specificity of the rules of the run (shift $>$ bound overall, at shot 2 and from shot 3). S2 and S3 are in the supplementary material.}
\label{tab:detection}
\footnotesize
\setlength{\tabcolsep}{3pt}
\begin{tabular}{lcccc}
\toprule
 & \multicolumn{2}{c}{S1, failure: error $>$ 5\,mm} & \multicolumn{2}{c}{S1, failure: error $>$ 2\,mm} \\
\cmidrule(lr){2-3}\cmidrule(lr){4-5}
 & all & bound $\le$ 2\,mm & all & bound $\le$ 2\,mm \\
\midrule
Fragment registrations (failures) & 14864 (2153) & 10846 (679) & 14864 (2890) & 10846 (840) \\
\midrule
\multicolumn{5}{l}{\emph{AUC}} \\
Prior distance & 0.84 [0.81, 0.86] & 0.91 [0.88, 0.94] & 0.77 [0.73, 0.80] & 0.87 [0.83, 0.90] \\
Ambiguity (objective gap) & 0.85 [0.83, 0.87] & 0.88 [0.86, 0.90] & 0.83 [0.81, 0.84] & 0.86 [0.83, 0.88] \\
Start disagreement & 0.83 [0.81, 0.85] & 0.87 [0.85, 0.89] & 0.82 [0.80, 0.84] & 0.85 [0.83, 0.88] \\
Shift (mm) & 0.91 [0.90, 0.92] & 0.94 [0.93, 0.95] & 0.90 [0.89, 0.91] & 0.93 [0.92, 0.94] \\
Shift / bound (instability) & 0.85 [0.83, 0.87] & 0.92 [0.91, 0.93] & 0.82 [0.80, 0.84] & 0.90 [0.88, 0.91] \\
Calibrated bound & 0.84 [0.82, 0.87] & 0.81 [0.78, 0.85] & 0.87 [0.85, 0.89] & 0.83 [0.79, 0.86] \\
Local image misfit & 0.75 [0.71, 0.79] & 0.84 [0.81, 0.87] & 0.74 [0.70, 0.78] & 0.83 [0.80, 0.86] \\
\midrule
\multicolumn{5}{l}{\emph{Rules of the run: sensitivity / specificity, \% (95\% CI)}} \\
Flags (implausible or ambiguous) & 41 [38, 44] / 98 [97, 98] & 32 [29, 35] / 100 [99, 100] & 34 [31, 37] / 99 [98, 99] & 27 [24, 29] / 100 [100, 100] \\
Shift $>$ bound (stability check) & 78 [76, 81] / 79 [78, 81] & 86 [83, 88] / 84 [82, 86] & 72 [69, 75] / 81 [80, 83] & 81 [78, 84] / 85 [83, 86] \\
Either & 86 [84, 88] / 79 [77, 81] & 89 [87, 91] / 84 [82, 85] & 78 [76, 81] / 81 [79, 82] & 84 [81, 87] / 85 [83, 86] \\
\quad shift $>$ bound at shot 2 & 85 / 31 & -- & 78 / 30 & -- \\
\quad shift $>$ bound from shot 3 & 75 / 89 & -- & 69 / 91 & -- \\
\bottomrule
\end{tabular}
\end{table*}

\subsection{Detecting failed registrations (exploratory)}
\label{sec:res-detect}
Every signal from beyond the local fit ranked gross failures above correct fragment registrations (Fig.~\ref{fig:detection}a, Table~\ref{tab:detection}; S1 AUC \FdSOneAucMin--\FdSOneAucMax; S3 and S2 alike, supplementary material), as did the calibrated bound itself (\FdSOneLocalBoundAuc); the local image misfit did least (\FdSOneLocalMisfitAuc). The strongest was a fragment's shift in millimetres when the newest view was added (\FdSOneShiftAuc{} \FdSOneShiftAucCI), stronger than the same shift divided by the bound (the instability, \FdSOneInstabilityAuc). Among the fragments with bounds of at most 2\,mm, which the bound passes (\FdSOneStratN, \FdSOneStratGrossN{} of them gross), the bound separated failures least (\FdSOneStratLocalBoundAuc), the image misfit next (\FdSOneStratLocalMisfitAuc), the signals from beyond the local fit better (\FdSOneStratAucMin--\FdSOneStratAucMax), the shift best (\FdSOneStratShiftAuc; Fig.~\ref{fig:detection}b). Registrations reached by minimax trajectories, the planner's own views, gave similar values (\FdSOneMmN{} fragment registrations, AUC \FdSOneMmAucMin--\FdSOneMmAucMax). With every error above 2\,mm counted as a failure (\FdTSOneFail{} of fragment registrations), the shift again separated best (\FdTSOneShiftAuc; among bounds of at most 2\,mm \FdTSOneStratShiftAuc, the bound itself \FdTSOneStratLocalBoundAuc).

The two rules of the run sit at different operating points. The flags caught \FdSOneFlagsSens{} \FdSOneFlagsSensCI{} of gross failures at \FdSOneFlagsSpec{} specificity (positive predictive value \FdSOneFlagsPpv); at that specificity the prior distance and the ambiguity caught \FdSOneAtFlagsSpecPriorDistance{} and \FdSOneAtFlagsSpecAmbiguity, the shift \FdSOneAtFlagsSpecShift. The stability check (shift $>$ bound) caught \FdSOneInstabSens{} \FdSOneInstabSensCI{} at \FdSOneInstabSpec{} \FdSOneInstabSpecCI{} specificity; at that specificity the shift caught \FdSOneAtInstabSpecShift{} and the ambiguity \FdSOneAtInstabSpecAmbiguity. Of the confident failures, the stability check caught \FdSOneStratInstabSens{} and the flags \FdSOneStratFlagsSens{} (stability check's specificity among correct fragments with such bounds: \FdSOneStratInstabSpec). Its false alarms concentrated at the second shot (specificity \FdSOneInstabSpecShotTwo{} there, \FdSOneInstabSpecLater{} from shot 3), when the previous estimate rested on the AP view alone. At the registration level, \FdSOneBpGross{} of the \FdSOneBpN{} S1 registrations whose case bound was at most 2\,mm contained a gross fragment; flags or instability caught \FdSOneBpCaught{} of these and blocked \FdSOneBpFalseAlarmN{} of the \FdSOneBpCorrectN{} others (\FdSOneBpFalseAlarmPct). In S2, where an implausible camera estimate flags every fragment of the shot, the flags were more sensitive and both rules less specific (supplementary material).

\begin{table*}[t]
\centering
\caption{Shots per policy at a 2\,mm threshold on a common five-shot horizon (mean over cases, random views averaged over seeds within a case; unreached endpoints count as 6). True: until the true worst-fragment mTRE is $\le$2\,mm (pre-specified primary endpoint). Local, stable: until certification under the two rules; false stops: trajectories whose first certificate was false, over those that certified; correct: until a correct first certificate under the stable rule (a false first certificate counts as none; exploratory). S1: 60 synthetic cases, tracked C-arm; S2: untracked C-arm, 15 cases; S3: lower noise ($\eta=0.5$), the same 15 cases; S4: real images, 5 specimens $\times$ 4 starting poses. PolyPose's view pairs are compared in the text.}
\label{tab:main}
\footnotesize
\begin{tabular}{llrrrrrr}
\toprule
 & & \multicolumn{1}{c}{True} & \multicolumn{2}{c}{Local rule} & \multicolumn{3}{c}{Stable rule} \\
\cmidrule(lr){3-3}\cmidrule(lr){4-5}\cmidrule(lr){6-8}
Setting & Policy & shots & shots & false stops & shots & false stops & correct \\
\midrule
S1 & Minimax (ours) & 3.68 & 4.20 & 11/38 & 4.87 & 1/29 & 4.88 \\
 & A-optimal & 4.02 & 4.23 & 10/39 & 4.95 & 1/27 & 4.98 \\
 & D-optimal & 3.50 & 4.32 & 6/41 & 4.77 & 1/37 & 4.80 \\
 & E-optimal & 3.50 & 4.60 & 8/32 & 5.07 & 2/24 & 5.17 \\
 & EIG (full) & 3.78 & 4.25 & 10/41 & 4.87 & 1/34 & 4.88 \\
 & EIG (diagonal) & 3.58 & 4.48 & 5/38 & 4.90 & 0/32 & 4.90 \\
 & Gradient heuristic & 3.32 & 4.67 & 4/37 & 4.97 & 1/29 & 4.98 \\
 & Clinical protocol & 3.70 & 4.72 & 9/35 & 5.17 & 3/24 & 5.25 \\
 & Max separation & 5.12 & 5.10 & 13/23 & 5.72 & 2/10 & 5.75 \\
 & Random & 3.98 & 4.92 & 30/144 & 5.19 & 11/122 & 5.26 \\
\midrule
S2 & Minimax (ours) & 4.60 & 5.07 & 1/6 & 5.47 & 0/5 & 5.47 \\
 & A-optimal & 4.93 & 5.07 & 3/7 & 5.40 & 1/5 & 5.47 \\
 & Clinical protocol & 4.47 & 5.67 & 0/3 & 5.80 & 0/2 & 5.80 \\
 & Max separation & 5.60 & 5.67 & 1/2 & 6.00 & 0/0 & 6.00 \\
 & Random & 4.76 & 5.49 & 1/11 & 5.58 & 1/10 & 5.62 \\
\midrule
S3 & Minimax (ours) & 3.13 & 3.60 & 1/11 & 4.13 & 1/11 & 4.27 \\
 & A-optimal & 3.07 & 3.67 & 2/12 & 4.20 & 1/11 & 4.33 \\
 & Clinical protocol & 2.87 & 4.33 & 0/10 & 4.60 & 0/10 & 4.60 \\
 & Max separation & 4.33 & 4.80 & 1/8 & 5.07 & 0/7 & 5.07 \\
 & Random & 3.24 & 4.56 & 3/27 & 4.96 & 0/23 & 4.96 \\
\midrule
S4 & Minimax (ours) & 1.40 & 2.20 & 0/20 & 2.70 & 0/20 & 2.70 \\
 & A-optimal & 1.40 & 2.20 & 0/20 & 2.90 & 0/20 & 2.90 \\
 & D-optimal & 1.40 & 2.55 & 0/20 & 3.05 & 0/20 & 3.05 \\
 & Gradient heuristic & 1.40 & 2.55 & 0/20 & 3.00 & 0/20 & 3.00 \\
 & Max separation & 1.55 & 3.00 & 0/20 & 3.40 & 0/20 & 3.40 \\
 & Random & 1.77 & 4.32 & 0/40 & 4.53 & 0/40 & 4.53 \\
\bottomrule
\end{tabular}
\end{table*}

\subsection{Stopping on the certificate}
\label{sec:res-stop}
A certificate matters when the system stops on it. Table~\ref{tab:stop} lists the first certificate of every S1 trajectory under each stop rule: correct, false (true worst-fragment mTRE 2--5\,mm or above 5\,mm) or none within five shots. Under the pre-specified local rule, \SrSOneLocalFalse{} of the \SrSOneLocalCert{} first certificates of the \SrSOneLocalTraj{} trajectories of all policies were false (\SrSOneLocalFalsePct, case-bootstrap CI \SrSOneLocalFalsePctCI; \SrSOneLocalEqFalsePct{} with equal weight per policy; per policy \SrSOneLocalPolicyRange, Table~\ref{tab:main}), \SrSOneLocalFalseGross{} of them more than 5\,mm off, and the true error at minimax's stops reached \HzSOneMinimaxTrueAtStopPNinety\,mm (90th percentile). At the registration level, of \SOneCertRegs{} certified registrations \SOneFalseRegs{} were false, and \SOneFalseViolGross{} of their \SOneFalseViolBodies{} violating fragments were more than 5\,mm off, a median \SOneFalseViolRatio{} times their bound. The flags had blocked \FlagsSOneBlocked{} certifications, \FlagsSOneBlockedFalse{} of them false. The second part of H3, that false certification is rare, was therefore not met under the local rule in S1, nor in S2 (\SrSTwoLocalFalsePct{} of first certificates false; S3 \SrSThreeLocalFalsePct).

The pre-specified stable rule, which adds the stability check, gave \SrSOneStableFalse{} false of \SrSOneStableCert{} first certificates (\SrSOneStableFalsePct, CI \SrSOneStableFalsePctCI; equal weight \SrSOneStableEqFalsePct; per policy \SrSOneStablePolicyRange), with \SrSOneStableCorrect{} correct first certificates against \SrSOneLocalCorrect{} and \SrSOneStableNone{} trajectories without a certificate against \SrSOneLocalNone{} (Fig.~\ref{fig:detection}c). It removed false stops of both kinds (2--5\,mm: \SrSOneLocalFalseMid{} to \SrSOneStableFalseMid; above 5\,mm: \SrSOneLocalFalseGross{} to \SrSOneStableFalseGross). For minimax, the planner we would deploy, the local rule stopped correctly in \SrSOneLocalMinimaxCorrect{} of 60 cases, falsely in \SrSOneLocalMinimaxFalseN{} and not at all in \SrSOneLocalMinimaxNone; the stable rule in \SrSOneStableMinimaxCorrect, \SrSOneStableMinimaxFalseN{} and \SrSOneStableMinimaxNone{} (false share \SrSOneStableMinimaxFalse, exact 95\% CI \SrSOneStableMinimaxFalseCP): the check traded false stops for later or no stops, not for correct ones. Random views contribute \SrSOneRandomTraj{} of the 840 trajectories (five seeds), which is why we also give equal-weight shares. In S3 \SrSThreeStableFalsePct{} and in S2 \SrSTwoStableFalsePct{} of first certificates were false under the stable rule, and on real images none under any rule.

The delay controls we evaluated did not reproduce this trade-off (exploratory): forbidding stops before the third shot, a delay at least as long as the stable rule's (which cannot stop at shot 1), left \SrSOneFromThreeFalsePct{} false, and requiring two consecutive local certificates left \SrSOneTwiceFalsePct{} false with only \SrSOneTwiceCorrect{} correct. At the stable rule's acceptance, a threshold on the unscaled shift gave \SrSOneMShiftFalsePct{} false (\SrSOneMShiftFalse{} of \SrSOneMShiftCert), and thresholds on the other signals \SrSOneMOtherRange{} (Table~\ref{tab:stop}): at equal acceptance, consistency across views was the most informative evidence the run provides. The protocol fixed both rules to be reported side by side; preferring the stable rule is a judgement made after the results.

\begin{table*}[t]
\centering
\caption{First 2\,mm certificates of the S1 trajectories under each stop rule (common five-shot horizon; exploratory except the local and stable rules): correct (c), false (f; true worst-fragment mTRE 2--5\,mm or above 5\,mm) and none (n) within five shots; false share of the first certificates with a case-bootstrap 95\% CI, and for minimax with an exact binomial CI. The lower rules add to the local rule a threshold on one run-time signal (largest value over the fragments), set on these data to certify as many trajectories as the stable rule.}
\label{tab:stop}
\footnotesize
\setlength{\tabcolsep}{3.5pt}
\begin{tabular}{lrrrrr@{\hspace{3pt}}lrr@{\hspace{3pt}}lrr}
\toprule
 & \multicolumn{6}{c}{All policies (840 trajectories)} & \multicolumn{3}{c}{Minimax (60)} & \multicolumn{2}{c}{Protocol (60)} \\
\cmidrule(lr){2-7}\cmidrule(lr){8-10}\cmidrule(lr){11-12}
 & & \multicolumn{2}{c}{false} & & \multicolumn{2}{c}{false share} & & \multicolumn{2}{c}{false share} & & \\
\cmidrule(lr){3-4}\cmidrule(lr){6-7}\cmidrule(lr){9-10}
Stop rule & correct & 2--5\,mm & $>$5\,mm & none & & [95\% CI] & c / f / n & & [exact CI] & c / f / n & false \\
\midrule
Local (bound and flags) & 362 & 28 & 78 & 372 & 23\% & [16, 29] & 27 / 11 / 22 & 29\% & [15.4, 45.9] & 26 / 9 / 25 & 26\% \\
\quad no stop before shot 3 & 369 & 25 & 74 & 372 & 21\% & [15, 28] & 29 / 9 / 22 & 24\% & [11.4, 40.2] & 26 / 9 / 25 & 26\% \\
\quad two local certificates in a row & 294 & 7 & 24 & 515 & 10\% & [6, 14] & 28 / 2 / 30 & 7\% & [0.8, 22.1] & 18 / 2 / 40 & 10\% \\
Stable (shift $\le$ bound) & 345 & 8 & 15 & 472 & 6\% & [4, 10] & 28 / 1 / 31 & 3\% & [0.1, 17.8] & 21 / 3 / 36 & 12\% \\
\multicolumn{12}{l}{\emph{Local rule plus a threshold on one signal, set to certify as many trajectories as the stable rule}} \\
\quad shift (mm) & 349 & 6 & 13 & 472 & 5\% & [3, 8] & 28 / 0 / 32 & 0\% & [0.0, 12.3] & 21 / 2 / 37 & 9\% \\
\quad ambiguity & 331 & 14 & 23 & 472 & 10\% & [6, 15] & 27 / 2 / 31 & 7\% & [0.8, 22.8] & 22 / 3 / 35 & 12\% \\
\quad prior distance & 320 & 25 & 23 & 472 & 13\% & [8, 19] & 27 / 4 / 29 & 13\% & [3.6, 29.8] & 21 / 3 / 36 & 12\% \\
\quad start disagreement & 322 & 16 & 30 & 472 & 12\% & [8, 18] & 25 / 4 / 31 & 14\% & [3.9, 31.7] & 19 / 5 / 36 & 21\% \\
\quad local image misfit & 314 & 17 & 37 & 472 & 15\% & [9, 20] & 24 / 6 / 30 & 20\% & [7.7, 38.6] & 23 / 2 / 35 & 8\% \\
\bottomrule
\end{tabular}
\end{table*}

\subsection{Model error shared by all views (exploratory)}
\label{sec:res-s5}
In S5 the registration model came from a segmentation whose fragment boundaries were off by about 2\,mm (median Dice \SFiveDiceMedian), an error that is the same in every view. Table~\ref{tab:s5} compares it with S1 on the same \SFiveCases{} cases and policies. By the reading fixed before the run, the certificate degraded: the stable rule's false share was \SFiveStableFalsePct{} (\SFiveStableFalse{} of \SFiveStableCert{} first certificates; S1 \SFiveRefStableFalsePct, \SFiveRefStableFalse{} of \SFiveRefStableCert), above the 10\% the reading allowed (difference \SFiveStableDiffPts{} points, case-bootstrap 95\% CI \SFiveStableDiffCI), and the local rule's \SFiveLocalFalsePct{} (S1 \SFiveRefLocalFalsePct; \SFiveLocalDiffPts{} points, CI \SFiveLocalDiffCI); coverage within 5\,mm stayed above its 85\% threshold (\SFiveCovNongross, S1 \SFiveRefCovNongross). Gross failures and their detection were unchanged: \SFiveGrossRate{} of fragment registrations from the second view on (S1 \SFiveRefGrossRate), shift AUC \SFiveShiftAuc{} (S1 \SFiveRefShiftAuc).

Where both settings acquired the same views (\SFivePairN{} fragment registrations from the second view on, mostly of the protocol, maximal separation and random views), the model error moved fragments between basins in both directions (\SFivePairToGross{} into a gross error, \SFivePairFromGross{} out of one; gross \SFiveRefPairGross{} and \SFivePairGross) and biased the others slightly: their median error rose from \SFiveRefPairMedian{} to \SFivePairMedian\,mm, the share beyond 2\,mm from \SFiveRefPairBeyond{} to \SFivePairBeyond, and coverage was lower than in S1 by \SFiveCovGapShotTwo{} points at the second shot and \SFiveCovGapLate{} points from the fourth on, consistent with an error that further views do not remove while the bounds keep shrinking. The false stops changed in kind: \SFiveStableFalseMid{} of the \SFiveStableFalse{} false first certificates under the stable rule were 2--5\,mm off (S1: \SFiveRefStableFalseMid{} of \SFiveRefStableFalse). One was a fragment that S1 registered within 2\,mm on the same views and S5 just beyond (\SFiveStableNewRefErr{} and \SFiveStableNewErr\,mm; Dice \SFiveStableNewDice), \SFiveStableClassPath{} were on planner paths that differed from S1's, and \SFiveStableClassPresent{} were errors that S1 made on the same views too, \SFiveStablePresentCaughtInRef{} of which led to no false stop in S1. At these counts the difference under the stable rule is within sampling variation (under the local rule its interval excludes zero), but its direction is the one expected: the stability check responds to an estimate that changes when a view is added, and an error that every view shares need not change it.

\begin{table}[t]
\centering
\caption{Persistent model error (S5, segmentation boundaries off by about 2\,mm in the registration model) against S1 on the same cases and policies (exploratory). Upper part: unique registrations; first 2\,mm certificates on the five-shot horizon, correct (c), false (f; split into 2--5\,mm and $>$5\,mm) and none (n), with the false share's case-bootstrap 95\% CI; AUCs for a gross failure, over all fragments and those with bounds $\le$2\,mm. Lower part: registrations from the second view on whose acquired views were identical in both settings.}
\label{tab:s5}
\footnotesize
\setlength{\tabcolsep}{3pt}
\begin{tabular}{lll}
\toprule
 & S1 & S5 \\
\midrule
\multicolumn{3}{l}{\emph{Whole trajectories (15 cases, 105 per setting)}} \\
Coverage, unflagged & 80.0\% & 76.7\% \\
\quad within 5\,mm & 90.1\% & 86.4\% \\
Gross ($>$5\,mm), shot $\ge$2 & 19\% & 18\% \\
Local rule: c / f / n & 35 / 8 / 62 & 28 / 12 / 65 \\
\quad f 2--5 / $>$5\,mm & 1 / 7 & 4 / 8 \\
\quad false share [CI] & 19\% [5, 35] & 30\% [17, 45] \\
Stable rule: c / f / n & 28 / 3 / 74 & 22 / 6 / 77 \\
\quad f 2--5 / $>$5\,mm & 0 / 3 & 4 / 2 \\
\quad false share [CI] & 10\% [0, 26] & 21\% [9, 39] \\
AUC of the shift / the bound & 0.88 / 0.80 & 0.88 / 0.81 \\
\quad bound $\le$2\,mm & 0.92 / 0.76 & 0.91 / 0.78 \\
\midrule
\multicolumn{3}{l}{\emph{Identical views (1384 fragment registrations)}} \\
Gross ($>$5\,mm) & 20.2\% & 20.1\% \\
\multicolumn{3}{l}{Within 5\,mm in both (1065):} \\
\quad median error (mm) & 0.63 & 0.69 \\
\quad error $>$2\,mm & 6.1\% & 8.2\% \\
\quad coverage & 91.5\% & 88.7\% \\
\quad at shots 2 / 4 / 6 (\%) & 94 / 91 / 92 & 92 / 88 / 89 \\
\bottomrule
\end{tabular}
\end{table}

\subsection{Components of the certificate (exploratory)}
Table~\ref{tab:predictors} compares five error predictors on the calibration split, each calibrated by the planner's rule: the sandwich~\eqref{eq:sandwich} with three noise models and the Bayesian Cram\'er--Rao bound $(F+P)^{-1}$ with Gaussian Fisher information $F$ for two. Without calibration the planner's predictor covered only \PredSpecCovAlphaOne{} of the correctly registered fragments. After calibration every predictor covered \PredCvCovMin--\PredCvCovMax{} of held-out fragments; the residual-spectrum sandwich gave the narrowest bounds at that coverage (median \PredSpecMedianBound\,mm) and the most certifications (\PredSpecCert, \PredSpecFalseCert{} false); the Cram\'er--Rao bounds were wider (\PredCrbWhiteMedianBound{} and \PredCrbKappaMedianBound\,mm) and certified fewer registrations (\PredCrbWhiteCert{} and \PredCrbKappaCert). The calibration factor, not the covariance form, brought the bounds near nominal coverage; the form set their sharpness.

\begin{table*}[t]
\centering
\caption{Error predictors on the calibration split (registrations of 2--3 views; accepted registrations; fragments within 5\,mm of the truth): calibration factor fitted on all cases, coverage without calibration, held-out coverage (five-fold by case), range of held-out coverage over four volume classes, median calibrated bound, and held-out certifications at 2\,mm (false ones in brackets). The planner uses the first row.}
\label{tab:predictors}
\footnotesize
\begin{tabular}{lrrrrrr}
\toprule
Predictor & $\alpha$ & coverage & held-out & coverage range & median & certified \\
 & & at $\alpha=1$ & coverage & over volumes & bound (mm) & (false) \\
\midrule
Sandwich, residual spectrum (ours) & 2.41 & 74\% & 94\% & 12 pts & 1.53 & 26 (6) \\
Sandwich, white noise & 0.826 & 96\% & 94\% & 12 pts & 1.55 & 23 (5) \\
Sandwich, white noise $\times$ correlation area & 0.00118 & 100\% & 94\% & 8 pts & 1.68 & 23 (5) \\
Cram\'er--Rao, white noise & 1.49 & 88\% & 93\% & 15 pts & 1.58 & 21 (5) \\
Cram\'er--Rao, white noise $\times$ correlation area & 0.00944 & 100\% & 94\% & 11 pts & 1.81 & 17 (4) \\
\bottomrule
\end{tabular}
\end{table*}

\IfFileExists{generated/supp_cert_ablation.tex}{%
On the S1 registrations (Table~\ref{tab:cabl}) we recomputed the posterior at every logged estimate with the planner run's own inputs; the recomputed bounds reproduced the logged ones (median relative difference \CabValMedian). Without calibration the bounds shrank to a median \CabAlphaOneRatio{} of their calibrated size and covered \CabAlphaOneCovNongross{} of unflagged fragments within 5\,mm of the truth instead of \CabFullCovNongross; under the local rule \CabAlphaOneFalseStop{} minimax trajectories stopped falsely (calibrated: \CabFullFalseStop). Under the stable rule, however, only \CabAlphaOneFalseStopStable{} did (calibrated: \CabFullFalseStopStable), and a correct certificate came after \CabAlphaOneMinimaxCorrect{} shots (calibrated: \CabFullMinimaxCorrect): the stability check largely protected the stop decision even without calibration. Dropping the coupling between fragments (block-diagonal $H$ and $V$, each fragment as if the others were known) changed no bound by more than \CabBlockRelMax{} (99\% within \CabBlockRelPNinetyNine), and coverage and certification were unchanged: in these scenes the fragments' image information was nearly independent, so the full blocks did not matter for the certificate.

\begin{table*}[t]
\centering
\caption{Certificate ablations on the S1 registrations (common five-shot horizon, 2\,mm): coverage of the calibrated 95\% bounds of unflagged fragments (all, and within 5\,mm of the truth) and their median; minimax's false stops (first certificate false, over trajectories that certified) under the local and stable rules; shots to a correct certification (stable rule) for minimax and the protocol. Estimates, flags and the planner's views are those of the run; the recomputation flips \CabFlips{} certification decisions whose case bound lies within \CabFlipMaxMm\,mm of the threshold, so the full model's scores can differ slightly from Table~\ref{tab:main}.}
\label{tab:cabl}
\footnotesize
\begin{tabular}{lrrrllrr}
\toprule
 & \multicolumn{3}{c}{Unflagged fragments} & \multicolumn{2}{c}{Minimax false stops} & \multicolumn{2}{c}{Correctly certified (stable)} \\
\cmidrule(lr){2-4}\cmidrule(lr){5-6}\cmidrule(lr){7-8}
Bound & coverage & non-gross & median (mm) & local & stable & minimax & fixed \\
\midrule
Full (recomputed) & 83\% & 91\% & 1.24 & 11/38 & 1/29 & 4.88 & 5.23 \\
No calibration ($\alpha=1$) & 65\% & 72\% & 0.82 & 25/52 & 3/37 & 4.87 & 5.05 \\
No coupling (block-diagonal) & 83\% & 91\% & 1.24 & 11/38 & 1/29 & 4.88 & 5.23 \\
\bottomrule
\end{tabular}
\end{table*}
}{}

\begin{table*}[t]
\centering
\caption{S1: minimax minus each baseline on the common five-shot horizon (negative: minimax needs fewer shots); mean over 60 cases, case-bootstrap 95\% confidence interval (unadjusted) and Wilcoxon signed-rank $p$ Holm-adjusted over the baselines. Correct certification is exploratory.}
\label{tab:paired}
\footnotesize
\setlength{\tabcolsep}{4pt}
\begin{tabular}{llrlrrlrrlr}
\toprule
 & & \multicolumn{3}{c}{True $\le$ 2\,mm} & \multicolumn{3}{c}{Certified (local rule)} & \multicolumn{3}{c}{Correctly certified (stable rule)} \\
\cmidrule(lr){3-5}\cmidrule(lr){6-8}\cmidrule(lr){9-11}
Setting & Baseline & diff. & 95\% CI & Holm $p$ & diff. & 95\% CI & Holm $p$ & diff. & 95\% CI & Holm $p$ \\
\midrule
S1 & A-optimal & $-$0.33 & [$-$0.63, $-$0.05] & 0.179 & $-$0.03 & [$-$0.22, 0.13] & 0.962 & $-$0.10 & [$-$0.32, 0.10] & 1.000 \\
 & D-optimal & +0.18 & [$-$0.13, 0.50] & 0.718 & $-$0.12 & [$-$0.28, 0.07] & 0.405 & +0.08 & [$-$0.12, 0.28] & 1.000 \\
 & E-optimal & +0.18 & [$-$0.15, 0.53] & 1.000 & $-$0.40 & [$-$0.62, $-$0.20] & 0.004 & $-$0.28 & [$-$0.47, $-$0.12] & 0.032 \\
 & EIG (full) & $-$0.10 & [$-$0.38, 0.18] & 1.000 & $-$0.05 & [$-$0.17, 0.07] & 0.877 & +0.00 & [$-$0.20, 0.18] & 1.000 \\
 & EIG (diagonal) & +0.10 & [$-$0.20, 0.38] & 1.000 & $-$0.28 & [$-$0.47, $-$0.10] & 0.012 & $-$0.02 & [$-$0.20, 0.15] & 1.000 \\
 & Gradient heuristic & +0.37 & [0.00, 0.73] & 0.405 & $-$0.47 & [$-$0.68, $-$0.25] & 0.001 & $-$0.10 & [$-$0.28, 0.08] & 1.000 \\
 & Clinical protocol & $-$0.02 & [$-$0.38, 0.37] & 1.000 & $-$0.52 & [$-$0.73, $-$0.32] & $<$0.001 & $-$0.37 & [$-$0.62, $-$0.13] & 0.043 \\
 & Max separation & $-$1.43 & [$-$1.80, $-$1.07] & $<$0.001 & $-$0.90 & [$-$1.18, $-$0.62] & $<$0.001 & $-$0.87 & [$-$1.17, $-$0.58] & $<$0.001 \\
 & Random & $-$0.30 & [$-$0.58, $-$0.03] & 0.565 & $-$0.72 & [$-$0.93, $-$0.52] & $<$0.001 & $-$0.38 & [$-$0.57, $-$0.19] & 0.039 \\
\bottomrule
\end{tabular}
\end{table*}

\begin{figure*}[t]
\centering
\includegraphics[width=\textwidth]{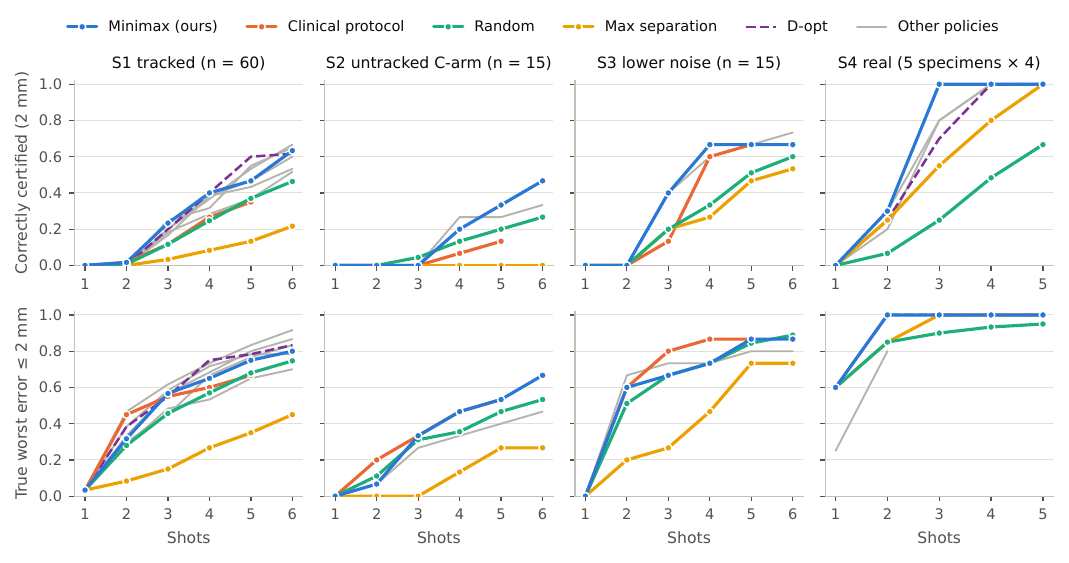}
\caption{Fraction of cases with a correct certificate under the stable rule (top) and with a true worst-fragment mTRE $\le$2\,mm (bottom) after each shot; each policy is drawn up to its last shot (protocol: 5; PolyPose pairs in S4: 2). D-optimal design dashed; grey: A- and E-optimal design, information gain (full and diagonal), gradient heuristic and PolyPose pairs.}
\label{fig:main}
\end{figure*}

\subsection{Choice of views: certification and accuracy}
\label{sec:res-views}
Planning on the calibrated posterior shortened the way to a certificate (Tables~\ref{tab:main} and~\ref{tab:paired}; Fig.~\ref{fig:main} shows correct certification under the stable rule). Under the local rule minimax certified after \HzSOneMinimaxCert{} shots against \HzSOneFixedCert{} for the protocol (\HzSOneDiffCertFixed{} \HzSOneDiffCertFixedCI, Holm $p$ \HzSOneDiffCertFixedHolm), sooner also than random views, maximal separation, the gradient heuristic, E-optimal design and diagonal information gain; it did not differ significantly from A- and D-optimal design and full information gain. Under the stable rule the difference to the protocol was \HzSOneDiffCertStableFixed{} (Holm $p$ \HzSOneDiffCertStableFixedHolm). Counting false stops as failures (exploratory), minimax reached a correct certificate within five shots in \SrSOneStableMinimaxCorrect{} of 60 cases and the protocol in \SrSOneStableFixedCorrect; on the composite score (a false or missing certificate counted as 6 shots) this was \HzSOneMinimaxCorrect{} against \HzSOneFixedCorrect{} (\HzSOneDiffCorrectFixed{} \HzSOneDiffCorrectFixedCI, Holm $p$ \HzSOneDiffCorrectFixedHolm), sooner also than random views, maximal separation and E-optimal design, and not significantly different from the other criteria (D-optimal design numerically ahead, \HzSOneDoptCorrect). In S2 and S3 the local-rule differences to the protocol were \HzSTwoDiffCertFixed{} (Holm $p$ \HzSTwoDiffCertFixedHolm) and \HzSThreeDiffCertFixed{} (\HzSThreeDiffCertFixedHolm), and the differences in correct certification (\HzSTwoDiffCorrectFixed{} and \HzSThreeDiffCorrectFixed{} shots) were not significant. In the pre-specified analysis (\ref{app:prespec}), minimax minus the protocol in S1 was \SOneDiffCertFixed{} (local) and \SuppSOneStableDiffFixed{} shots (stable; Holm $p$ \SOneDiffCertFixedHolm{} and \SuppSOneStableDiffFixedHolm).

On the pre-specified primary endpoint the policies differed little. In S1 minimax brought the true worst-fragment error to at most 2\,mm after \HzSOneMinimaxTrue{} shots and the clinical protocol after \HzSOneFixedTrue{} (difference \HzSOneDiffTrueFixed{} \HzSOneDiffTrueFixedCI, Holm $p$ \HzSOneDiffTrueFixedHolm); only maximal angular separation was clearly worse. H1 is thus supported against maximal angular separation, not against the protocol. H2 is not supported: no whole-model criterion was significantly behind minimax, and the gradient heuristic (\HzSOneGradientTrue) and D-optimal design (\HzSOneDoptTrue) were numerically ahead. After Holm correction no criterion reached the endpoint significantly sooner than the protocol (exploratory, Holm $p\ge$ \HzSOneVsFixedTrueMinHolm), although the gradient heuristic and D-optimal design were numerically ahead (gradient heuristic minus protocol \HzSOneVsFixedTrueGradient{} shots, unadjusted CI \HzSOneVsFixedTrueGradientCI); the data do not show equivalence either. In S2 and S3, minimax minus the protocol was \HzSTwoDiffTrueFixed{} and \HzSThreeDiffTrueFixed{} shots.

\begin{figure*}[t]
\centering
\includegraphics[width=\textwidth]{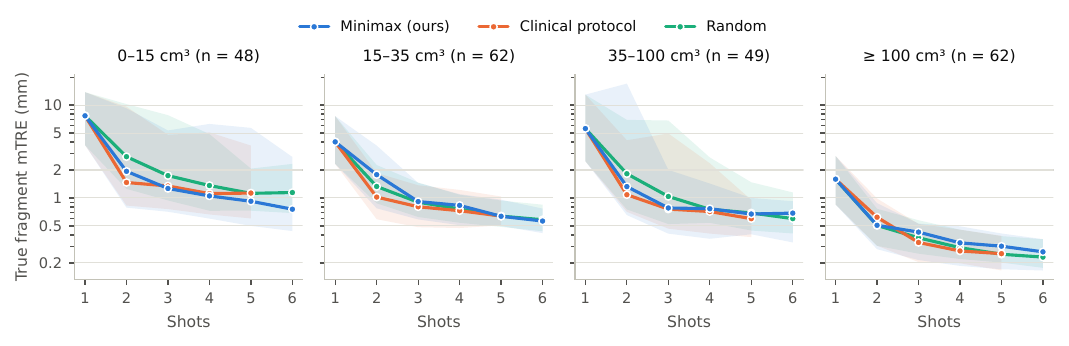}
\caption{True fragment mTRE after each shot by fragment volume (S1; median and interquartile range over fragments). Horizontal line: 2\,mm.}
\label{fig:volume}
\end{figure*}

Small and overlapped fragments dominate the worst-fragment error (Fig.~\ref{fig:volume}). Among fragments under 15\,cm$^3$, the shots until a fragment's own error was within 2\,mm were \SOneVolAMinimax{} for minimax and \SOneVolADopt{} for D-optimal design (\SOneVolDMinimax{} and \SOneVolDDopt{} from 100\,cm$^3$). We measured a fragment's projected overlap as the share of other bone in its footprint in the AP view. Fragments in the most overlapped third failed grossly in \SuppOvHighGross{} of registrations, against \SuppOvLowGross{} in the least overlapped third (\SuppOvLargeGrossHigh{} and \SuppOvLargeGrossLow{} from 35\,cm$^3$), and minimax did not bring them within 2\,mm sooner than D-optimal design (\SuppOvHighDiffDopt{} shots, Holm $p$ \SuppOvHighDiffDoptHolm; exploratory, supplementary material). The worst-fragment criterion thus did not help small or overlapped fragments most, contrary to H2.

Kept in the true basin (Table~\ref{tab:oracle}), every policy reached a true error of 2\,mm sooner, by \HzObCostMin{} to \HzObCostMax{} shots (minimax \HzObCostMinimax, protocol \HzObCostFixed), but the policies still did not differ significantly, apart from maximal separation: minimax needed \HzObMinimaxTrue{} shots and the protocol \HzObFixedTrue{} (\HzObDiffTrueFixed{} \HzObDiffTrueFixedCI, Holm $p$ \HzObDiffTrueFixedHolm), and D-optimal design was numerically fastest (\HzObDoptTrue; Holm $p$ \HzObDiffTrueDoptHolm{} against minimax, \HzObVsFixedTrueDoptHolm{} against the protocol). False certifications became rarer (\ObFalsePct{} of certified registrations, deployable \ObDepFalsePct).

\begin{table*}[t]
\centering
\caption{Oracle ablations on the first \OrcCases{} S1 test cases, common five-shot horizon: mean shots until the true worst-fragment mTRE is $\le$2\,mm with the deployable registration, the oracle start and the oracle basin; with the oracle basin, shots to a correct certification (stable rule) and minimax minus each policy on the true-error endpoint (case-bootstrap 95\% CI, Holm-adjusted Wilcoxon $p$).}
\label{tab:oracle}
\footnotesize
\begin{tabular}{lrrrrrlr}
\toprule
 & \multicolumn{3}{c}{Shots to true $\le$ 2\,mm} & \multicolumn{4}{c}{Oracle basin} \\
\cmidrule(lr){2-4}\cmidrule(lr){5-8}
Policy & deployable & oracle start & oracle basin & correctly & \multicolumn{3}{c}{minimax $-$ policy (true)} \\
 & & & & certified & diff. & 95\% CI & Holm $p$ \\
\midrule
Minimax (ours) & 3.93 & 4.23 & 3.07 & 4.67 & -- & &  \\
A-optimal & 4.27 & 4.20 & 3.03 & 4.67 & +0.03 & [$-$0.47, 0.50] & 1.000 \\
D-optimal & 3.73 & 3.93 & 2.57 & 4.50 & +0.50 & [0.13, 0.90] & 0.160 \\
E-optimal & 3.60 & 3.83 & 2.67 & 4.93 & +0.40 & [$-$0.03, 0.83] & 0.600 \\
EIG (full) & 3.90 & 4.10 & 3.00 & 4.60 & +0.07 & [$-$0.27, 0.43] & 1.000 \\
EIG (diagonal) & 3.70 & 4.07 & 2.93 & 4.67 & +0.13 & [$-$0.17, 0.43] & 1.000 \\
Gradient heuristic & 3.23 & 3.83 & 2.80 & 4.93 & +0.27 & [$-$0.17, 0.70] & 1.000 \\
Clinical protocol & 3.80 & 3.83 & 2.97 & 5.17 & +0.10 & [$-$0.47, 0.63] & 1.000 \\
Max separation & 5.20 & 5.10 & 3.77 & 5.40 & $-$0.70 & [$-$1.20, $-$0.23] & 0.046 \\
Random & 4.01 & 4.14 & 3.06 & 5.11 & +0.01 & [$-$0.39, 0.43] & 1.000 \\
\bottomrule
\end{tabular}
\end{table*}

\begin{figure*}[t]
\centering
\includegraphics[width=\textwidth]{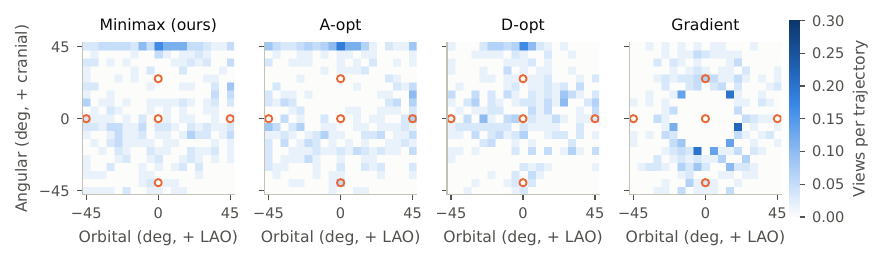}
\caption{Views acquired at shots 2--6 in S1 per policy (orbital and angular angle, 5$^\circ$ bins; mean number per trajectory). Circles: the protocol's views (AP, inlet, outlet, LAO and RAO 45$^\circ$).}
\label{fig:views}
\end{figure*}

After a chosen view was acquired, the posterior's worst-fragment RMS error was within a few percent of its prediction (median ratio \PredMinimaxPostRatio{} for minimax, \PredDoptPostRatio{} for D-optimal design): the prediction is self-consistent. The true error, however, exceeded the prediction more for minimax (median true/predicted \PredMinimaxTrueRatio) and A-optimal design (\PredAoptTrueRatio) than for D-optimal design (\PredDoptTrueRatio; exploratory). Minimax chose \PredMinimaxEdgeFrac{} of its views at shots 2--6 at the edge of the candidate range ($\pm45^\circ$; Fig.~\ref{fig:views}), where the ratio was \PredMinimaxEdgeTrueRatio{} against \PredMinimaxCenterTrueRatio{} elsewhere (in the true basin: \ObPredMinimaxEdgeTrueRatio{} and \ObPredMinimaxCenterTrueRatio).

Every trajectory starts with AP, so each case has many registered two-view sets, one per policy and random seed (median \SvPerCase{} distinct second views). After AP and its second view, the worst fragment was within 2\,mm in \SvMinimax{} of cases for minimax, \SvFixed{} for the protocol's inlet view and \SvGradient{} for the gradient heuristic (Table~\ref{tab:secondview}). Over all registered second views, those 15--30$^\circ$ from AP reached 2\,mm in \SvSepMid{} of registrations and those $45^\circ$ or more away in \SvSepFar. The best registered second view of a case reached 2\,mm in \SvBest{} of cases, but the best of the five random draws, chosen with the truth, already did in \SvBestRandom: a single registration's outcome depends largely on chance, so the best of several outcomes is an optimistic bound on what view choice can achieve.

\begin{table}[t]
\centering
\caption{S1: fraction of cases whose true worst-fragment mTRE was within 2 and 1\,mm after AP and the second view of each policy (random: mean over seeds), and for the best registered second view of each case, chosen with the truth.}
\label{tab:secondview}
\scriptsize
\setlength{\tabcolsep}{3pt}
\begin{tabular}{lrr}
\toprule
Second view (after AP) & true $\le$ 2\,mm & true $\le$ 1\,mm \\
\midrule
Minimax (ours) & 32\% & 13\% \\
A-optimal & 27\% & 17\% \\
D-optimal & 38\% & 23\% \\
E-optimal & 38\% & 20\% \\
EIG (full) & 30\% & 23\% \\
EIG (diagonal) & 33\% & 13\% \\
Gradient heuristic & 47\% & 13\% \\
Clinical protocol & 43\% & 17\% \\
Max separation & 7\% & 3\% \\
Random & 28\% & 12\% \\
\midrule
Best registered view (with the truth) & 68\% & 45\% \\
Best of 5 random views (with the truth) & 53\% & 32\% \\
\bottomrule
\end{tabular}
\end{table}

\subsection{Real images}
On real images of intact pelves, every policy but random views and the two-view PolyPose pairs certified every trajectory within the budget (random views: \RealRandomCertified{} of trajectories), and no certification was false (Tables~\ref{tab:main} and~\ref{tab:deepfluoro}). Minimax certified after \HzSFourMinimaxCert{} shots (stable rule \HzSFourMinimaxCertStable), random views after \HzSFourRandomCert; minimax needed fewer shots than random views on \RealSpecRandomFewer{} of \RealSpecN{} specimens, fewer than maximal separation and D-optimal design on \RealSpecMaxAngleFewer{} and \RealSpecDoptFewer, and more than a baseline on none. By the second shot, minimax had certified \RealTwoShotMinimaxCert{} of trajectories and PolyPose's view pairs, registered with our pipeline, \RealTwoShotPolyPoseCert{} (true error $\le$2\,mm: \RealTwoShotMinimaxTrue{} and \RealTwoShotPolyPoseTrue). The hemipelves were mostly within 2\,mm after the first, frontal view (\HzSFourMinimaxTrue{} shots on average for the information policies), so certification was the bottleneck. With the two-start registration of the planning studies, \SuppRealRunSetGrossN{} of the real-image calibration registrations left a hemipelvis \SuppRealRunSetGrossMin--\SuppRealRunSetGrossMax\,mm from the truth; the four-start registration with a translation grid of our earlier experiments recovered all of them (supplementary material).

\begin{table*}[t]
\centering
\caption{Real images (DeepFluoro) per specimen: registration of 2--3 real views (median fragment mTRE, median of the per-registration maximum landmark error), held-out coverage of the leave-one-specimen-out bounds, and mean shots to certification at 2\,mm in S4 (specimen 03 excluded from planning; PolyPose pairs: 6 = not certified).}
\label{tab:deepfluoro}
\scriptsize
\setlength{\tabcolsep}{3pt}
\begin{tabular}{lrrrrrr}
\toprule
Specimen & mTRE (mm) & landmark (mm) & coverage & \multicolumn{3}{c}{Shots to certification at 2\,mm} \\
 & median & median max & held-out & minimax & random & PolyPose \\
\midrule
01 & 0.67 & 0.72 & 95\% & 2.00 & 4.58 & 2.00 \\
02 & 0.41 & 0.59 & 100\% & 2.00 & 5.75 & 6.00 \\
03 & 1.18 & 1.26 & 86\% & -- & -- & -- \\
04 & 0.54 & 0.88 & 95\% & 3.00 & 5.58 & 6.00 \\
05 & 0.36 & 0.43 & 95\% & 2.00 & 2.50 & 2.00 \\
06 & 0.39 & 0.52 & 95\% & 2.00 & 3.17 & 6.00 \\
\bottomrule
\end{tabular}
\end{table*}

\subsection{Reproducibility and decision time}
One test case run on both GPU types reached the same shots-to-endpoint in \HwSameTrueN{} of \HwTraj{} trajectories (true error) and \HwSameCertN{} of \HwTraj{} (certification); single shots differed by up to \HwMaxDiff\,mm where floating-point differences tipped a fragment into another basin. On one RTX 5070, a decision took a median of \TimeRegMedian\,s to register all acquired views, \TimePostMedian\,s for the posterior and bounds and \TimePlanMedian\,s to score the 361 candidates.
\section{Discussion}
\label{sec:discussion}

\paragraph*{When the registration can be trusted}
A calibrated local bound is necessary but not sufficient. Calibrated, the bounds came close to nominal coverage for registrations within 5\,mm of the truth (about 90\% against 95\%) but covered only \SOneCovUnflagged{} of all unflagged ones, because under realistic per-view uncertainty about one in seven fragment registrations from the second view on failed grossly, mostly in a wrong basin, and a third of these passed the 2\,mm threshold. The covariance describes the curvature of the objective at the estimate and says nothing about optima elsewhere: the bound still ranked failures above correct registrations overall, but among fragments with tight bounds it separated them least. Evidence from beyond the local fit did better: among fragments with tight bounds every such signal separated failures better than the bound did, the shift best; over all fragments, the optimiser's competing optima and the distance from the prior were the most sensitive at high specificity, and a fragment's shift as views are added over most of the range. That a wrong basin rarely survives a further view unchanged is what makes the shift informative, and it is what the stability check tests. It lowered false stops from \SrSOneLocalFalsePct{} to \SrSOneStableFalsePct{} of first certificates over all view policies, at little cost in correct certificates but with more trajectories left uncertified, whereas waiting longer did not help, requiring two consecutive certificates helped less and cost more, and, at equal acceptance, thresholds on the optimiser's evidence, the prior distance or the image misfit left two to three times as many false stops as the shift. Our answer to the title is therefore threefold: trust a multi-fragment registration when its calibrated bound is small, when the optimiser found no competing pose, and when the estimate held as a further view was added. This carries verification by agreement, used between independently registered vertebrae \citep{varnavas2015fully}, and the forward--backward consistency of visual tracking \citep{kalal2010forward} over to a sequence of views. Four cautions apply. The rules' thresholds were fixed before the test data, but the comparison of signals and the preference for the stable rule were made after the results and should be confirmed on new cases, where a detector that combines the signals could also be fitted on a calibration split. The 95\% bound is a property of the local error model, not a guarantee for the error at a stop that a sequential procedure chooses: the stop decisions' error rates have to be measured, as here, and with \SrSOneStableMinimaxFalse{} false first certificates for the planner the exact 95\% interval of its false share is still \SrSOneStableMinimaxFalseCP. The check sees only errors that change between views: with a segmentation error shared by all views, its false share was \SFiveStableFalsePct{} against \SFiveRefStableFalsePct{} on \SFiveCases{} cases (within sampling variation) and coverage was lower, so a deployed certificate has to be calibrated, and its stop decisions measured, under the model errors it will meet, as the real-image calibration was. And the stability check needs a further view, so a planner could value views that test the current estimate, not only views that shrink its bound.

\paragraph*{What makes the certificate reliable}
The ablations separate the certificate's parts. The calibration factor brought the bounds near nominal coverage: uncalibrated, the planner's bounds covered about three quarters of the correctly registered fragments (\PredSpecCovAlphaOne{} on the calibration split, \CabAlphaOneCovNongross{} on the test split), and once calibrated every covariance form covered alike, the form mainly setting their sharpness. The stability check made the stop decision more reliable: it removed most false stops under the calibrated bounds and kept them rare even without calibration. The full inter-fragment covariance, which we expected to matter where projections overlap, changed the bounds by a few percent at most; with the weak coupling of these scenes, a block-diagonal model would have served the certificate equally. For a practical system this ranks the components: a calibration on representative data and a check of consistency across views matter more than a refined local covariance.

\paragraph*{Wrong basins limit accuracy and certification}
The wrong basins that defeat the bound also delayed the true 2\,mm error of every policy and struck overlapped fragments most. About half of the sampled failures were optimisation failures, which a better search would remove, and these caused most sampled false certifications; most of the rest were errors of the estimator, mostly with two or three views, which further views remove but a better optimiser cannot. Kept in the true basin, every policy reached 2\,mm sooner and false certifications became rarer. Carrying several pose hypotheses through the sequence, and choosing views that discriminate between them, is a natural next step for both accuracy and certification.

\paragraph*{What planning buys}
Planning on the calibrated posterior bought earlier certification, not earlier accuracy. Within the protocol's five shots, minimax certified about half a shot sooner than the clinical protocol under the local rule; under the stable rule the difference was smaller and not significant, although minimax reached a correct certificate in more cases than the protocol (\SrSOneStableMinimaxCorrect{} against \SrSOneStableFixedCorrect{} of 60; exploratory), and on real images it certified sooner than random views on every specimen. Minimax was one reasonable criterion among several: A- and D-optimal design and information gain with full blocks did as well, and D-optimal design was numerically ahead on correct certification. The true error fell to 2\,mm or less after about as many shots for every reasonable policy, even in the true basin: most fragments are within 2\,mm after two views, but the worst fragment needed three to four views on average, and planned views did not shorten this. Whether a given two-view registration lands within 2\,mm depends largely on chance, the image noise and the basin it ends in: view geometry changes these odds, but less than the predicted information suggests.

\paragraph*{Optimal criteria and model trust}
The rendered-derivative prediction was self-consistent, yet the worst-case and trace criteria favoured views at the edge of the candidate range, where the true error exceeded the prediction most, also in the true basin. Minimax greedily minimises the quantity that the certificate thresholds, so wherever the local model is optimistic it is drawn to exactly those views, the optimiser's curse of choosing by estimated values \citep{smith2006optimizer}; registered second views far from AP were indeed accurate less often. A robust criterion that discounts views where the model is least trustworthy (large view changes, heavy overlap, truncation), and a wider candidate range, are directions we did not pursue.

\paragraph*{Clinical relevance}
A fraction of a shot per case is small against the dozens to hundreds of shots of conventional screw fixation \citep{alemdaroglu2014pedicle}; the clinical value of such a system lies in a stop signal of known reliability, which says when the registration may guide the next step and when another view is needed, rather than in the shots it saves. A decision took about a minute on a consumer GPU, almost all of it registration, which would have to be faster for routine use. The mTRE averages over the fragment surface and can hide larger errors at a fragment's extremities, and our thresholds follow radiographic reduction grades, not screw-corridor tolerances.

\paragraph*{Limitations}
(i)~The fractures and their images are synthetic: the fragments are real, but each received one random displacement of up to 10$^\circ$ and 15\,mm, which excludes complete lateral compression injuries, and the fluoroscopy was simulated; segmentation was exact except in S5, and the images showed no instruments. (ii)~DeepFluoro has no fractures; the real-image study tests intact hemipelves of five specimens, on which no certificate was false, so the failure-detection results rest on the synthetic studies. (iii)~The synthetic per-view uncertainty is stationary noise, independent between views, matched to real residuals in spectrum and amplitude; real model mismatch (segmentation, anatomy, instruments, geometry) is tied to the patient and persists across views. S5 simulated one such error, a segmentation off by about 2\,mm, under which the stability check stopped falsely more often (\SFiveStableFalsePct{} against \SFiveRefStableFalsePct, within sampling variation at \SFiveCases{} cases); its size was chosen rather than measured on clinical segmentations, and persistent errors of anatomy, instruments and geometry were not simulated. (iv)~One calibration factor serves all fragments; it was fitted on fixed and random view sets rather than planner-chosen ones, assuming that the bound scales with $\sqrt\alpha$ (the deployed bound was a median \AsRatioMedianLater{} narrower), and the synthetic factor did not transfer to real images (a real-data factor did). (v)~The failure-detection analysis and the stop-rule controls were defined after the results and evaluated on the test split; the operating rules were fixed in advance, but we fitted no combined detector, and registrations and trajectories of one case share fragments and views, which the case bootstrap accounts for only in the intervals. (vi)~The untracked camera estimation often failed; the camera flag made such shots non-certifying rather than repairing them. (vii)~The real-image registration used a start set chosen on synthetic data that missed three large-fragment basins a translation-grid search recovers. (viii)~S2 and S3 had 15 cases each; we did not ablate the tempering or plan with block-diagonal covariances (H4, that full blocks beat diagonal approximations, is not supported: block-diagonal bounds equalled the full ones, and information gain with full and diagonal blocks did not differ notably), and several analyses were defined after the results. (ix)~All 100 PENGWIN cases are used for development, calibration or testing, so the post hoc findings, the preference for the stable rule and the strength of the shift, need confirmation on new cases; and a phantom or cadaver study with known fragment displacements would test failure detection on real fractured anatomy, which DeepFluoro cannot.

\section{Conclusion}
We asked when multi-fragment 2D/3D registration can be trusted, in a closed loop that registers all fragments after each C-arm shot, bounds each fragment's error, and certifies the case or chooses the next view. A sandwich covariance calibrated by one factor gave bounds close to nominal coverage for registrations in the true basin, but under realistic image uncertainty the dominant failure was a wrong basin, often with a bound as tight as a correct one. Evidence from beyond the local fit separated these failures, the consistency of each fragment's estimate as views were added best, and a check of that consistency lowered the share of false stops from \SrSOneLocalFalsePct{} to \SrSOneStableFalsePct{} at the price of more uncertified trajectories; waiting longer did not, two consecutive certificates did less at a higher cost, and rules on the other signals at equal acceptance left more false stops. With a segmentation error shared by all views, both the bounds and the check performed worse (exploratory, \SFiveCases{} cases). Planning views on the calibrated posterior certified sooner than the clinical protocol under the bound and flags alone, but did not reach the pre-specified accuracy endpoint significantly sooner, even in the true basin. Trusting such a registration takes evidence across views, not only a calibrated local bound, and a calibration under the model errors it will meet; reaching accuracy sooner will need registration and planning that reason about competing pose hypotheses.

\FloatBarrier
\balance %

\section*{Data and code availability}
The PENGWIN fracture CTs \citep{sang2024pengwin} and the DeepFluoro cadaver data \citep{grupp2020deepfluorodata,grupp2020automatic} are publicly available from their repositories. The code, configurations and logged results that regenerate every number, table and figure of this paper will be released in a public repository on acceptance, in the version that matches the published article, and are available to the editors and reviewers on request until then; images derived from the datasets are not redistributed, in line with their licences.

\section*{Declaration of competing interest}
The author declares that they have no known competing financial interests or personal relationships that could have appeared to influence the work reported in this paper.

\section*{CRediT authorship contribution statement}
\textbf{Bowen Xiang:} Conceptualization, Methodology, Software, Validation, Formal analysis, Investigation, Data curation, Visualization, Writing -- original draft, Writing -- review \& editing.

\section*{Declaration of generative AI and AI-assisted technologies in the manuscript preparation process}
During the preparation of this work the author used Claude (Anthropic; model Claude Opus 5.5, through Claude Code) to assist with writing the research and analysis software, with the plotting code that produces the tables and figures from the logged results, with drafting and editing the text, and with collating and checking the references against their sources. No figure was generated or altered by an AI image tool. After using this tool, the author reviewed and edited the content as needed and takes full responsibility for the content of the published article.

\section*{Funding}
This work was supported by Vanderbilt University's Innovation Catalyst Fund. The content is solely the responsibility of the authors and does not necessarily represent the official views of Vanderbilt University.

\onecolumn
\appendix
\section{Protocol record and deviations}
\label{app:protocol}
\setcounter{table}{0}
All design decisions were recorded, with their reasons, in a dated decision log under version control, and every experiment records the code commit, configuration, seeds and start time. Table~\ref{tab:protocol} lists the decisions that define the evaluation with the commit that introduced each. The study protocol (settings, policies, endpoints, statistical analysis and readings), the amended calibration rule and the stable certification rule were committed before the first run on the test split (\SuppFirstTestRun{} UTC, commit \SuppFirstTestCommit). Of all \SuppRunsScanned{} logged runs, including pilots and stopped runs, \SuppTestUseBefore{} used a test case before that time.

\begin{table}[H]
\centering
\caption{Protocol record: decisions defining the evaluation, by the commit of the decision log that introduced them. Rows above the marked line were fixed before any test case was used.}
\label{tab:protocol}
\footnotesize
\begin{tabular}{lll}
\toprule
Time (UTC) & Commit & Decision (decision log) \\
\midrule
2026-10-28 23:53 & \texttt{2f5c329} & Case split: development, calibration, test \\
2026-10-29 13:55 & \texttt{921c82d} & Realistic per-view uncertainty (redesign) \\
2026-10-29 14:58 & \texttt{1d76d94} & Noise amplitude $\eta=1$ ($\eta=0.5$ sensitivity) \\
2026-10-29 16:47 & \texttt{1653bbd} & Registration for realistic noise (tempering, search) \\
2026-10-29 17:59 & \texttt{26f603d} & Study protocol: settings, policies, endpoints, statistics \\
2026-10-29 18:29 & \texttt{b422ae6} & Calibration rule amended \\
2026-10-29 19:03 & \texttt{db54e3d} & Stable certification rule added \\
\midrule
2026-10-29 20:04 & \texttt{b8f2d20} & \emph{first run on the test split (S1, cases 1--3)} \\
\midrule
2026-10-29 21:36 & \texttt{5d8f9ea} & S1 reduced to 20 cases (compute) \\
2026-10-29 23:29 & \texttt{d349646} & S1 restored to 60 cases (rented GPUs) \\
2026-10-30 00:39 & \texttt{9d2d20f} & Gradient baseline fix; S4 re-run from scratch \\
2026-10-30 00:40 & \texttt{eae9946} & Real-image start set: post hoc diagnostic \\
2026-10-30 05:07 & \texttt{9d9edc3} & S2 deviation found; re-run \\
2026-10-30 13:50 & \texttt{e25e0e5} & Oracle-start ablation (pre-specified) \\
2026-10-31 12:35 & \texttt{909596f} & Oracle-basin ablation (post hoc) \\
2026-10-31 22:35 & \texttt{8202cfc} & Common five-shot horizon; false stops (post hoc) \\
2026-11-01 04:10 & \texttt{d140444} & S5 persistent model error: design and readings \\
\bottomrule
\end{tabular}
\end{table}

Besides fitting $\alpha$ and the ambiguity threshold, the 30 calibration cases served two earlier development checks, the registration accuracy across simulators and the calibration of the error predictor across simulators. The first attempt of the registration check was stopped after one calibration case showed a failure mode not seen on the development cases (a small fragment moved away by the coarse translation search under the physics-based simulator); the registration was revised, the revision checked on development cases, and the check re-run from scratch.

Deviations from the original study design: (i)~it planned, besides the PENGWIN fractures with several displacement draws per case, 50--100 intact CTs with synthetic fracture lines; the studies used the PENGWIN fracture segmentations with one displacement draw per case. (ii)~Random views were run with 5 seeds (S1) and 3 (S2--S4) instead of 10; S2 and S3 used 15 cases and five policies. (iii)~S1 was specified for the first 40 test cases, with the remaining 20 ``if time remains''; it was cut to 20 cases when the cost on one local GPU was measured and restored to 60 when rented GPUs became available, when only cases 1--3 had been run and none analysed. (iv)~DeepDRR \citep{unberath2018deepdrr} was not used in the planning studies, because the realistic per-view noise dominates the measured images; it validated the registration and the calibration. (v)~Not run: an oracle view selection, planning with block-diagonal Fisher information, the prior strength, finite-difference versus gradient-volume Jacobians in planning, an overlap penalty and the displacement range. (vi)~The first S2 run did not restart the untracked registration from the prior mean and did not confine camera corrections before the residual temperature was known; both were fixed and S2 re-run on the same cases (supplementary material). (vii)~The gradient heuristic crashed on the clustered real views of S4 (no candidate 20$^\circ$ from all acquired views); its separation threshold now halves until a candidate qualifies, and S4 was re-run from scratch (S1--S3 are unaffected). (viii)~An internal review after the results found the censoring asymmetry of the pre-specified horizon (Section~\ref{sec:experiments}); all comparisons use the common five-shot horizon, and the pre-specified results are in \ref{app:prespec}. (ix)~Defined after the results and exploratory: false stops, shots to a correct certification, the comparison of every policy with the protocol, the certificate ablations, the detection of wrong registrations, the stop-rule controls, the preference for the stable rule, the second-view analysis, the overlap stratification, the wrong-basin analysis, the real-image start-set comparison and the oracle-basin ablation. (x)~S5 was added after an outside review of the results; its design, the comparison with S1 and its readings were fixed before it ran (Table~\ref{tab:protocol}), and the paired analysis on identical views was defined after its results.

The ambiguity threshold is the largest value on a fixed grid ($\ln 20$, 4, 5, 6, 8, 10, 12, 15, 20, 30 nats) whose rate of flagging correctly registered calibration fragments ($\le$5\,mm) is at most 2\%, and $\ln 20$ if none qualifies; none did ($\ln 20$ flagged \SuppCalibNonGrossFlag).

\section{Pre-specified analysis}
\label{app:prespec}
\setcounter{table}{0}
Tables~\ref{tab:prespec}--\ref{tab:stable} give the main comparisons on the pre-specified horizon (the budget, unreached endpoints counted as budget plus one). Because the clinical protocol has five views, this counts its unreached endpoints as 7 while the other policies get a sixth shot, so minimax's certification advantages over the protocol are larger here than on the common horizon; the stable-rule advantage in S1 is significant here but not on the common horizon. On the true-error endpoint both analyses agree.

\begin{table}[H]
\centering
\caption{Pre-specified horizon, 2\,mm: shots until the true worst-fragment mTRE is $\le$2\,mm; to certification under the local rule with false certifications over certified shots and the fraction of trajectories with a false certification; and under the stable rule.}
\label{tab:prespec}
\footnotesize
\begin{tabular}{llrrrrrr}
\toprule
 & & \multicolumn{1}{c}{True $\le$ 2\,mm} & \multicolumn{3}{c}{Local rule} & \multicolumn{2}{c}{Stable rule} \\
\cmidrule(lr){3-3}\cmidrule(lr){4-6}\cmidrule(lr){7-8}
Setting & Policy & shots & shots & false/cert. & false traj. & shots & false traj. \\
\midrule
S1 tracked, synthetic ($n=60$) & Minimax (ours) & 3.88 & 4.47 & 16/148 & 20\% & 5.20 & 3\% \\
 & A-optimal & 4.32 & 4.55 & 20/138 & 20\% & 5.40 & 2\% \\
 & D-optimal & 3.67 & 4.62 & 10/141 & 10\% & 5.13 & 2\% \\
 & E-optimal & 3.72 & 4.98 & 10/118 & 13\% & 5.52 & 3\% \\
 & EIG (full) & 3.95 & 4.53 & 14/147 & 18\% & 5.20 & 2\% \\
 & EIG (diagonal) & 3.72 & 4.77 & 8/132 & 10\% & 5.22 & 2\% \\
 & Gradient heuristic & 3.40 & 4.97 & 6/121 & 10\% & 5.33 & 3\% \\
 & Clinical protocol & 4.03 & 5.13 & 13/74 & 17\% & 5.77 & 5\% \\
 & Max separation & 5.67 & 5.62 & 34/74 & 35\% & 6.45 & 5\% \\
 & Random & 4.23 & 5.32 & 60/499 & 15\% & 5.68 & 5\% \\
\midrule
S2 untracked, synthetic ($n=15$) & Minimax (ours) & 4.93 & 5.47 & 1/22 & 7\% & 6.00 & 0\% \\
 & A-optimal & 5.47 & 5.53 & 4/20 & 20\% & 5.93 & 13\% \\
 & Clinical protocol & 4.93 & 6.47 & 0/5 & 0\% & 6.67 & 0\% \\
 & Max separation & 6.33 & 6.47 & 2/3 & 13\% & 7.00 & 0\% \\
 & Random & 5.22 & 6.18 & 1/36 & 2\% & 6.29 & 2\% \\
\midrule
S3 lower noise, synthetic ($n=15$) & Minimax (ours) & 3.27 & 3.80 & 3/47 & 20\% & 4.33 & 13\% \\
 & A-optimal & 3.27 & 3.73 & 4/49 & 20\% & 4.33 & 13\% \\
 & Clinical protocol & 3.00 & 4.67 & 0/25 & 0\% & 4.93 & 0\% \\
 & Max separation & 4.60 & 5.13 & 2/25 & 13\% & 5.47 & 7\% \\
 & Random & 3.36 & 4.89 & 6/92 & 9\% & 5.33 & 2\% \\
\midrule
S4 real images ($n=20$) & Minimax (ours) & 1.40 & 2.20 & 0/76 & 0\% & 2.70 & 0\% \\
 & A-optimal & 1.40 & 2.20 & 0/76 & 0\% & 2.90 & 0\% \\
 & D-optimal & 1.40 & 2.55 & 0/69 & 0\% & 3.05 & 0\% \\
 & Gradient heuristic & 1.40 & 2.55 & 0/69 & 0\% & 3.00 & 0\% \\
 & Max separation & 1.55 & 3.00 & 0/60 & 0\% & 3.40 & 0\% \\
 & Random & 1.77 & 4.32 & 0/101 & 0\% & 4.53 & 0\% \\
 & PolyPose pairs & 2.55 & 4.40 & 0/8 & 0\% & 5.00 & 0\% \\
\bottomrule
\end{tabular}
\end{table}

\begin{table}[H]
\centering
\caption{Pre-specified horizon, 2\,mm: minimax minus each baseline (S1 and S4). S4 shows differences only: its 20 starting poses come from 5 specimens and are not independent cases.}
\label{tab:prespecpaired}
\footnotesize
\begin{tabular}{llrlrrlr}
\toprule
 & & \multicolumn{3}{c}{Shots to true $\le$ 2\,mm} & \multicolumn{3}{c}{Shots to certification} \\
Setting & Baseline & diff. & 95\% CI & Holm $p$ & diff. & 95\% CI & Holm $p$ \\
\midrule
S1 tracked, synthetic & A-optimal & $-$0.43 & [$-$0.82, $-$0.07] & 0.230 & $-$0.08 & [$-$0.33, 0.15] & 0.657 \\
 & D-optimal & +0.22 & [$-$0.13, 0.58] & 0.732 & $-$0.15 & [$-$0.33, 0.05] & 0.140 \\
 & E-optimal & +0.17 & [$-$0.23, 0.57] & 1.000 & $-$0.52 & [$-$0.80, $-$0.27] & $<$0.001 \\
 & EIG (full) & $-$0.07 & [$-$0.40, 0.27] & 1.000 & $-$0.07 & [$-$0.22, 0.10] & 0.386 \\
 & EIG (diagonal) & +0.17 & [$-$0.17, 0.52] & 1.000 & $-$0.30 & [$-$0.50, $-$0.08] & 0.005 \\
 & Gradient heuristic & +0.48 & [0.10, 0.88] & 0.230 & $-$0.50 & [$-$0.73, $-$0.27] & $<$0.001 \\
 & Clinical protocol & $-$0.15 & [$-$0.62, 0.32] & 1.000 & $-$0.67 & [$-$0.92, $-$0.43] & $<$0.001 \\
 & Max separation & $-$1.78 & [$-$2.22, $-$1.37] & $<$0.001 & $-$1.15 & [$-$1.48, $-$0.83] & $<$0.001 \\
 & Random & $-$0.35 & [$-$0.67, $-$0.04] & 0.413 & $-$0.85 & [$-$1.07, $-$0.64] & $<$0.001 \\
\midrule
S4 real images & A-optimal & +0.00 & -- & -- & +0.00 & -- & -- \\
 & D-optimal & +0.00 & -- & -- & $-$0.35 & -- & -- \\
 & Gradient heuristic & +0.00 & -- & -- & $-$0.35 & -- & -- \\
 & Max separation & $-$0.15 & -- & -- & $-$0.80 & -- & -- \\
 & Random & $-$0.37 & -- & -- & $-$2.12 & -- & -- \\
 & PolyPose pairs & $-$1.15 & -- & -- & $-$2.20 & -- & -- \\
\bottomrule
\end{tabular}
\end{table}

\begin{table}[H]
\centering
\caption{Pre-specified horizon, 2\,mm: shots to certification under the stable rule, means and minimax minus each baseline (S4: differences only, as in Table~\ref{tab:prespecpaired}).}
\label{tab:stable}
\footnotesize
\begin{tabular}{llrrrlr}
\toprule
Setting & Baseline & minimax & baseline & diff. & 95\% CI & Holm $p$ \\
\midrule
S1 & A-optimal & 5.20 & 5.40 & $-$0.20 & [$-$0.47, 0.03] & 0.907 \\
 & D-optimal & 5.20 & 5.13 & +0.07 & [$-$0.13, 0.28] & 1.000 \\
 & E-optimal & 5.20 & 5.52 & $-$0.32 & [$-$0.57, $-$0.08] & 0.115 \\
 & EIG (full) & 5.20 & 5.20 & +0.00 & [$-$0.23, 0.23] & 1.000 \\
 & EIG (diagonal) & 5.20 & 5.22 & $-$0.02 & [$-$0.23, 0.20] & 1.000 \\
 & Gradient heuristic & 5.20 & 5.33 & $-$0.13 & [$-$0.38, 0.10] & 0.907 \\
 & Clinical protocol & 5.20 & 5.77 & $-$0.57 & [$-$0.90, $-$0.25] & 0.006 \\
 & Max separation & 5.20 & 6.45 & $-$1.25 & [$-$1.62, $-$0.92] & $<$0.001 \\
 & Random & 5.20 & 5.68 & $-$0.48 & [$-$0.69, $-$0.26] & 0.002 \\
\midrule
S2 & A-optimal & 6.00 & 5.93 & +0.07 & [$-$0.27, 0.40] & 0.988 \\
 & Clinical protocol & 6.00 & 6.67 & $-$0.67 & [$-$1.27, $-$0.20] & 0.078 \\
 & Max separation & 6.00 & 7.00 & $-$1.00 & [$-$1.60, $-$0.40] & 0.036 \\
 & Random & 6.00 & 6.29 & $-$0.29 & [$-$0.80, 0.11] & 0.988 \\
\midrule
S3 & A-optimal & 4.33 & 4.33 & +0.00 & [$-$0.33, 0.40] & 1.000 \\
 & Clinical protocol & 4.33 & 4.93 & $-$0.60 & [$-$1.07, $-$0.20] & 0.060 \\
 & Max separation & 4.33 & 5.47 & $-$1.13 & [$-$1.87, $-$0.40] & 0.060 \\
 & Random & 4.33 & 5.33 & $-$1.00 & [$-$1.71, $-$0.33] & 0.060 \\
\midrule
S4 & A-optimal & 2.70 & 2.90 & $-$0.20 & -- & -- \\
 & D-optimal & 2.70 & 3.05 & $-$0.35 & -- & -- \\
 & Gradient heuristic & 2.70 & 3.00 & $-$0.30 & -- & -- \\
 & Max separation & 2.70 & 3.40 & $-$0.70 & -- & -- \\
 & Random & 2.70 & 4.53 & $-$1.83 & -- & -- \\
 & PolyPose pairs & 2.70 & 5.00 & $-$2.30 & -- & -- \\
\bottomrule
\end{tabular}
\end{table}

\section{Study configuration}
\label{app:config}
\setcounter{table}{0}
Table~\ref{tab:config} lists the settings of the planning studies.

\begin{table}[H]
\centering
\caption{Registration, certification and simulator settings of the planning studies (S1; S2--S4 differ where stated in Section~\ref{sec:experiments}).}
\label{tab:config}
\small
\begin{tabular}{p{0.4\linewidth}p{0.55\linewidth}}
\toprule
Setting & Value \\
\midrule
Pose prior (per axis) & 5$^\circ$, 10\,mm \\
Working resolutions & 32\,px, 64\,px, 128\,px \\
Adam step sizes (rotation, translation) & 0.01\,rad, 1\,mm (scaled by pixel size per resolution) \\
Iterations per stage (max.), plateau patience & 300, 10 \\
Trust region per stage (coarsest) & 30$^\circ$, 20\,mm (halved per finer resolution) \\
Starts & 2 (previous estimate, prior mean); selection by the polished MAP objective \\
Per-fragment translation search & $\pm$30\,mm, 5\,mm grid at 64\,px; 2 alternatives $\ge$7.5\,mm apart \\
Tempering factor $\gamma$ & 4 \\
Quasi-Newton polish (float64) & tolerance 0.0001\,mm, at most 100 iterations \\
Calibration factor $\alpha$ & 2.411 \\
Ambiguity threshold & 2.996 nats ($\ln 20$) \\
Plausibility threshold & 4.10 prior standard deviations \\
Monte-Carlo samples per bound & 2000 \\
Surface targets per fragment & 300 \\
Simulator: photons per pixel, scatter & 10000; 0.5 $\times$ Gaussian blur ($\sigma=50$\,mm) of the primary \\
Simulator: detector blur, electronic noise & 0.5\,px, $\sigma=2$ counts \\
Simulator: gain, offset, gamma & U[0.5, 2.0], U[$-$1.0, 1.0], U[0.9, 1.1] \\
Realistic noise amplitude $\eta$ & 1 (S3: 0.5) \\
Budget (shots) & 6 (S4: 5) \\
\bottomrule
\end{tabular}
\end{table}
\twocolumn

\clearpage
\onecolumn
\phantomsection\pdfbookmark[0]{Supplementary material}{supplementary}
\begin{center}{\Large\bfseries Supplementary material}\end{center}
\vspace{0.5em}
\makeatletter
\setcounter{section}{0}\setcounter{subsection}{0}\setcounter{table}{0}\setcounter{figure}{0}\setcounter{equation}{0}
\gdef\thesection{S\arabic{section}}
\gdef\thetable{S\arabic{table}}
\gdef\thefigure{S\arabic{figure}}
\gdef\theequation{S\arabic{section}.\arabic{equation}}
\gdef\theHsection{supp.\arabic{section}}
\gdef\theHtable{supp.\arabic{table}}
\gdef\theHfigure{supp.\arabic{figure}}
\gdef\theHequation{supp.\arabic{section}.\arabic{equation}}
\makeatother
\setlength{\tabcolsep}{4pt}
\noindent Settings S1--S5, policies, endpoints and statistics are as defined in the main paper; the protocol record, the pre-specified analysis and the study configuration are in its appendices. Every number below is generated from a logged run (code version, configuration, seeds and time recorded) by the same table generator as the paper's numbers. ``Minimax $-$ baseline'' differences are means over cases (random views averaged over seeds within a case; negative: minimax needs fewer shots) with case-bootstrap 95\% confidence intervals (unadjusted) and Wilcoxon signed-rank $p$-values Holm-adjusted over the baselines of a setting and endpoint. \emph{Common horizon}: all policies truncated at five shots, unreached endpoints counted as 6 (the paper's analysis). \emph{Pre-specified horizon}: the budget (6; S4: 5), unreached endpoints counted as budget plus one, which counts the five-view clinical protocol's unreached endpoints as 7.

\FloatBarrier
\section{Untracked C-arm and lower noise: paired comparisons}
\begin{table}[!htbp]
\centering
\caption{S2 (untracked C-arm) and S3 (lower noise), 15 cases each, common horizon: minimax minus each baseline. The per-policy means are in the paper's Table~2.}
\label{tab:pairedh5}
\scriptsize
\begin{tabular}{llrlrrlrrlr}
\toprule
 & & \multicolumn{3}{c}{True $\le$ 2\,mm} & \multicolumn{3}{c}{Certified (local rule)} & \multicolumn{3}{c}{Correctly certified (stable rule)} \\
\cmidrule(lr){3-5}\cmidrule(lr){6-8}\cmidrule(lr){9-11}
Setting & Baseline & diff. & 95\% CI & Holm $p$ & diff. & 95\% CI & Holm $p$ & diff. & 95\% CI & Holm $p$ \\
\midrule
S2 & A-optimal & $-$0.33 & [$-$0.80, 0.00] & 0.251 & +0.00 & [$-$0.27, 0.27] & 1.000 & +0.00 & [$-$0.40, 0.40] & 1.000 \\
 & Clinical protocol & +0.13 & [$-$0.60, 0.93] & 1.000 & $-$0.60 & [$-$1.00, $-$0.20] & 0.060 & $-$0.33 & [$-$0.73, 0.00] & 0.251 \\
 & Max separation & $-$1.00 & [$-$1.67, $-$0.47] & 0.022 & $-$0.60 & [$-$1.20, $-$0.07] & 0.160 & $-$0.53 & [$-$0.93, $-$0.13] & 0.104 \\
 & Random & $-$0.16 & [$-$0.58, 0.27] & 1.000 & $-$0.42 & [$-$0.80, $-$0.11] & 0.060 & $-$0.16 & [$-$0.51, 0.13] & 1.000 \\
\midrule
S3 & A-optimal & +0.07 & [$-$0.67, 0.73] & 1.000 & $-$0.07 & [$-$0.27, 0.13] & 0.564 & $-$0.07 & [$-$0.47, 0.33] & 0.922 \\
 & Clinical protocol & +0.27 & [$-$0.40, 1.00] & 1.000 & $-$0.73 & [$-$1.13, $-$0.33] & 0.011 & $-$0.33 & [$-$0.67, $-$0.07] & 0.179 \\
 & Max separation & $-$1.20 & [$-$1.80, $-$0.67] & 0.014 & $-$1.20 & [$-$1.73, $-$0.67] & 0.009 & $-$0.80 & [$-$1.47, $-$0.13] & 0.179 \\
 & Random & $-$0.11 & [$-$0.62, 0.47] & 1.000 & $-$0.96 & [$-$1.42, $-$0.51] & 0.010 & $-$0.69 & [$-$1.27, $-$0.13] & 0.179 \\
\bottomrule
\end{tabular}
\end{table}

\FloatBarrier
\section{Results at 1 and 3\,mm}
Table~\ref{tab:thresholds} gives the endpoints at the 1 and 3\,mm thresholds for every setting and policy, and Tables~\ref{tab:paired1}--\ref{tab:paired3} the paired comparisons, all on the pre-specified horizon. At 1\,mm few cases certify within the budget in the synthetic settings, so most certification endpoints are censored.

\begin{table}[!htbp]
\centering
\caption{Pre-specified horizon: mean shots until the true worst-fragment mTRE is $\le t$ (``true'') and until certification under the local rule (``certified''), and false certifications over certified shots, at $t=1$ and 3\,mm.}
\label{tab:thresholds}
\small
\begin{tabular}{llrrrrrr}
\toprule
 & & \multicolumn{3}{c}{$t=1$\,mm} & \multicolumn{3}{c}{$t=3$\,mm} \\
\cmidrule(lr){3-5}\cmidrule(lr){6-8}
Setting & Policy & true & certified & false/cert. & true & certified & false/cert. \\
\midrule
S1 & Minimax (ours) & 4.92 & 6.52 & 0/29 & 3.55 & 3.38 & 36/208 \\
 & A-optimal & 5.10 & 6.57 & 0/26 & 3.98 & 3.52 & 37/192 \\
 & D-optimal & 4.65 & 6.60 & 0/24 & 3.48 & 3.48 & 26/207 \\
 & E-optimal & 5.13 & 6.63 & 0/22 & 3.43 & 3.73 & 19/189 \\
 & EIG (full) & 4.83 & 6.62 & 0/23 & 3.77 & 3.50 & 34/207 \\
 & EIG (diagonal) & 4.97 & 6.70 & 0/18 & 3.52 & 3.60 & 25/198 \\
 & Gradient heuristic & 4.75 & 6.65 & 0/21 & 3.17 & 3.83 & 20/186 \\
 & Clinical protocol & 5.08 & 6.77 & 0/8 & 3.78 & 3.92 & 28/130 \\
 & Max separation & 6.33 & 6.88 & 1/7 & 5.48 & 4.70 & 59/122 \\
 & Random & 5.42 & 6.82 & 0/53 & 3.85 & 4.16 & 126/810 \\
\midrule
S2 & Minimax (ours) & 5.53 & 6.80 & 0/3 & 4.27 & 4.20 & 8/36 \\
 & A-optimal & 6.13 & 6.87 & 0/2 & 5.00 & 4.53 & 11/33 \\
 & Clinical protocol & 6.20 & 6.87 & 0/1 & 4.87 & 5.20 & 3/19 \\
 & Max separation & 6.80 & 7.00 & 0/0 & 6.00 & 5.47 & 7/12 \\
 & Random & 6.09 & 7.00 & 0/0 & 4.73 & 5.42 & 11/68 \\
\midrule
S3 & Minimax (ours) & 4.13 & 6.40 & 0/9 & 3.00 & 2.93 & 5/59 \\
 & A-optimal & 4.73 & 6.40 & 0/9 & 3.20 & 2.73 & 10/64 \\
 & Clinical protocol & 4.13 & 6.73 & 0/3 & 2.80 & 3.40 & 3/40 \\
 & Max separation & 5.20 & 6.93 & 0/1 & 4.20 & 4.07 & 9/40 \\
 & Random & 4.71 & 6.78 & 0/10 & 3.04 & 3.73 & 13/136 \\
\midrule
S4 & Minimax (ours) & 1.80 & 3.60 & 0/48 & 1.35 & 1.90 & 0/82 \\
 & A-optimal & 1.80 & 3.60 & 0/48 & 1.35 & 1.90 & 0/82 \\
 & D-optimal & 2.10 & 4.00 & 0/40 & 1.35 & 1.90 & 0/82 \\
 & Gradient heuristic & 1.80 & 3.80 & 0/44 & 1.35 & 2.25 & 0/75 \\
 & Max separation & 2.20 & 4.60 & 0/28 & 1.50 & 2.70 & 0/66 \\
 & Random & 2.83 & 5.05 & 0/57 & 1.52 & 3.52 & 1/149 \\
 & PolyPose pairs & 3.65 & 6.00 & 0/0 & 2.55 & 4.40 & 0/8 \\
\bottomrule
\end{tabular}
\end{table}

\begin{table}[!htbp]
\centering
\caption{Pre-specified horizon: minimax minus each baseline at 1\,mm.}
\label{tab:paired1}
\small
\begin{tabular}{llrlrrlr}
\toprule
 & & \multicolumn{3}{c}{Shots to true $\le t$} & \multicolumn{3}{c}{Shots to certification (local)} \\
\cmidrule(lr){3-5}\cmidrule(lr){6-8}
Setting & Baseline & diff. & 95\% CI & Holm $p$ & diff. & 95\% CI & Holm $p$ \\
\midrule
S1 & A-optimal & $-$0.18 & [$-$0.55, 0.20] & 0.978 & $-$0.05 & [$-$0.13, 0.03] & 0.308 \\
 & D-optimal & +0.27 & [$-$0.13, 0.65] & 0.978 & $-$0.08 & [$-$0.20, 0.02] & 0.308 \\
 & E-optimal & $-$0.22 & [$-$0.65, 0.20] & 1.000 & $-$0.12 & [$-$0.22, $-$0.03] & 0.072 \\
 & EIG (full) & +0.08 & [$-$0.28, 0.45] & 1.000 & $-$0.10 & [$-$0.22, 0.00] & 0.280 \\
 & EIG (diagonal) & $-$0.05 & [$-$0.38, 0.27] & 1.000 & $-$0.18 & [$-$0.33, $-$0.07] & 0.066 \\
 & Gradient heuristic & +0.17 & [$-$0.17, 0.50] & 0.978 & $-$0.13 & [$-$0.27, $-$0.03] & 0.132 \\
 & Clinical protocol & $-$0.17 & [$-$0.58, 0.27] & 0.978 & $-$0.25 & [$-$0.42, $-$0.12] & 0.011 \\
 & Max separation & $-$1.42 & [$-$1.80, $-$1.05] & $<$0.001 & $-$0.37 & [$-$0.58, $-$0.18] & 0.004 \\
 & Random & $-$0.51 & [$-$0.79, $-$0.22] & 0.010 & $-$0.31 & [$-$0.48, $-$0.15] & 0.003 \\
\midrule
S2 & A-optimal & $-$0.60 & [$-$1.20, $-$0.13] & 0.078 & $-$0.07 & [$-$0.20, 0.00] & 1.000 \\
 & Clinical protocol & $-$0.67 & [$-$1.67, 0.27] & 0.631 & $-$0.07 & [$-$0.20, 0.00] & 1.000 \\
 & Max separation & $-$1.27 & [$-$2.07, $-$0.53] & 0.036 & $-$0.20 & [$-$0.60, 0.00] & 1.000 \\
 & Random & $-$0.56 & [$-$1.36, 0.16] & 0.631 & $-$0.20 & [$-$0.60, 0.00] & 1.000 \\
\midrule
S3 & A-optimal & $-$0.60 & [$-$1.40, 0.07] & 0.821 & +0.00 & [0.00, 0.00] & 1.000 \\
 & Clinical protocol & +0.00 & [$-$0.53, 0.60] & 0.821 & $-$0.33 & [$-$0.73, 0.00] & 0.185 \\
 & Max separation & $-$1.07 & [$-$2.07, $-$0.20] & 0.149 & $-$0.53 & [$-$1.07, $-$0.13] & 0.185 \\
 & Random & $-$0.58 & [$-$1.42, 0.11] & 0.821 & $-$0.38 & [$-$0.73, $-$0.07] & 0.185 \\
\midrule
S4 & A-optimal & +0.00 & -- & -- & +0.00 & -- & -- \\
 & D-optimal & $-$0.30 & -- & -- & $-$0.40 & -- & -- \\
 & Gradient heuristic & +0.00 & -- & -- & $-$0.20 & -- & -- \\
 & Max separation & $-$0.40 & -- & -- & $-$1.00 & -- & -- \\
 & Random & $-$1.03 & -- & -- & $-$1.45 & -- & -- \\
 & PolyPose pairs & $-$1.85 & -- & -- & $-$2.40 & -- & -- \\
\bottomrule
\end{tabular}
\end{table}

\begin{table}[!htbp]
\centering
\caption{Pre-specified horizon: minimax minus each baseline at 3\,mm.}
\label{tab:paired3}
\small
\begin{tabular}{llrlrrlr}
\toprule
 & & \multicolumn{3}{c}{Shots to true $\le t$} & \multicolumn{3}{c}{Shots to certification (local)} \\
\cmidrule(lr){3-5}\cmidrule(lr){6-8}
Setting & Baseline & diff. & 95\% CI & Holm $p$ & diff. & 95\% CI & Holm $p$ \\
\midrule
S1 & A-optimal & $-$0.43 & [$-$0.82, $-$0.07] & 0.391 & $-$0.13 & [$-$0.43, 0.15] & 0.609 \\
 & D-optimal & +0.07 & [$-$0.28, 0.43] & 1.000 & $-$0.10 & [$-$0.27, 0.07] & 0.497 \\
 & E-optimal & +0.12 & [$-$0.30, 0.53] & 1.000 & $-$0.35 & [$-$0.62, $-$0.08] & 0.032 \\
 & EIG (full) & $-$0.22 & [$-$0.58, 0.15] & 0.889 & $-$0.12 & [$-$0.30, 0.07] & 0.336 \\
 & EIG (diagonal) & +0.03 & [$-$0.28, 0.37] & 1.000 & $-$0.22 & [$-$0.47, 0.05] & 0.089 \\
 & Gradient heuristic & +0.38 & [0.02, 0.77] & 0.544 & $-$0.45 & [$-$0.72, $-$0.17] & 0.002 \\
 & Clinical protocol & $-$0.23 & [$-$0.67, 0.18] & 1.000 & $-$0.53 & [$-$0.83, $-$0.23] & 0.001 \\
 & Max separation & $-$1.93 & [$-$2.40, $-$1.47] & $<$0.001 & $-$1.32 & [$-$1.70, $-$0.93] & $<$0.001 \\
 & Random & $-$0.30 & [$-$0.65, 0.04] & 0.513 & $-$0.78 & [$-$1.02, $-$0.54] & $<$0.001 \\
\midrule
S2 & A-optimal & $-$0.73 & [$-$1.33, $-$0.20] & 0.055 & $-$0.33 & [$-$1.00, 0.33] & 0.554 \\
 & Clinical protocol & $-$0.60 & [$-$1.60, 0.40] & 0.509 & $-$1.00 & [$-$1.80, $-$0.20] & 0.079 \\
 & Max separation & $-$1.73 & [$-$2.40, $-$1.13] & 0.003 & $-$1.27 & [$-$2.20, $-$0.40] & 0.079 \\
 & Random & $-$0.47 & [$-$1.22, 0.29] & 0.509 & $-$1.22 & [$-$1.76, $-$0.73] & 0.006 \\
\midrule
S3 & A-optimal & $-$0.20 & [$-$0.87, 0.40] & 1.000 & +0.20 & [0.00, 0.40] & 0.167 \\
 & Clinical protocol & +0.20 & [$-$0.53, 1.07] & 1.000 & $-$0.47 & [$-$1.13, 0.07] & 0.212 \\
 & Max separation & $-$1.20 & [$-$2.00, $-$0.40] & 0.094 & $-$1.13 & [$-$1.87, $-$0.53] & 0.016 \\
 & Random & $-$0.04 & [$-$0.76, 0.84] & 0.960 & $-$0.80 & [$-$1.33, $-$0.38] & 0.014 \\
\midrule
S4 & A-optimal & +0.00 & -- & -- & +0.00 & -- & -- \\
 & D-optimal & +0.00 & -- & -- & +0.00 & -- & -- \\
 & Gradient heuristic & +0.00 & -- & -- & $-$0.35 & -- & -- \\
 & Max separation & $-$0.15 & -- & -- & $-$0.80 & -- & -- \\
 & Random & $-$0.17 & -- & -- & $-$1.62 & -- & -- \\
 & PolyPose pairs & $-$1.20 & -- & -- & $-$2.50 & -- & -- \\
\bottomrule
\end{tabular}
\end{table}

\FloatBarrier
\section{Projected overlap, fragment volume and fragment count}
\label{sec:strata}
\IfFileExists{generated/supp_overlap.tex}{%
\paragraph{Projected overlap}
We define the overlap of a fragment in a view as the share of bone attenuation inside its projected footprint (pixels where its own projection exceeds 5\% of its maximum) that belongs to other bone: the other fragments at their true poses and the static bone, without soft tissue. It is 0 when nothing else projects onto the fragment and approaches 1 when the fragment is buried under other bone. Table~\ref{tab:overlap} splits the \SuppOvBodies{} S1 fragments into tertiles of their overlap in the AP view, the first shot of every trajectory. Overlap and volume were correlated (Spearman $\rho=\SuppOvSpearman$): large fragments dominate their own footprints. Gross failures were more frequent for overlapped fragments (\SuppOvHighGross{} in the highest tertile against \SuppOvLowGross{} in the lowest), most clearly among fragments of 35\,cm$^3$ or more (\SuppOvLargeGrossHigh{} against \SuppOvLargeGrossLow; under 35\,cm$^3$: \SuppOvSmallGrossHigh{} against \SuppOvSmallGrossLow). Minimax did not bring overlapped fragments below 2\,mm sooner than D-optimal design (highest tertile: \SuppOvHighDiffDopt{} shots, Holm $p$ \SuppOvHighDiffDoptHolm).

\begin{table}[!htbp]
\centering
\caption{S1 fragments by tertile of projected overlap in the AP view: mean shots until the fragment's own true mTRE is $\le$2\,mm (per case the mean over its fragments in the tertile; pre-specified horizon, every listed policy has six shots), minimax minus D-optimal design (Holm over the baselines within the tertile), and over unique registrations the rate of gross failures and the coverage of the calibrated 95\% bounds (unflagged fragments; all, and those within 5\,mm of the truth).}
\label{tab:overlap}
\small
\begin{tabular}{lrrrrrlrrr}
\toprule
Overlap (AP) & fragments & minimax & D-opt. & A-opt. & \multicolumn{2}{c}{minimax $-$ D-opt.} & gross & \multicolumn{2}{c}{coverage} \\
\cmidrule(lr){6-7}\cmidrule(lr){9-10}
 & & & & & diff. & Holm $p$ & $>$5\,mm & unflagged & non-gross \\
\midrule
low (0.00--0.09) & 74 & 1.63 & 1.67 & 1.68 & $-$0.04 & 1.000 & 5\% & 89\% & 91\% \\
middle (0.09--0.34) & 73 & 2.42 & 2.17 & 2.49 & +0.25 & 0.177 & 14\% & 83\% & 91\% \\
high (0.34--0.85) & 74 & 3.04 & 2.75 & 3.32 & +0.28 & 0.562 & 26\% & 77\% & 92\% \\
\bottomrule
\end{tabular}
\end{table}
}{}

\paragraph{Coverage by fragment volume}
Table~\ref{tab:covvol} gives the coverage of the bounds by fragment volume: gross failures were most frequent below 15\,cm$^3$ and at 35--100\,cm$^3$ and rare from 100\,cm$^3$, and among fragments registered within 5\,mm the single calibration factor is slightly optimistic for the largest fragments.

\begin{table}[!htbp]
\centering
\caption{Unflagged fragment registrations by volume (unique registrations per case and acquired view set): rate of gross failures and coverage of the calibrated 95\% bounds, all and within 5\,mm of the truth.}
\label{tab:covvol}
\small
\begin{tabular}{llrrrr}
\toprule
Setting & Volume (cm$^3$) & registrations & gross ($>$5\,mm) & coverage & coverage, non-gross \\
\midrule
S1 & 0--15 & 2690 & 19\% & 78\% & 95\% \\
 & 15--35 & 3981 & 8\% & 89\% & 96\% \\
 & 35--100 & 3007 & 15\% & 76\% & 89\% \\
 & $\ge$100 & 4199 & 1\% & 86\% & 88\% \\
\midrule
S2 & 0--15 & 223 & 24\% & 71\% & 90\% \\
 & 15--35 & 450 & 11\% & 84\% & 93\% \\
 & 35--100 & 164 & 34\% & 57\% & 83\% \\
 & $\ge$100 & 352 & 6\% & 78\% & 82\% \\
\midrule
S3 & 0--15 & 321 & 17\% & 81\% & 96\% \\
 & 15--35 & 632 & 2\% & 96\% & 98\% \\
 & 35--100 & 268 & 15\% & 77\% & 90\% \\
 & $\ge$100 & 510 & 0\% & 88\% & 88\% \\
\bottomrule
\end{tabular}
\end{table}

\paragraph{Number of moving fragments}
Table~\ref{tab:k} splits the S1 cases by the number $K$ of moving fragments (pre-specified horizon). The certification advantage over the fixed protocol was clear for $K\le4$ (for $K\le3$: \SuppKLowCertDiffFixed{} shots, \SuppKLowCertDiffFixedCI) and smaller for $K\ge5$ (\SuppKHighCertDiffFixed{} shots, \SuppKHighCertDiffFixedCI; \SuppKHighCases{} cases). On the true error no stratum showed a clear difference. The strata are small and were not pre-specified; we draw no conclusion from them.

\begin{table}[!htbp]
\centering
\caption{S1 by the number of moving fragments $K$ (2\,mm, local rule, pre-specified horizon): mean shots and minimax minus the fixed protocol and minus D-optimal design (case-bootstrap 95\% CI; exploratory, no multiplicity correction).}
\label{tab:k}
\small
\begin{tabular}{llrrrrlrl}
\toprule
Stratum & Endpoint & minimax & fixed & D-opt. & \multicolumn{2}{c}{minimax $-$ fixed} & \multicolumn{2}{c}{minimax $-$ D-opt.} \\
\cmidrule(lr){6-7}\cmidrule(lr){8-9}
 & & & & & diff. & 95\% CI & diff. & 95\% CI \\
\midrule
$K\le$3 ($n=26$) & true & 3.38 & 4.00 & 3.19 & $-$0.62 & [$-$1.38, 0.15] & +0.19 & [$-$0.31, 0.69] \\
 & certified & 3.54 & 4.35 & 3.77 & $-$0.81 & [$-$1.19, $-$0.46] & $-$0.23 & [$-$0.46, 0.00] \\
\midrule
$K$=4 ($n=21$) & true & 3.95 & 3.86 & 3.81 & +0.10 & [$-$0.67, 0.90] & +0.14 & [$-$0.48, 0.71] \\
 & certified & 5.24 & 6.00 & 5.24 & $-$0.76 & [$-$1.14, $-$0.43] & +0.00 & [$-$0.38, 0.43] \\
\midrule
$K\ge$5 ($n=13$) & true & 4.77 & 4.38 & 4.38 & +0.38 & [$-$0.31, 1.08] & +0.38 & [$-$0.54, 1.31] \\
 & certified & 5.08 & 5.31 & 5.31 & $-$0.23 & [$-$0.77, 0.31] & $-$0.23 & [$-$0.54, 0.08] \\
\bottomrule
\end{tabular}
\end{table}

\FloatBarrier
\section{Detecting failed registrations in the synthetic settings}
\label{sec:detection}
Table~\ref{tab:suppdetection} repeats the main paper's detection analysis (exploratory; fragment registrations from the second view on, gross failure: true mTRE $>$5\,mm) for S1, S3 and S2. In S2 an implausible camera estimate flags every fragment of the shot, which makes the flags more sensitive and both rules less specific.

\begin{table}[!htbp]
\centering
\caption{AUC of each run-time signal and sensitivity / specificity of the rules of the run, for all fragments and for those with a calibrated bound $\le$2\,mm, in S1, S3 and S2 (failure: true mTRE $>$5\,mm).}
\label{tab:suppdetection}
\footnotesize
\setlength{\tabcolsep}{3pt}
\begin{tabular}{lcccccc}
\toprule
 & \multicolumn{2}{c}{S1 ($>$ 5\,mm)} & \multicolumn{2}{c}{S3 ($>$ 5\,mm)} & \multicolumn{2}{c}{S2 ($>$ 5\,mm)} \\
\cmidrule(lr){2-3}\cmidrule(lr){4-5}\cmidrule(lr){6-7}
 & all & bound $\le$ 2\,mm & all & bound $\le$ 2\,mm & all & bound $\le$ 2\,mm \\
\midrule
Fragment registrations (failures) & 14864 (2153) & 10846 (679) & 1788 (169) & 1445 (63) & 1790 (495) & 1123 (153) \\
\midrule
\multicolumn{7}{l}{\emph{AUC}} \\
Prior distance & 0.84 & 0.91 & 0.88 & 0.90 & 0.88 & 0.90 \\
Ambiguity (objective gap) & 0.85 & 0.88 & 0.88 & 0.86 & 0.80 & 0.83 \\
Start disagreement & 0.83 & 0.87 & 0.86 & 0.85 & 0.82 & 0.86 \\
Shift (mm) & 0.91 & 0.94 & 0.91 & 0.93 & 0.88 & 0.91 \\
Shift / bound (instability) & 0.85 & 0.92 & 0.84 & 0.90 & 0.83 & 0.89 \\
Calibrated bound & 0.84 & 0.81 & 0.84 & 0.78 & 0.79 & 0.71 \\
Local image misfit & 0.75 & 0.84 & 0.71 & 0.81 & 0.78 & 0.89 \\
\midrule
\multicolumn{7}{l}{\emph{Rules of the run: sensitivity / specificity, \%}} \\
Flags (implausible or ambiguous) & 41 / 98 & 32 / 100 & 41 / 98 & 37 / 99 & 66 / 76 & 72 / 79 \\
Shift $>$ bound (stability check) & 78 / 79 & 86 / 84 & 76 / 83 & 81 / 85 & 87 / 71 & 90 / 76 \\
Either & 86 / 79 & 89 / 84 & 82 / 82 & 84 / 85 & 95 / 57 & 95 / 63 \\
\quad shift $>$ bound at shot 2 & 85 / 31 & -- & 90 / 36 & -- & 94 / 43 & -- \\
\quad shift $>$ bound from shot 3 & 75 / 89 & -- & 69 / 93 & -- & 84 / 77 & -- \\
\bottomrule
\end{tabular}
\end{table}

\FloatBarrier
\section{Model error shared by all views (S5)}
\label{sec:s5}
Table~\ref{tab:supps5false} lists every false first certificate of S5 at 2\,mm (five-shot horizon) with the violating fragment's volume, the Dice of its perturbed segmentation, its true error, calibrated bound and shift when the newest view was added, and its error in S1 when S1 acquired the same views. Classes: present in S1 (above 2\,mm in S1 on the same views), new (at most 2\,mm in S1; 2--5\,mm or above 5\,mm in S5), path differs (planner paths that differed from S1's). Table~\ref{tab:supps5paired} compares the errors of S1 and S5 where both acquired the same views (post hoc, exploratory), for fragments within 5\,mm in both settings.

\begin{table}[!htbp]
\centering
\caption{S5: false first certificates at 2\,mm under the stable and the local rule (errors, bounds and shifts in mm).}
\label{tab:supps5false}
\footnotesize
\setlength{\tabcolsep}{3pt}
\begin{tabular}{lllrrrrrrrl}
\toprule
Rule & Case & Policy & Shot & cm$^3$ & Dice & Error & Bound & Shift & S1 error & Class \\
\midrule
Stable & 019 & Random (2) & 4 & 43 & 0.87 & 6.46 & 1.70 & 1.07 & 6.26 & present in S1 \\
 & 031 & Minimax & 4 & 45 & 0.88 & 2.51 & 1.40 & 0.26 & -- & path differs \\
 & 031 & Random (2) & 4 & 45 & 0.88 & 4.75 & 1.39 & 1.35 & 5.91 & present in S1 \\
 & 074 & Minimax & 5 & 40 & 0.87 & 28.22 & 1.53 & 1.43 & -- & path differs \\
 & 092 & Max separation & 4 & 32 & 0.77 & 2.06 & 1.93 & 0.53 & 1.16 & new, 2--5\,mm \\
 & 093 & Protocol & 5 & 22 & 0.85 & 3.63 & 1.77 & 1.64 & 3.40 & present in S1 \\
\midrule
Local & 019 & Protocol & 3 & 43 & 0.87 & 23.67 & 1.76 & 5.23 & 23.54 & present in S1 \\
 & 019 & Random (1) & 4 & 43 & 0.87 & 15.83 & 1.99 & 19.88 & 16.40 & present in S1 \\
 & 019 & Random (2) & 3 & 43 & 0.87 & 7.17 & 1.94 & 2.15 & 7.28 & present in S1 \\
 & 031 & Minimax & 3 & 45 & 0.88 & 2.49 & 1.46 & 4.49 & -- & path differs \\
 & 031 & Random (2) & 4 & 45 & 0.88 & 4.75 & 1.39 & 1.35 & 5.91 & present in S1 \\
 & 074 & A-optimal & 4 & 40 & 0.87 & 15.52 & 1.67 & 28.83 & -- & path differs \\
 & 074 & Minimax & 4 & 40 & 0.87 & 28.28 & 1.77 & 25.58 & -- & path differs \\
 & 074 & Random (2) & 5 & 40 & 0.87 & 27.41 & 1.82 & 1.73 & 22.61 & present in S1 \\
 & 090 & Max separation & 3 & 32 & 0.93 & 7.13 & 1.61 & 11.23 & 1.17 & new, $>$5\,mm \\
 & 092 & Max separation & 4 & 32 & 0.77 & 2.06 & 1.93 & 0.53 & 1.16 & new, 2--5\,mm \\
 & 093 & Protocol & 5 & 22 & 0.85 & 3.63 & 1.77 & 1.64 & 3.40 & present in S1 \\
 & 097 & Protocol & 5 & 14 & 0.89 & 7.91 & 1.99 & 8.02 & 7.42 & present in S1 \\
\bottomrule
\end{tabular}
\end{table}

\begin{table}[!htbp]
\centering
\caption{S1 against S5 on identical acquired views (fragment registrations from the second view on, within 5\,mm in both settings), by the Dice of the fragment's perturbed segmentation, by fragment volume and by shot.}
\label{tab:supps5paired}
\footnotesize
\begin{tabular}{lrrrrrrr}
\toprule
 & & \multicolumn{2}{c}{median error (mm)} & \multicolumn{2}{c}{error $>$2\,mm} & \multicolumn{2}{c}{coverage} \\
\cmidrule(lr){3-4}\cmidrule(lr){5-6}\cmidrule(lr){7-8}
Fragments & $n$ & S1 & S5 & S1 & S5 & S1 & S5 \\
\midrule
\multicolumn{8}{l}{\emph{Dice}} \\
\quad $\le$0.80 & 63 & 0.73 & 0.99 & 9.5\% & 22.2\% & 100.0\% & 90.5\% \\
\quad 0.80--0.85 & 179 & 0.92 & 1.05 & 12.8\% & 12.3\% & 96.6\% & 96.1\% \\
\quad 0.85--0.90 & 423 & 0.64 & 0.70 & 4.5\% & 7.3\% & 89.4\% & 84.6\% \\
\quad 0.90--0.95 & 378 & 0.49 & 0.51 & 4.5\% & 5.3\% & 89.4\% & 88.9\% \\
\quad 0.95--1.00 & 22 & 0.74 & 0.72 & 0.0\% & 0.0\% & 100.0\% & 100.0\% \\
\multicolumn{8}{l}{\emph{Volume (cm$^3$)}} \\
\quad $<$15 & 152 & 0.91 & 1.07 & 13.2\% & 19.7\% & 98.0\% & 95.4\% \\
\quad 15--35 & 434 & 0.75 & 0.81 & 7.1\% & 8.1\% & 96.3\% & 94.2\% \\
\quad 35--100 & 111 & 0.98 & 1.03 & 12.6\% & 18.9\% & 83.8\% & 76.6\% \\
\quad $\ge$100 & 368 & 0.36 & 0.37 & 0.0\% & 0.3\% & 85.3\% & 83.2\% \\
\multicolumn{8}{l}{\emph{Shot}} \\
\quad 2 & 218 & 0.80 & 0.92 & 11.5\% & 12.4\% & 93.6\% & 91.7\% \\
\quad 3 & 219 & 0.69 & 0.75 & 7.8\% & 11.0\% & 89.5\% & 88.6\% \\
\quad 4 & 213 & 0.59 & 0.64 & 5.2\% & 7.5\% & 91.1\% & 87.8\% \\
\quad 5 & 228 & 0.56 & 0.60 & 3.5\% & 6.1\% & 91.2\% & 86.8\% \\
\quad 6 & 187 & 0.52 & 0.56 & 2.1\% & 3.2\% & 92.0\% & 88.8\% \\
\bottomrule
\end{tabular}
\end{table}

\FloatBarrier
\section{Real images}
\label{sec:real}
\begin{figure}[!htbp]
\centering
\includegraphics[width=\textwidth]{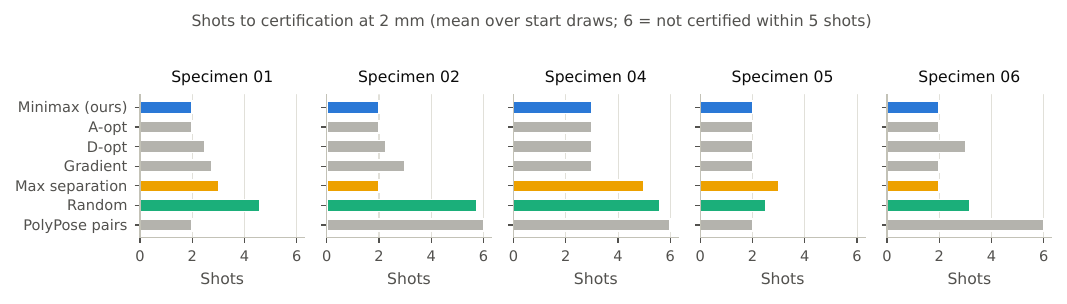}
\caption{Shots to certification at 2\,mm per real specimen and policy (S4; mean over the four starting poses, random views also over seeds; 6: not certified within 5 shots).}
\label{fig:deepfluoro}
\end{figure}

Fig.~\ref{fig:deepfluoro} shows the shots to certification per specimen.

\paragraph{Start set on real images (post hoc)}
The planning studies used a registration with two starts (the previous estimate and the prior mean) and a per-fragment translation search, chosen on synthetic development cases. On the 57 real view sets used for the real-data calibration, this registration left three large hemipelves 35--63\,mm from the truth (maximum \SuppRealRunSetMax\,mm; median error \SuppRealRunSetMedian\,mm). The start set of our earlier calibration experiments (four starts including a translation grid search) recovered all three (maximum \SuppRealGridSetMax\,mm, median \SuppRealGridSetMedian\,mm) at about 1.5 times the registration time, and lowered the real-data calibration factor from \SuppRealRunSetAlpha{} to \SuppRealGridSetAlpha. S4 is reported as specified; for deployment we recommend the larger start set.

\FloatBarrier
\section{Validation of the registration and the calibration}
\label{sec:validation}
\paragraph{Registration across simulators}
Before the planning studies, the registration was checked on \SuppMFourCases{} calibration cases with three clinical views and measured images from four sources: the registration's own renderer (inverse crime), the light simulator without and with noise, and DeepDRR \citep{unberath2018deepdrr} with label-based bone materials. The median fragment mTRE of converged registrations was \SuppMFourInverseMedian, \SuppMFourLightCleanMedian, \SuppMFourLightNoisyMedian{} and \SuppMFourDeepdrrMedian\,mm, respectively, and \SuppMFourDeepdrrSuccess{} of fragments ended within 2\,mm under DeepDRR (median \SuppMFourSeconds\,s per registration).

\paragraph{Calibration factors and transfer to real images}
Without the realistic per-view noise, the calibration factor of the error predictor was \SuppAlphaDeepdrr{} for DeepDRR images and \SuppAlphaLight{} for the light simulator, against \SuppAlphaReal{} on real DeepFluoro images (the four-start registration of our earlier experiments, fitted over all specimens; the planning study's leave-one-specimen-out factors are given in the paper); the DeepDRR factor covered only \SuppRealCovSynthAlpha{} of real fragment registrations. Synthetic simulators thus understate the per-view uncertainty of real fluoroscopy several-fold, which motivated the realistic noise model of the planning studies. With the noise at $\eta=1$, the relative residual variance was \RealismRelVar{} (real images \RealismRealRelVar).

\paragraph{Calibration split under realistic noise}
The planner's calibration used \SuppCalibRegs{} registrations (\SuppCalibBodies{} fragment registrations) of 2--3 views on the 30 calibration cases. \SuppCalibGrossRate{} of fragment registrations were gross failures (by volume: \SuppCalibGrossVolA{} under 10\,cm$^3$, \SuppCalibGrossVolB{} for 10--30, \SuppCalibGrossVolC{} for 30--100 and \SuppCalibGrossVolD{} from 100\,cm$^3$); the flags caught \SuppCalibGrossFlagged{} of them. Cross-validated by case, the calibrated bounds covered \SuppCalibCvCov{} (95\% CI \SuppCalibCvCovCI) of the correctly registered fragments of accepted registrations, with a median bound of \SuppCalibMedianBound\,mm over all fragments of accepted registrations. Of \SuppCalibCert{} registrations certified at 2\,mm, \SuppCalibFalseCert{} were false, which motivated the stable rule.

\FloatBarrier
\section{First run of the untracked setting}
Table~\ref{tab:s2v1} compares the first S2 run, which deviated from the protocol (main paper, Appendix~A), with the re-run reported in the paper. The fix reduced false certifications (\SuppSTwoRunOneFalsePct{} to \SuppSTwoRunTwoFalsePct{} of certified shots) and raised the coverage of unflagged fragments (\SuppSTwoRunOneCov{} to \SuppSTwoRunTwoCov); it turned wrong camera estimates into non-certifying shots rather than repairing them.

\begin{table}[!htbp]
\centering
\caption{S2 (untracked C-arm, 15 cases): first run against the re-run (pre-specified horizon). Gross: fragment registrations more than 5\,mm off; false/cert.: false certifications at 2\,mm over certified shots, all policies; last two columns: minimax's mean shots to certification at 2\,mm and its difference to the fixed protocol.}
\label{tab:s2v1}
\small
\begin{tabular}{lrrrrrl}
\toprule
S2 run & gross & coverage & coverage, & false/cert. & minimax & minimax $-$ fixed \\
 & $>$5\,mm & unflagged & non-gross & (all policies) & certified & (Holm $p$) \\
\midrule
first run (deviation) & 30\% & 65\% & 82\% & 23/105 & 5.53 & $-$0.87 (0.061) \\
re-run (as specified) & 28\% & 76\% & 88\% & 8/86 & 5.47 & $-$1.00 (0.013) \\
\bottomrule
\end{tabular}
\end{table}

\FloatBarrier
\section{Oracle ablations on the pre-specified horizon}
Table~\ref{tab:prespecoracle} repeats the paper's oracle comparison on the pre-specified horizon (budget plus one for unreached endpoints).
\begin{table}[!htbp]
\centering
\caption{Pre-specified horizon: oracle ablations on the first 30 S1 cases (as the paper's oracle table, with shots to certification under the local rule).}
\label{tab:prespecoracle}
\small
\begin{tabular}{lrrrrrlr}
\toprule
 & \multicolumn{3}{c}{Shots to true $\le$ 2\,mm} & \multicolumn{4}{c}{Oracle basin} \\
\cmidrule(lr){2-4}\cmidrule(lr){5-8}
Policy & deployable & oracle start & oracle basin & certified & \multicolumn{3}{c}{minimax $-$ policy (true)} \\
 & & & & shots & diff. & 95\% CI & Holm $p$ \\
\midrule
Minimax (ours) & 4.20 & 4.57 & 3.17 & 4.43 & -- & &  \\
A-optimal & 4.67 & 4.50 & 3.03 & 4.40 & +0.13 & [$-$0.40, 0.63] & 1.000 \\
D-optimal & 3.93 & 4.27 & 2.60 & 4.47 & +0.57 & [0.13, 1.03] & 0.150 \\
E-optimal & 3.83 & 4.13 & 2.70 & 4.97 & +0.47 & [0.00, 0.93] & 0.572 \\
EIG (full) & 4.13 & 4.33 & 3.10 & 4.50 & +0.07 & [$-$0.33, 0.53] & 1.000 \\
EIG (diagonal) & 3.83 & 4.33 & 2.93 & 4.63 & +0.23 & [$-$0.10, 0.60] & 1.000 \\
Gradient heuristic & 3.33 & 4.03 & 2.87 & 5.13 & +0.30 & [$-$0.20, 0.80] & 1.000 \\
Clinical protocol & 4.17 & 4.20 & 3.17 & 5.17 & +0.00 & [$-$0.67, 0.67] & 1.000 \\
Max separation & 5.80 & 5.73 & 4.00 & 5.47 & $-$0.83 & [$-$1.43, $-$0.30] & 0.062 \\
Random & 4.28 & 4.47 & 3.16 & 5.24 & +0.01 & [$-$0.43, 0.50] & 1.000 \\
\bottomrule
\end{tabular}
\end{table}
\FloatBarrier


\begin{thebibliography}{42}
\expandafter\ifx\csname natexlab\endcsname\relax\def\natexlab#1{#1}\fi
\providecommand{\url}[1]{\texttt{#1}}
\providecommand{\href}[2]{#2}
\providecommand{\path}[1]{#1}
\providecommand{\DOIprefix}{doi:}
\providecommand{\ArXivprefix}{arXiv:}
\providecommand{\URLprefix}{URL: }
\providecommand{\Pubmedprefix}{pmid:}
\providecommand{\doi}[1]{\href{http://dx.doi.org/#1}{\path{#1}}}
\providecommand{\Pubmed}[1]{\href{pmid:#1}{\path{#1}}}
\providecommand{\bibinfo}[2]{#2}
\ifx\xfnm\relax \def\xfnm[#1]{\unskip,\space#1}\fi
\bibitem[{Alemdaro{\u{g}}lu et~al.(2014)Alemdaro{\u{g}}lu, Y{\"u}cens, Kara,
  G{\"u}l and Aydo{\u{g}}an}]{alemdaroglu2014pedicle}
\bibinfo{author}{Alemdaro{\u{g}}lu, K.B.}, \bibinfo{author}{Y{\"u}cens, M.},
  \bibinfo{author}{Kara, T.}, \bibinfo{author}{G{\"u}l, D.},
  \bibinfo{author}{Aydo{\u{g}}an, N.H.}, \bibinfo{year}{2014}.
\newblock \bibinfo{title}{Pedicle axis view combined by sacral mapping can
  decrease fluoroscopic shot count in percutaneous iliosacral screw placement}.
\newblock \bibinfo{journal}{Injury} \bibinfo{volume}{45},
  \bibinfo{pages}{1921--1927}.
\newblock \DOIprefix\doi{10.1016/j.injury.2014.10.048}.
\bibitem[{Arrabi et~al.(2025)Arrabi, Jung, Luo, Franssen, Raymond and
  Wshah}]{arrabi2025automated}
\bibinfo{author}{Arrabi, A.}, \bibinfo{author}{Jung, J.H.},
  \bibinfo{author}{Luo, J.}, \bibinfo{author}{Franssen, N.},
  \bibinfo{author}{Raymond, S.}, \bibinfo{author}{Wshah, S.},
  \bibinfo{year}{2025}.
\newblock \bibinfo{title}{Automated {C}-arm positioning via conformal landmark
  localization}, in: \bibinfo{booktitle}{2025 IEEE/CVF International Conference
  on Computer Vision Workshops (ICCVW)}, \bibinfo{publisher}{IEEE}. pp.
  \bibinfo{pages}{4451--4460}.
\newblock \DOIprefix\doi{10.1109/ICCVW69036.2025.00461},
  \href{http://arxiv.org/abs/2510.16160}{{\tt arXiv:2510.16160}}.
\bibitem[{Capostagno et~al.(2019)Capostagno, Stayman, Jacobson, Ehtiati, Weiss
  and Siewerdsen}]{capostagno2019taskdriven}
\bibinfo{author}{Capostagno, S.}, \bibinfo{author}{Stayman, J.W.},
  \bibinfo{author}{Jacobson, M.W.}, \bibinfo{author}{Ehtiati, T.},
  \bibinfo{author}{Weiss, C.R.}, \bibinfo{author}{Siewerdsen, J.H.},
  \bibinfo{year}{2019}.
\newblock \bibinfo{title}{Task-driven source--detector trajectories in
  cone-beam computed tomography: {II}. {Application} to neuroradiology}.
\newblock \bibinfo{journal}{J. Med. Imaging} \bibinfo{volume}{6},
  \bibinfo{pages}{025004}.
\newblock \DOIprefix\doi{10.1117/1.JMI.6.2.025004}.
\bibitem[{Chaloner and Verdinelli(1995)}]{chaloner1995bayesian}
\bibinfo{author}{Chaloner, K.}, \bibinfo{author}{Verdinelli, I.},
  \bibinfo{year}{1995}.
\newblock \bibinfo{title}{{Bayesian} experimental design: A review}.
\newblock \bibinfo{journal}{Stat. Sci.} \bibinfo{volume}{10},
  \bibinfo{pages}{273--304}.
\newblock \DOIprefix\doi{10.1214/ss/1177009939}.
\bibitem[{Danilchenko and Fitzpatrick(2011)}]{danilchenko2011general}
\bibinfo{author}{Danilchenko, A.}, \bibinfo{author}{Fitzpatrick, J.M.},
  \bibinfo{year}{2011}.
\newblock \bibinfo{title}{General approach to first-order error prediction in
  rigid point registration}.
\newblock \bibinfo{journal}{IEEE Trans. Med. Imaging} \bibinfo{volume}{30},
  \bibinfo{pages}{679--693}.
\newblock \DOIprefix\doi{10.1109/TMI.2010.2091513}.
\bibitem[{Fitzpatrick et~al.(1998)Fitzpatrick, West and
  Maurer}]{fitzpatrick1998predicting}
\bibinfo{author}{Fitzpatrick, J.M.}, \bibinfo{author}{West, J.B.},
  \bibinfo{author}{Maurer, C.R.}, \bibinfo{year}{1998}.
\newblock \bibinfo{title}{Predicting error in rigid-body point-based
  registration}.
\newblock \bibinfo{journal}{IEEE Trans. Med. Imaging} \bibinfo{volume}{17},
  \bibinfo{pages}{694--702}.
\newblock \DOIprefix\doi{10.1109/42.736021}.
\bibitem[{Gopalakrishnan et~al.(2026)Gopalakrishnan, Chlorogiannis, Abumoussa,
  Larson, Haouchine, Orbach, Frisken, Dey and
  Golland}]{gopalakrishnan2026rapid}
\bibinfo{author}{Gopalakrishnan, V.}, \bibinfo{author}{Chlorogiannis, D.D.},
  \bibinfo{author}{Abumoussa, A.}, \bibinfo{author}{Larson, A.M.},
  \bibinfo{author}{Haouchine, N.}, \bibinfo{author}{Orbach, D.B.},
  \bibinfo{author}{Frisken, S.}, \bibinfo{author}{Dey, N.},
  \bibinfo{author}{Golland, P.}, \bibinfo{year}{2026}.
\newblock \bibinfo{title}{Rapid patient-specific neural networks for {X}-ray to
  volume registration}.
\newblock \bibinfo{journal}{Nature} \DOIprefix\doi{10.1038/s41586-026-11045-x},
  \href{http://arxiv.org/abs/2503.16309}{{\tt arXiv:2503.16309}}.
\bibitem[{Gopalakrishnan et~al.(2024)Gopalakrishnan, Dey and
  Golland}]{gopalakrishnan2024intraoperative}
\bibinfo{author}{Gopalakrishnan, V.}, \bibinfo{author}{Dey, N.},
  \bibinfo{author}{Golland, P.}, \bibinfo{year}{2024}.
\newblock \bibinfo{title}{Intraoperative {2D}/{3D} image registration via
  differentiable {X}-ray rendering}, in: \bibinfo{booktitle}{2024 IEEE/CVF
  Conference on Computer Vision and Pattern Recognition (CVPR)},
  \bibinfo{publisher}{IEEE}. pp. \bibinfo{pages}{11662--11672}.
\newblock \DOIprefix\doi{10.1109/CVPR52733.2024.01108},
  \href{http://arxiv.org/abs/2312.06358}{{\tt arXiv:2312.06358}}.
\bibitem[{Gopalakrishnan et~al.(2025)Gopalakrishnan, Dey and
  Golland}]{gopalakrishnan2025polypose}
\bibinfo{author}{Gopalakrishnan, V.}, \bibinfo{author}{Dey, N.},
  \bibinfo{author}{Golland, P.}, \bibinfo{year}{2025}.
\newblock \bibinfo{title}{{PolyPose}: Deformable {2D}/{3D} registration via
  polyrigid transformations}, in: \bibinfo{booktitle}{Advances in Neural
  Information Processing Systems}, pp. \bibinfo{pages}{67454--67480}.
\newblock \DOIprefix\doi{10.52202/085713-2026},
  \href{http://arxiv.org/abs/2505.19256}{{\tt arXiv:2505.19256}}.
\bibitem[{Gopalakrishnan and Golland(2023)}]{gopalakrishnan2023fast}
\bibinfo{author}{Gopalakrishnan, V.}, \bibinfo{author}{Golland, P.},
  \bibinfo{year}{2023}.
\newblock \bibinfo{title}{Fast auto-differentiable digitally reconstructed
  radiographs for solving inverse problems in intraoperative imaging}, in:
  \bibinfo{booktitle}{Clinical Image-Based Procedures: 11th Workshop, {CLIP}
  2022, Held in Conjunction with {MICCAI} 2022}, \bibinfo{publisher}{Springer
  Nature Switzerland}. pp. \bibinfo{pages}{1--11}.
\newblock \DOIprefix\doi{10.1007/978-3-031-23179-7_1},
  \href{http://arxiv.org/abs/2208.12737}{{\tt arXiv:2208.12737}}.
\bibitem[{Grupp et~al.(2020a)Grupp, Unberath, Gao, Hegeman, Murphy, Alexander,
  Otake, McArthur, Armand and Taylor}]{grupp2020deepfluorodata}
\bibinfo{author}{Grupp, R.}, \bibinfo{author}{Unberath, M.},
  \bibinfo{author}{Gao, C.}, \bibinfo{author}{Hegeman, R.},
  \bibinfo{author}{Murphy, R.}, \bibinfo{author}{Alexander, C.},
  \bibinfo{author}{Otake, Y.}, \bibinfo{author}{McArthur, B.},
  \bibinfo{author}{Armand, M.}, \bibinfo{author}{Taylor, R.},
  \bibinfo{year}{2020}a.
\newblock \bibinfo{title}{Data and code associated with the publication:
  Automatic annotation of hip anatomy in fluoroscopy for robust and efficient
  {2D}/{3D} registration}.
\newblock \bibinfo{howpublished}{[dataset]. Johns Hopkins Research Data
  Repository, V1}.
\newblock \DOIprefix\doi{10.7281/T1/IFSXNV}.
\bibitem[{Grupp et~al.(2020b)Grupp, Hegeman, Murphy, Alexander, Otake,
  McArthur, Armand and Taylor}]{grupp2020pose}
\bibinfo{author}{Grupp, R.B.}, \bibinfo{author}{Hegeman, R.A.},
  \bibinfo{author}{Murphy, R.J.}, \bibinfo{author}{Alexander, C.P.},
  \bibinfo{author}{Otake, Y.}, \bibinfo{author}{McArthur, B.A.},
  \bibinfo{author}{Armand, M.}, \bibinfo{author}{Taylor, R.H.},
  \bibinfo{year}{2020}b.
\newblock \bibinfo{title}{Pose estimation of periacetabular osteotomy fragments
  with intraoperative {X}-ray navigation}.
\newblock \bibinfo{journal}{IEEE Trans. Biomed. Eng.} \bibinfo{volume}{67},
  \bibinfo{pages}{441--452}.
\newblock \DOIprefix\doi{10.1109/TBME.2019.2915165}.
\bibitem[{Grupp et~al.(2020c)Grupp, Unberath, Gao, Hegeman, Murphy, Alexander,
  Otake, McArthur, Armand and Taylor}]{grupp2020automatic}
\bibinfo{author}{Grupp, R.B.}, \bibinfo{author}{Unberath, M.},
  \bibinfo{author}{Gao, C.}, \bibinfo{author}{Hegeman, R.A.},
  \bibinfo{author}{Murphy, R.J.}, \bibinfo{author}{Alexander, C.P.},
  \bibinfo{author}{Otake, Y.}, \bibinfo{author}{McArthur, B.A.},
  \bibinfo{author}{Armand, M.}, \bibinfo{author}{Taylor, R.H.},
  \bibinfo{year}{2020}c.
\newblock \bibinfo{title}{Automatic annotation of hip anatomy in fluoroscopy
  for robust and efficient {2D}/{3D} registration}.
\newblock \bibinfo{journal}{Int. J. Comput. Assist. Radiol. Surg.}
  \bibinfo{volume}{15}, \bibinfo{pages}{759--769}.
\newblock \DOIprefix\doi{10.1007/s11548-020-02162-7}.
\bibitem[{Han et~al.(2020)Han, Uneri, Ketcha, Vijayan, Sheth, Wu, Vagdargi,
  Vogt, Kleinszig, Osgood and Siewerdsen}]{han2020multibody}
\bibinfo{author}{Han, R.}, \bibinfo{author}{Uneri, A.},
  \bibinfo{author}{Ketcha, M.}, \bibinfo{author}{Vijayan, R.},
  \bibinfo{author}{Sheth, N.}, \bibinfo{author}{Wu, P.},
  \bibinfo{author}{Vagdargi, P.}, \bibinfo{author}{Vogt, S.},
  \bibinfo{author}{Kleinszig, G.}, \bibinfo{author}{Osgood, G.M.},
  \bibinfo{author}{Siewerdsen, J.H.}, \bibinfo{year}{2020}.
\newblock \bibinfo{title}{Multi-body {3D}--{2D} registration for image-guided
  reduction of pelvic dislocation in orthopaedic trauma surgery}.
\newblock \bibinfo{journal}{Phys. Med. Biol.} \bibinfo{volume}{65},
  \bibinfo{pages}{135009}.
\newblock \DOIprefix\doi{10.1088/1361-6560/ab843c}.
\bibitem[{Han et~al.(2021)Han, Uneri, Vijayan, Wu, Vagdargi, Sheth, Vogt,
  Kleinszig, Osgood and Siewerdsen}]{han2021fracture}
\bibinfo{author}{Han, R.}, \bibinfo{author}{Uneri, A.},
  \bibinfo{author}{Vijayan, R.C.}, \bibinfo{author}{Wu, P.},
  \bibinfo{author}{Vagdargi, P.}, \bibinfo{author}{Sheth, N.},
  \bibinfo{author}{Vogt, S.}, \bibinfo{author}{Kleinszig, G.},
  \bibinfo{author}{Osgood, G.M.}, \bibinfo{author}{Siewerdsen, J.H.},
  \bibinfo{year}{2021}.
\newblock \bibinfo{title}{Fracture reduction planning and guidance in
  orthopaedic trauma surgery via multi-body image registration}.
\newblock \bibinfo{journal}{Med. Image Anal.} \bibinfo{volume}{68},
  \bibinfo{pages}{101917}.
\newblock \DOIprefix\doi{10.1016/j.media.2020.101917}.
\bibitem[{Heiselman et~al.(2022)Heiselman, Jarnagin and
  Miga}]{heiselman2022registration}
\bibinfo{author}{Heiselman, J.S.}, \bibinfo{author}{Jarnagin, W.R.},
  \bibinfo{author}{Miga, M.I.}, \bibinfo{year}{2022}.
\newblock \bibinfo{title}{Registration uncertainty in deforming organs: a novel
  approach for ensuring navigational confidence during image-guided
  procedures}, in: \bibinfo{booktitle}{Medical Imaging 2022: Image-Guided
  Procedures, Robotic Interventions, and Modeling}, p.
  \bibinfo{pages}{120340Z}.
\newblock \DOIprefix\doi{10.1117/12.2612938}.
\bibitem[{Hu et~al.(2016)Hu, Bonmati, Gibson, Hipwell, Hawkes, Bandula, Pereira
  and Barratt}]{hu2016registration}
\bibinfo{author}{Hu, Y.}, \bibinfo{author}{Bonmati, E.},
  \bibinfo{author}{Gibson, E.}, \bibinfo{author}{Hipwell, J.H.},
  \bibinfo{author}{Hawkes, D.J.}, \bibinfo{author}{Bandula, S.},
  \bibinfo{author}{Pereira, S.P.}, \bibinfo{author}{Barratt, D.C.},
  \bibinfo{year}{2016}.
\newblock \bibinfo{title}{{2D}-{3D} registration accuracy estimation for
  optimised planning of image-guided pancreatobiliary interventions}, in:
  \bibinfo{booktitle}{Medical Image Computing and Computer-Assisted
  Intervention -- {MICCAI} 2016}, \bibinfo{publisher}{Springer International
  Publishing}. pp. \bibinfo{pages}{516--524}.
\newblock \DOIprefix\doi{10.1007/978-3-319-46720-7_60}.
\bibitem[{Jiang et~al.(2024)Jiang, Lei and Daniilidis}]{jiang2024fisherrf}
\bibinfo{author}{Jiang, W.}, \bibinfo{author}{Lei, B.},
  \bibinfo{author}{Daniilidis, K.}, \bibinfo{year}{2024}.
\newblock \bibinfo{title}{{FisherRF}: Active view selection and mapping with
  radiance fields using {Fisher} information}, in: \bibinfo{booktitle}{Computer
  Vision -- {ECCV} 2024}, \bibinfo{publisher}{Springer Nature Switzerland}. pp.
  \bibinfo{pages}{422--440}.
\newblock \DOIprefix\doi{10.1007/978-3-031-72624-8_24},
  \href{http://arxiv.org/abs/2311.17874}{{\tt arXiv:2311.17874}}.
\bibitem[{Kalal et~al.(2010)Kalal, Mikolajczyk and Matas}]{kalal2010forward}
\bibinfo{author}{Kalal, Z.}, \bibinfo{author}{Mikolajczyk, K.},
  \bibinfo{author}{Matas, J.}, \bibinfo{year}{2010}.
\newblock \bibinfo{title}{Forward-backward error: Automatic detection of
  tracking failures}, in: \bibinfo{booktitle}{Proceedings of the 20th
  International Conference on Pattern Recognition (ICPR)}, pp.
  \bibinfo{pages}{2756--2759}.
\newblock \DOIprefix\doi{10.1109/ICPR.2010.675}.
\bibitem[{Ketcha et~al.(2017)Ketcha, De~Silva, Han, Uneri, Goerres, Jacobson,
  Vogt, Kleinszig and Siewerdsen}]{ketcha2017effects}
\bibinfo{author}{Ketcha, M.D.}, \bibinfo{author}{De~Silva, T.},
  \bibinfo{author}{Han, R.}, \bibinfo{author}{Uneri, A.},
  \bibinfo{author}{Goerres, J.}, \bibinfo{author}{Jacobson, M.W.},
  \bibinfo{author}{Vogt, S.}, \bibinfo{author}{Kleinszig, G.},
  \bibinfo{author}{Siewerdsen, J.H.}, \bibinfo{year}{2017}.
\newblock \bibinfo{title}{Effects of image quality on the fundamental limits of
  image registration accuracy}.
\newblock \bibinfo{journal}{IEEE Trans. Med. Imaging} \bibinfo{volume}{36},
  \bibinfo{pages}{1997--2009}.
\newblock \DOIprefix\doi{10.1109/TMI.2017.2725644}.
\bibitem[{Killeen et~al.(2023)Killeen, Gao, Oguine, Darcy, Armand, Taylor,
  Osgood and Unberath}]{killeen2023autonomous}
\bibinfo{author}{Killeen, B.D.}, \bibinfo{author}{Gao, C.},
  \bibinfo{author}{Oguine, K.J.}, \bibinfo{author}{Darcy, S.},
  \bibinfo{author}{Armand, M.}, \bibinfo{author}{Taylor, R.H.},
  \bibinfo{author}{Osgood, G.}, \bibinfo{author}{Unberath, M.},
  \bibinfo{year}{2023}.
\newblock \bibinfo{title}{An autonomous {X}-ray image acquisition and
  interpretation system for assisting percutaneous pelvic fracture fixation}.
\newblock \bibinfo{journal}{Int. J. Comput. Assist. Radiol. Surg.}
  \bibinfo{volume}{18}, \bibinfo{pages}{1201--1208}.
\newblock \DOIprefix\doi{10.1007/s11548-023-02941-y}.
\bibitem[{Kybic(2010)}]{kybic2010bootstrap}
\bibinfo{author}{Kybic, J.}, \bibinfo{year}{2010}.
\newblock \bibinfo{title}{Bootstrap resampling for image registration
  uncertainty estimation without ground truth}.
\newblock \bibinfo{journal}{IEEE Trans. Image Process.} \bibinfo{volume}{19},
  \bibinfo{pages}{64--73}.
\newblock \DOIprefix\doi{10.1109/TIP.2009.2030955}.
\bibitem[{Markelj et~al.(2012)Markelj, Toma{\v{z}}evi{\v{c}}, Likar and
  Pernu{\v{s}}}]{markelj2012review}
\bibinfo{author}{Markelj, P.}, \bibinfo{author}{Toma{\v{z}}evi{\v{c}}, D.},
  \bibinfo{author}{Likar, B.}, \bibinfo{author}{Pernu{\v{s}}, F.},
  \bibinfo{year}{2012}.
\newblock \bibinfo{title}{A review of {3D}/{2D} registration methods for
  image-guided interventions}.
\newblock \bibinfo{journal}{Med. Image Anal.} \bibinfo{volume}{16},
  \bibinfo{pages}{642--661}.
\newblock \DOIprefix\doi{10.1016/j.media.2010.03.005}.
\bibitem[{Matta(1996)}]{matta1996fractures}
\bibinfo{author}{Matta, J.M.}, \bibinfo{year}{1996}.
\newblock \bibinfo{title}{Fractures of the acetabulum: accuracy of reduction
  and clinical results in patients managed operatively within three weeks after
  the injury}.
\newblock \bibinfo{journal}{J. Bone Joint Surg. Am.} \bibinfo{volume}{78},
  \bibinfo{pages}{1632--1645}.
\newblock \DOIprefix\doi{10.2106/00004623-199611000-00002}.
\bibitem[{Matta and {Tornetta III}(1996)}]{matta1996pelvic}
\bibinfo{author}{Matta, J.M.}, \bibinfo{author}{{Tornetta III}, P.},
  \bibinfo{year}{1996}.
\newblock \bibinfo{title}{Internal fixation of unstable pelvic ring injuries}.
\newblock \bibinfo{journal}{Clin. Orthop. Relat. Res.} \bibinfo{volume}{329},
  \bibinfo{pages}{129--140}.
\newblock \DOIprefix\doi{10.1097/00003086-199608000-00016}.
\bibitem[{Prost et~al.(2024)Prost, Taday, Beyersdorf, Latz, Windolf, Scheyerer
  and Konieczny}]{prost2024navigation}
\bibinfo{author}{Prost, M.}, \bibinfo{author}{Taday, R.},
  \bibinfo{author}{Beyersdorf, C.C.P.}, \bibinfo{author}{Latz, D.},
  \bibinfo{author}{Windolf, J.}, \bibinfo{author}{Scheyerer, M.J.},
  \bibinfo{author}{Konieczny, M.R.}, \bibinfo{year}{2024}.
\newblock \bibinfo{title}{Navigation versus fluoroscopy in minimalinvasive
  iliosacral screw placement}.
\newblock \bibinfo{journal}{J. Orthop. Surg. Res.} \bibinfo{volume}{19},
  \bibinfo{pages}{185}.
\newblock \DOIprefix\doi{10.1186/s13018-024-04669-0}.
\bibitem[{Pukelsheim(2006)}]{pukelsheim2006optimal}
\bibinfo{author}{Pukelsheim, F.}, \bibinfo{year}{2006}.
\newblock \bibinfo{title}{Optimal Design of Experiments}.
  volume~\bibinfo{volume}{50} of \textit{\bibinfo{series}{Classics in Applied
  Mathematics}}.
\newblock \bibinfo{publisher}{Society for Industrial and Applied Mathematics},
  \bibinfo{address}{Philadelphia, PA}.
\newblock \DOIprefix\doi{10.1137/1.9780898719109}.
\bibitem[{Risholm et~al.(2013)Risholm, Janoos, Norton, Golby and {Wells
  III}}]{risholm2013bayesian}
\bibinfo{author}{Risholm, P.}, \bibinfo{author}{Janoos, F.},
  \bibinfo{author}{Norton, I.}, \bibinfo{author}{Golby, A.J.},
  \bibinfo{author}{{Wells III}, W.M.}, \bibinfo{year}{2013}.
\newblock \bibinfo{title}{{B}ayesian characterization of uncertainty in
  intra-subject non-rigid registration}.
\newblock \bibinfo{journal}{Med. Image Anal.} \bibinfo{volume}{17},
  \bibinfo{pages}{538--555}.
\newblock \DOIprefix\doi{10.1016/j.media.2013.03.002}.
\bibitem[{Robinson and Milanfar(2004)}]{robinson2004fundamental}
\bibinfo{author}{Robinson, D.}, \bibinfo{author}{Milanfar, P.},
  \bibinfo{year}{2004}.
\newblock \bibinfo{title}{Fundamental performance limits in image
  registration}.
\newblock \bibinfo{journal}{IEEE Trans. Image Process.} \bibinfo{volume}{13},
  \bibinfo{pages}{1185--1199}.
\newblock \DOIprefix\doi{10.1109/TIP.2004.832923}.
\bibitem[{Sang et~al.(2026)Sang, Liu, Yibulayimu, Wang, Killeen, Liu, Ku,
  Johannsen, Gotkowski, Zenk, Maier-Hein, Isensee, Yue, Wang, Yu, Pan, He,
  Liang, Liu, Fan, Jurgas, Skalski, Ma, Yang, P{\l}otka, Litka, Zhu, Song,
  Unberath, Armand, Ruan, Zhou, Cao, Zhao, Wu and Wang}]{sang2026benchmark}
\bibinfo{author}{Sang, Y.}, \bibinfo{author}{Liu, Y.},
  \bibinfo{author}{Yibulayimu, S.}, \bibinfo{author}{Wang, Y.},
  \bibinfo{author}{Killeen, B.D.}, \bibinfo{author}{Liu, M.},
  \bibinfo{author}{Ku, P.C.}, \bibinfo{author}{Johannsen, O.},
  \bibinfo{author}{Gotkowski, K.}, \bibinfo{author}{Zenk, M.},
  \bibinfo{author}{Maier-Hein, K.}, \bibinfo{author}{Isensee, F.},
  \bibinfo{author}{Yue, P.}, \bibinfo{author}{Wang, Y.}, \bibinfo{author}{Yu,
  H.}, \bibinfo{author}{Pan, Z.}, \bibinfo{author}{He, Y.},
  \bibinfo{author}{Liang, X.}, \bibinfo{author}{Liu, D.}, \bibinfo{author}{Fan,
  F.}, \bibinfo{author}{Jurgas, A.}, \bibinfo{author}{Skalski, A.},
  \bibinfo{author}{Ma, Y.}, \bibinfo{author}{Yang, J.},
  \bibinfo{author}{P{\l}otka, S.}, \bibinfo{author}{Litka, R.},
  \bibinfo{author}{Zhu, G.}, \bibinfo{author}{Song, Y.},
  \bibinfo{author}{Unberath, M.}, \bibinfo{author}{Armand, M.},
  \bibinfo{author}{Ruan, D.}, \bibinfo{author}{Zhou, S.K.},
  \bibinfo{author}{Cao, Q.}, \bibinfo{author}{Zhao, C.}, \bibinfo{author}{Wu,
  X.}, \bibinfo{author}{Wang, Y.}, \bibinfo{year}{2026}.
\newblock \bibinfo{title}{Benchmark of segmentation techniques for pelvic
  fracture in {CT} and {X}-ray: Summary of the {PENGWIN} 2024 challenge}.
\newblock \bibinfo{journal}{IEEE Trans. Med. Imaging} \bibinfo{volume}{45},
  \bibinfo{pages}{2212--2228}.
\newblock \DOIprefix\doi{10.1109/TMI.2025.3650126},
  \href{http://arxiv.org/abs/2504.02382}{{\tt arXiv:2504.02382}}.
\bibitem[{Sang et~al.(2024)Sang, Liu, Yibulayimu, Zhu, Wang, Killeen, Liu, Ku,
  Armand, Unberath, Wu and Zhao}]{sang2024pengwin}
\bibinfo{author}{Sang, Y.}, \bibinfo{author}{Liu, Y.},
  \bibinfo{author}{Yibulayimu, S.}, \bibinfo{author}{Zhu, G.},
  \bibinfo{author}{Wang, Y.}, \bibinfo{author}{Killeen, B.},
  \bibinfo{author}{Liu, M.}, \bibinfo{author}{Ku, P.C.},
  \bibinfo{author}{Armand, M.}, \bibinfo{author}{Unberath, M.},
  \bibinfo{author}{Wu, X.}, \bibinfo{author}{Zhao, C.}, \bibinfo{year}{2024}.
\newblock \bibinfo{title}{{PENGWIN} task 1: Pelvic fracture segmentation on
  {CT}}.
\newblock \bibinfo{howpublished}{[dataset]. Zenodo}.
\newblock \DOIprefix\doi{10.5281/zenodo.10927452}.
\bibitem[{Smith and Winkler(2006)}]{smith2006optimizer}
\bibinfo{author}{Smith, J.E.}, \bibinfo{author}{Winkler, R.L.},
  \bibinfo{year}{2006}.
\newblock \bibinfo{title}{The optimizer's curse: Skepticism and postdecision
  surprise in decision analysis}.
\newblock \bibinfo{journal}{Manag. Sci.} \bibinfo{volume}{52},
  \bibinfo{pages}{311--322}.
\newblock \DOIprefix\doi{10.1287/mnsc.1050.0451}.
\bibitem[{Sokooti et~al.(2019)Sokooti, Saygili, Glocker, Lelieveldt and
  Staring}]{sokooti2019quantitative}
\bibinfo{author}{Sokooti, H.}, \bibinfo{author}{Saygili, G.},
  \bibinfo{author}{Glocker, B.}, \bibinfo{author}{Lelieveldt, B.P.F.},
  \bibinfo{author}{Staring, M.}, \bibinfo{year}{2019}.
\newblock \bibinfo{title}{Quantitative error prediction of medical image
  registration using regression forests}.
\newblock \bibinfo{journal}{Med. Image Anal.} \bibinfo{volume}{56},
  \bibinfo{pages}{110--121}.
\newblock \DOIprefix\doi{10.1016/j.media.2019.05.005}.
\bibitem[{Stayman et~al.(2019)Stayman, Capostagno, Gang and
  Siewerdsen}]{stayman2019taskdriven}
\bibinfo{author}{Stayman, J.W.}, \bibinfo{author}{Capostagno, S.},
  \bibinfo{author}{Gang, G.J.}, \bibinfo{author}{Siewerdsen, J.H.},
  \bibinfo{year}{2019}.
\newblock \bibinfo{title}{Task-driven source--detector trajectories in
  cone-beam computed tomography: {I}. {Theory} and methods}.
\newblock \bibinfo{journal}{J. Med. Imaging} \bibinfo{volume}{6},
  \bibinfo{pages}{025002}.
\newblock \DOIprefix\doi{10.1117/1.JMI.6.2.025002}.
\bibitem[{Syring and Martin(2019)}]{syring2019calibrating}
\bibinfo{author}{Syring, N.}, \bibinfo{author}{Martin, R.},
  \bibinfo{year}{2019}.
\newblock \bibinfo{title}{Calibrating general posterior credible regions}.
\newblock \bibinfo{journal}{Biometrika} \bibinfo{volume}{106},
  \bibinfo{pages}{479--486}.
\newblock \DOIprefix\doi{10.1093/biomet/asy054}.
\bibitem[{Thusini et~al.(2020)Thusini, Reyneke, Aellen, Forster, Fouefack,
  Nbonsou~Tegang, Vetter, Douglas and Mutsvangwa}]{thusini2020uncertainty}
\bibinfo{author}{Thusini, X.O.}, \bibinfo{author}{Reyneke, C.J.F.},
  \bibinfo{author}{Aellen, J.}, \bibinfo{author}{Forster, A.},
  \bibinfo{author}{Fouefack, J.R.}, \bibinfo{author}{Nbonsou~Tegang, N.H.},
  \bibinfo{author}{Vetter, T.}, \bibinfo{author}{Douglas, T.S.},
  \bibinfo{author}{Mutsvangwa, T.E.M.}, \bibinfo{year}{2020}.
\newblock \bibinfo{title}{Uncertainty reduction in contour-based {3D}/{2D}
  registration of bone surfaces}, in: \bibinfo{booktitle}{Shape in Medical
  Imaging ({ShapeMI} 2020, held in conjunction with {MICCAI} 2020)},
  \bibinfo{publisher}{Springer International Publishing}. pp.
  \bibinfo{pages}{18--29}.
\newblock \DOIprefix\doi{10.1007/978-3-030-61056-2_2}.
\bibitem[{Unberath et~al.(2018)Unberath, Zaech, Lee, Bier, Fotouhi, Armand and
  Navab}]{unberath2018deepdrr}
\bibinfo{author}{Unberath, M.}, \bibinfo{author}{Zaech, J.N.},
  \bibinfo{author}{Lee, S.C.}, \bibinfo{author}{Bier, B.},
  \bibinfo{author}{Fotouhi, J.}, \bibinfo{author}{Armand, M.},
  \bibinfo{author}{Navab, N.}, \bibinfo{year}{2018}.
\newblock \bibinfo{title}{{DeepDRR} -- a catalyst for machine learning in
  fluoroscopy-guided procedures}, in: \bibinfo{booktitle}{Medical Image
  Computing and Computer Assisted Intervention -- {MICCAI} 2018},
  \bibinfo{publisher}{Springer International Publishing}. pp.
  \bibinfo{pages}{98--106}.
\newblock \DOIprefix\doi{10.1007/978-3-030-00937-3_12}.
\bibitem[{Uneri et~al.(2014)Uneri, Otake, Wang, Kleinszig, Vogt, Khanna and
  Siewerdsen}]{uneri2014registration}
\bibinfo{author}{Uneri, A.}, \bibinfo{author}{Otake, Y.},
  \bibinfo{author}{Wang, A.S.}, \bibinfo{author}{Kleinszig, G.},
  \bibinfo{author}{Vogt, S.}, \bibinfo{author}{Khanna, A.J.},
  \bibinfo{author}{Siewerdsen, J.H.}, \bibinfo{year}{2014}.
\newblock \bibinfo{title}{{3D}--{2D} registration for surgical guidance: effect
  of projection view angles on registration accuracy}.
\newblock \bibinfo{journal}{Phys. Med. Biol.} \bibinfo{volume}{59},
  \bibinfo{pages}{271--287}.
\newblock \DOIprefix\doi{10.1088/0031-9155/59/2/271}.
\bibitem[{Varnavas et~al.(2015)Varnavas, Carrell and
  Penney}]{varnavas2015fully}
\bibinfo{author}{Varnavas, A.}, \bibinfo{author}{Carrell, T.},
  \bibinfo{author}{Penney, G.}, \bibinfo{year}{2015}.
\newblock \bibinfo{title}{Fully automated {2D}--{3D} registration and
  verification}.
\newblock \bibinfo{journal}{Med. Image Anal.} \bibinfo{volume}{26},
  \bibinfo{pages}{108--119}.
\newblock \DOIprefix\doi{10.1016/j.media.2015.08.005}.
\bibitem[{Verbeek et~al.(2017)Verbeek, van~der List, Villa, Wellman and
  Helfet}]{verbeek2017ct}
\bibinfo{author}{Verbeek, D.O.}, \bibinfo{author}{van~der List, J.P.},
  \bibinfo{author}{Villa, J.C.}, \bibinfo{author}{Wellman, D.S.},
  \bibinfo{author}{Helfet, D.L.}, \bibinfo{year}{2017}.
\newblock \bibinfo{title}{Postoperative {CT} is superior for acetabular
  fracture reduction assessment and reliably predicts hip survivorship}.
\newblock \bibinfo{journal}{J. Bone Joint Surg. Am.} \bibinfo{volume}{99},
  \bibinfo{pages}{1745--1752}.
\newblock \DOIprefix\doi{10.2106/JBJS.16.01446}.
\bibitem[{Vijayan et~al.(2024)Vijayan, Sheth, Wei, Venkataraman, Ghanem,
  Shafiq, Siewerdsen, Zbijewski, Li, Cleary and Uneri}]{vijayan2024robot}
\bibinfo{author}{Vijayan, R.C.}, \bibinfo{author}{Sheth, N.M.},
  \bibinfo{author}{Wei, J.}, \bibinfo{author}{Venkataraman, K.},
  \bibinfo{author}{Ghanem, D.}, \bibinfo{author}{Shafiq, B.},
  \bibinfo{author}{Siewerdsen, J.H.}, \bibinfo{author}{Zbijewski, W.},
  \bibinfo{author}{Li, G.}, \bibinfo{author}{Cleary, K.},
  \bibinfo{author}{Uneri, A.}, \bibinfo{year}{2024}.
\newblock \bibinfo{title}{Robot-assisted reduction of the ankle joint via
  multi-body {3D}--{2D} image registration}.
\newblock \bibinfo{journal}{IEEE Trans. Med. Robot. Bionics}
  \bibinfo{volume}{6}, \bibinfo{pages}{1591--1602}.
\newblock \DOIprefix\doi{10.1109/TMRB.2024.3464095}.
\bibitem[{Wu et~al.(2026)Wu, Zha, Cao, Li, Cai and Liu}]{wu2026active}
\bibinfo{author}{Wu, Y.}, \bibinfo{author}{Zha, R.}, \bibinfo{author}{Cao, W.},
  \bibinfo{author}{Li, Y.}, \bibinfo{author}{Cai, Y.}, \bibinfo{author}{Liu,
  Y.}, \bibinfo{year}{2026}.
\newblock \bibinfo{title}{Active view selection with perturbed {G}aussian
  ensemble for tomographic reconstruction}, in: \bibinfo{booktitle}{Computer
  Vision -- {ECCV} 2026}, \bibinfo{publisher}{Springer Nature Switzerland}. pp.
  \bibinfo{pages}{405--424}.
\newblock \DOIprefix\doi{10.1007/978-3-032-37235-2_24},
  \href{http://arxiv.org/abs/2603.06852}{{\tt arXiv:2603.06852}}.

\end{thebibliography}
\end{document}